\documentclass[12pt]{iopart}
\usepackage{color,hyperref}
\usepackage{graphicx}
\DeclareUnicodeCharacter{2009}{\,} 
\usepackage{tabularx,ragged2e,booktabs,caption,multirow}
\usepackage{iopams}
\usepackage{bm}
\begin{document}
	\title[Towards improved loading, cooling, and trapping of molecules in magneto-optical traps]{Towards improved loading, cooling, and trapping of molecules in magneto-optical traps}
	
	\author{T K Langin, D DeMille}
	
	\address{University of Chicago, Department of Physics, Chicago, IL, 60637}
	\ead{tlangin@uchicago.edu}
	
	\begin{abstract}
		Recent experiments have demonstrated direct cooling and trapping of diatomic and triatomic molecules in magneto-optical traps (MOTs). However, even the best molecular MOTs to date still have density $10^{-5}$ times smaller than in typical atomic MOTs.  The main limiting factors are: (i) inefficiencies in slowing molecules to velocities low enough to be captured by the MOT, (ii) low MOT capture velocities, and (iii) limits on density within the MOT resulting from sub-Doppler heating~[J. A. Devlin and M. R. Tarbutt, Phys. Rev. A \textbf{90}, 063415 (2018)].  All of these are consequences of the need to drive `Type-II' optical cycling transitions, where dark states appear in Zeeman sublevels, in order to avoid rotational branching.  We present simulations demonstrating ways to mitigate each of these limitations.  This should pave the way towards loading molecules into conservative traps with sufficiently high density and number to evaporatively cool them to quantum degeneracy.
	\end{abstract}
	\maketitle
	
	\section{Introduction}
	Over the last decade, considerable progress has been made in direct laser-cooling and trapping of diatomic~\cite{bmd2014,aad2017,twt2017,cdy2018} and even triatomic~\cite{vhd2022} molecules in magneto-optical traps (MOTs).  This has increased the variety of molecular gases that can be cooled and trapped at ultracold temperatures beyond those that can be assembled from two laser-coolable atoms~\cite{noy2008,kgc2014,gzw2016,pwz2015}.  However, while particle numbers of $N_{MOT}\sim10^{10}$, densities $n_{MOT}\sim10^{12}$\,cm$^{-3}$, temperatures $T_{MOT}\sim 30\,\mu$K, and phase-space densities $\phi_{MOT}\sim 10^{-5}$ are achievable in alkali-atom MOTs~\cite{rcc2011,rwf2013,jdt2018}, the best molecular MOTs to date have $N_{MOT}\sim 10^{6}$, $n_{MOT}\sim10^{7}$\,cm$^{-3}$, $T_{MOT}\sim 400\,\mu$K, and $\phi_{MOT}\sim 10^{-11}$~\cite{aad2017}.  In this paper, we report on simulations of techniques for improving the current state-of-the-art in molecule MOTs.  
	
	Denser MOTs with higher molecule number would be especially beneficial for experiments seeking to subsequently load into optical dipole traps (ODTs)~\cite{cad2018,ljd2021,wbd2021,lhc2022,hvd2022} for the purpose of evaporative cooling to quantum degeneracy.  Such cooling has recently been demonstrated in ultracold assembled bi-alkali molecules~\cite{vmy2020,lty2021,sbl2022}, but is yet to be achieved for directly laser-cooled molecules.  ODTs of directly cooled molecules, loaded from MOTs, are thus far limited to molecule numbers of $N_{ODT}\sim 10^{3}$ and initial phase space densities of $\phi_{ODT}\sim 10^{-6}$~\cite{cad2018,ljd2021,wbd2021}.  Both of these are insufficient for evaporative cooling, which `sacrifices' energetic molecules to reach degeneracy ($\phi\sim 1$).  Increasing $N_{ODT}$ would (assuming the same ODT temperature) increase the initial phase space density while also allowing for faster evaporative cooling, as the rate for the necessary rethermalization scales linearly with $N_{ODT}$.  This is particularly important for molecular evaporative cooling, as light-induced and chemically-reactive inelastic collisions~\cite{iju2010,cad2020}, losses due to phase-noise in microwaves used to shield from said collisions~\cite{abd2021,sbl2022}, and vibrational transitions induced by black body radiation~\cite{vdu2007} have, to date, combined to limit the ODT lifetime to $\sim 1$\,s.

	Here, we present simulations of new techniques for improving the MOT molecule number and density.  The simulations are based on numerically solving the Optical Bloch Equations (OBEs) for the combination of lasers used for molecular cooling and/or slowing (Sec.~\ref{sec:simTechniques})~\cite{dta2016,dta2018}.  We find in these simulations that the number of trapped molecules in the MOT can be improved by increasing the capture velocity of the MOT (Sec.~\ref{sec:bichromTrapping}) and/or by increasing the flux of slowed molecules reaching the MOT (Sec.~\ref{sec:slowingImprovements}).  The density can then be further enhanced, and the MOT temperature lowered, by use of a blue detuned MOT, in which molecules undergo gray-molasses cooling while trapped, as has been recently demonstrated in atoms~\cite{jdt2018} (Sec.~\ref{sec:blueMOT}).  We also discuss special considerations for simulating MOTs of molecules with nearly degenerate hyperfine levels in the ground state--- examples include SrOH, CaOH, and MgF (Sec.~\ref{sec:ZeeMOT})--- and demonstrate that these techniques will work for them as well.

	\section{Overview: Laser Slowing, Cooling, and Trapping of $^{2}\Sigma$ molecules}
	\label{sec:level}
	
	In this paper, we focus on molecules with a X$^{2}\Sigma$ ground state comprised of an alkaline-earth metal with a ligand (here, either F or OH); to date, molecules with this ground state are the only ones that have been slowed and loaded into MOTs, although there are several groups working towards cooling and trapping for systems with a X$^{1}\Sigma$ ground state~\cite{hpd2012,tmm2019,dwk2021,shm2021}.  These molecules have nearly diagonal matrices of Franck-Condon factors for electronic excitations to the A$^{2}\Pi_{1/2}$ and/or B$^{2}\Sigma$ electronic excited states; this Franck-Condon structure is necessary to limit the number of vibrational repump lasers required for laser-cooling~\cite{dir2004}.  To avoid rotational branching, laser-cooling is performed by coupling the $|\textrm{X}^{2}\Sigma,N=1\rangle$ state to either the $|\textrm{A}^{2}\Pi_{1/2},J'=1/2\rangle$ or $|\textrm{B}^{2}\Sigma,N'=0\rangle$ states, where $\bm{J}$ is the total angular momentum without nuclear spin and $\bm{N}=\bm{J}-\bm{S}$ is the rotational angular momentum where $\bm{S}$ is the electronic spin~\cite{sbd2010}.  Throughout the paper, unless otherwise noted, we implicitly assume states are sublevels of these specific rotational states when referring to a given electronic state; quantum numbers associated with an electronic excited state are indicated by primes.

\begin{figure}
		\centering
		\includegraphics{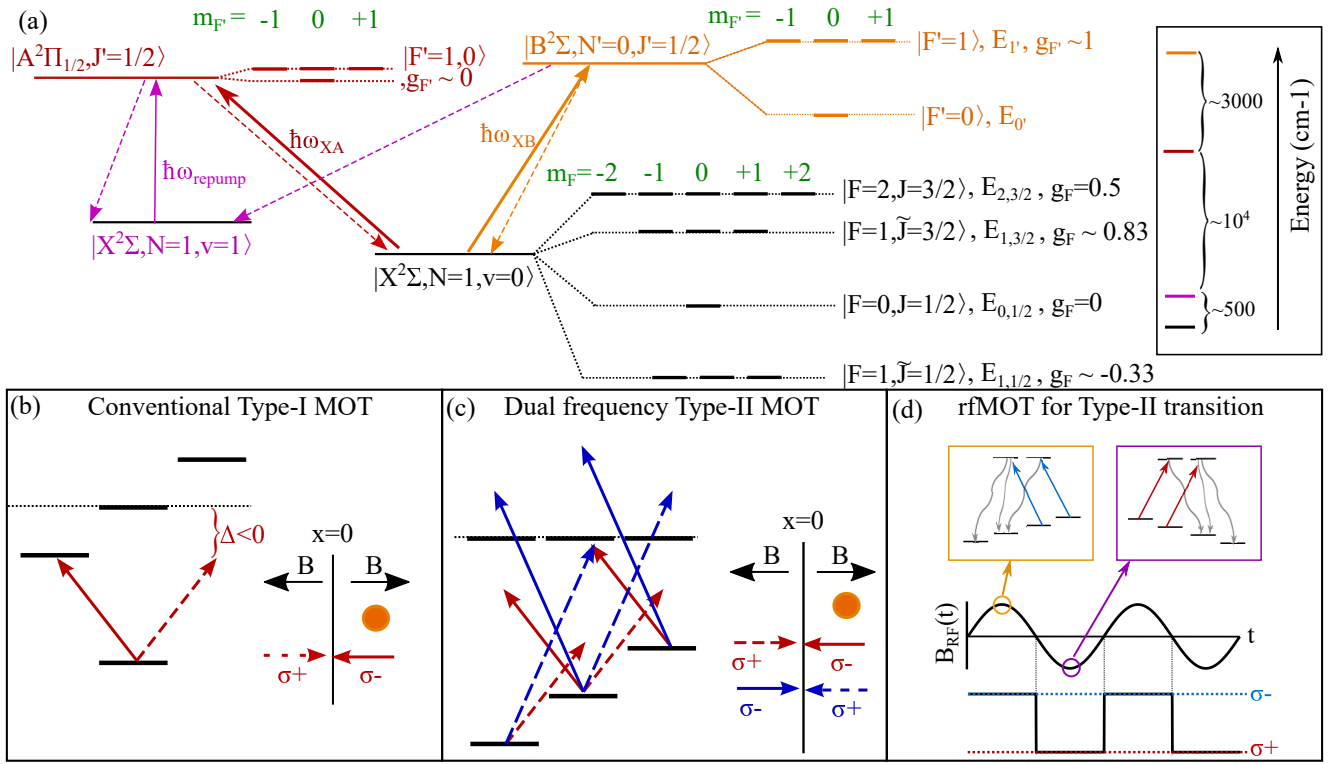}%
		\caption{(a): Level diagram for the laser excitations described in the paper.  All transitions are rotationally closed.  The $\tilde{J}$ labels for the two $|F=1\rangle$ levels of the ground state indicates that these levels are not pure $J$-states, as described in the text.  The level of $J$-mixing also affects the $g$-factor for these states; here, the un-mixed $g$-factor is shown.  Energy differences between adjacent hyperfine manifolds, $E_{F1,J1}-E_{F2,J2}$, for the species discussed in this paper range from $0$ to $7.5\Gamma$.  This splitting is small enough that a single-frequency laser frequency can drive transitions involving multiple ground state hyperfine manifolds.  Electronically excited states have small (for molecules in this paper, $\sim1$\%~\cite{wkt2008,smo1977,kbd2015,ksd2019,gxz2022}) likelihood to undergo a vibrationally off-diagonal decay to $|\textrm{X}^{2}\Sigma,N=1,v=1\rangle$ (whose hyperfine structure, which is very similar to that of $|\textrm{X}^{2}\Sigma,N=1,v=0\rangle$, is not shown).  Molecules decaying to this vibrationally excited state can be repumped via laser excitation to $|\textrm{A}^{2}\Pi_{1/2},J'=1/2,v'=0\rangle$.  Inset shows typical magnitude of energy differences between the states.  (b-d): Diagrams for the MOT configurations discussed in the text. \label{fig:levelStructure}}
	\end{figure}

	The level structure of these molecules, including the relevant hyperfine and spin-rotation structure, is illustrated in Fig.~\ref{fig:levelStructure}(a)~\cite{cgr1981,cgg1981}.  States of the same $F$, where $\bm{F}=\bm{J}+\bm{I}$ refers to the total angular momentum and $\bm{I}$ is the nuclear spin, but different $J$ are mixed via the molecular hyperfine interaction~\cite{brown,swh1996}.  The level of this `$J$-mixing' depends on the molecule in question; $J$-mixing is described in greater detail in ~\ref{sec:CpAndUDeriv}.  We refer to the `mixed' state using the label $\tilde{J}$.  The Lande-g factors $g_{F}$ are also modified by the $J$-mixing; again, the modified values will depend on the molecule in question.  In Fig.~\ref{fig:levelStructure}(a), we list the `un-mixed' $g_{F}$ values.  The energy of state $|F,J\rangle$, $E_{F,J}$, is expressed in units of $E/(\hbar\Gamma)$, where $\Gamma\equiv\Gamma_{XA}$ is the natural linewidth of the $X\rightarrow A$ transition; the same is done for all energies throughout this paper.  We note that, for all molecules considered here, $\Gamma_{XB}\approx\Gamma$~\cite{dcz1974,nwh1983}.
	
	Conventional atom MOTs drive transitions of the form $F\rightarrow F'=F+1$ (referred to as `Type-I' transitions), with a combination of magnetic fields and polarizations chosen such that atoms primarily absorb photons from the laser propagating in the opposite direction of their displacement from the trap center~\cite{mvd}; this is illustrated in Fig.~\ref{fig:levelStructure}B.  This works primarily because transitions between stretched states $|F,m_{F}=\pm F\rangle$ and $|F'=F+1,m_{F'}=\pm F'\rangle$, where $m_{F}$ is the magnetic quantum number, can be driven continuously with $\sigma^{\pm}$ light; we refer to these cases, where the excited state can only decay to only a single $|F,m_{F}\rangle$ state, as `true' cycling transitions.
	
	In contrast, molecular MOTs require driving transitions where $F\ge F'$ (`Type-II' transitions).  In these `quasi-cycling' transitions, decays to several states with different values of $m_{F}$ and sometimes also $F$ are possible.  Moreover, some linear combination of $m_{F}$ levels will be dark in any given polarization.  This lack of a `true' cycling transition limits the force that can be applied in a Type-II MOT to about 1/10 of that in Type-I MOTs, for transitions in which the ground state and excited state have comparable $g$-factors of order unity~\cite{tar2015,dta2016}.  Even worse, however, molecular MOTs thus far all have used the $\textrm{A}^{2}\Pi_{1/2}$ state for the optical transition, which has a negligible $g$-factor ($|g_{F'}|\le 0.1$) compared to the ground electronic state (see Fig.~\ref{fig:levelStructure}(a))~\cite{tar2015}.  This reduces the achievable force to $\sim 1/100$ that of a Type-I MOT, as shown in~\cite{dta2016,tar2015,tst2015}.
	
	Two techniques have been used thus far to overcome this.  The first is `dual-frequency' trapping~\cite{tst2015,twt2017}, in which one of the hyperfine levels in Fig.~\ref{fig:levelStructure}(a) is addressed by a pair of lasers with opposite detuning and polarization.  This effect is illustrated in Fig.~\ref{fig:levelStructure}(c) for $F=1\rightarrow F'=1$, for a case where $g_{F}>0$ and $g_{F'}=0$.  Here, molecules absorb preferentially from the restoring laser in either $m=\pm 1$, while they are equally likely to absorb from either direction for $m=0$; thus, on balance, a restoring force is felt by the molecule.  The second technique is called the radio-frequency MOT (rfMOT)~\cite{hyy2013,nmd2016,aad2017}.  Here, the polarization and magnetic field orientation are switched synchronously at a rate comparable to the photon scattering rate (typically $\omega_{rf}\sim\Gamma/5$), such that, after molecules have fallen into an optically dark Zeeman state for a given field and polarization, the field and polarization reverse, allowing for continuous scatter primarily from the restoring laser (Fig.~\ref{fig:levelStructure}(d)).
	
	The lack of a `true' cycling transition in Type-II transitions also inhibits the use of `standard' Zeeman slowing~\cite{pme1982}, where a circularly polarized, red-detuned laser slows the beam while a magnetic field whose strength varies along the beam is applied to compensate for the changing Doppler shift as the beam is slowed~\cite{mvd}.  Recently, there have been a few proposals reported for Zeeman slowing of molecules~\cite{pko2018,pko2018PRA,lby2019}, along with a demonstration of Zeeman slowing using a Type-II transition in an atom~\cite{pko2018,pko2018PRA}.  In Sec.~\ref{sec:slowingImprovements} we discuss additional prospects for Zeeman slowing with molecules.
	
	Sub-Doppler `Sisyphus' forces are also strikingly different between Type-I and Type-II transitions.  For Type-I transitions, there is sub-Doppler \textit{cooling} for red-detuned light~\cite{dct1989,lwm1988,uwc1989}, while for Type-II transitions, red detuned light results in sub-Doppler \textit{heating}~\cite{dta2016,weh1994}.  Thus, type-II (red-detuned) MOTs, including molecular MOTs, tend to be hotter than Type-I MOTs, as the temperature is dictated by the balance between Doppler cooling and sub-Doppler heating~\cite{jst2018}.  The sign of the sub-Doppler force is reversed for blue-detuned light; this Type-II `gray-molasses' cooling has been used to produce $T\sim10$\,$\mu$K gases of atoms~\cite{gfs2013,skc2015,rbm2018,bvr2014,sfb2013} and molecules~\cite{cad2018,ljd2021,dwy2020,cdt2019}.  In principle, a blue-detuned type-II transition can provide both sub-Doppler cooling \textit{and} a restoring force; such a `blueMOT' can be colder and denser than a Type-II `redMOT'~\cite{jdt2018,jst2018}.  To date, a Type-II blueMOT has only been demonstrated in atoms~\cite{jdt2018}; later in this paper, we discuss potential implementations in molecular systems (Secs~\ref{sec:blueMOT} and~\ref{sec:ZeeMOT}). 
	
	Another important factor for molecule MOTs is the need for vibrational repumping, particularly from the $v=1$ state (magenta, Fig.~\ref{fig:levelStructure}(a)), where $v$ is the vibrational quantum number~\cite{brown}.  In $^{2}\Sigma$ ground state molecules, this state is typically repumped by a laser that couples it to $|$A$^{2}\Pi_{1/2},v'=0\rangle$.  Hence, in MOTs or slowers using the X-A transition, both $|$X$^{2}\Sigma,v=0\rangle$ and $|$X$^{2}\Sigma,v=1\rangle$ are coupled to this single excited simultaneously.  This creates a so-called $\Lambda$-system, whereby light coherently couples states in $|$X$^{2}\Sigma,v=0\rangle$ and $|$X$^{2}\Sigma,v=1\rangle$ via $|$A$^{2}\Pi_{1/2},v'=0\rangle$~\cite{tar2015}.  This results in significant population accruing in $|$X$^{2}\Sigma,v=1\rangle$ (up to $\sim 50$\% of the total).  The corresponding substantial decrease in the optical force~\cite{dta2018,tar2015} means that simulations must include this $v=1$ repumping state in order to capture the relevant physics. 
	
	\subsection{Typical Experimental setup} 
	\label{subsec:expSetup}
	
\begin{figure}
		\centering
		\includegraphics{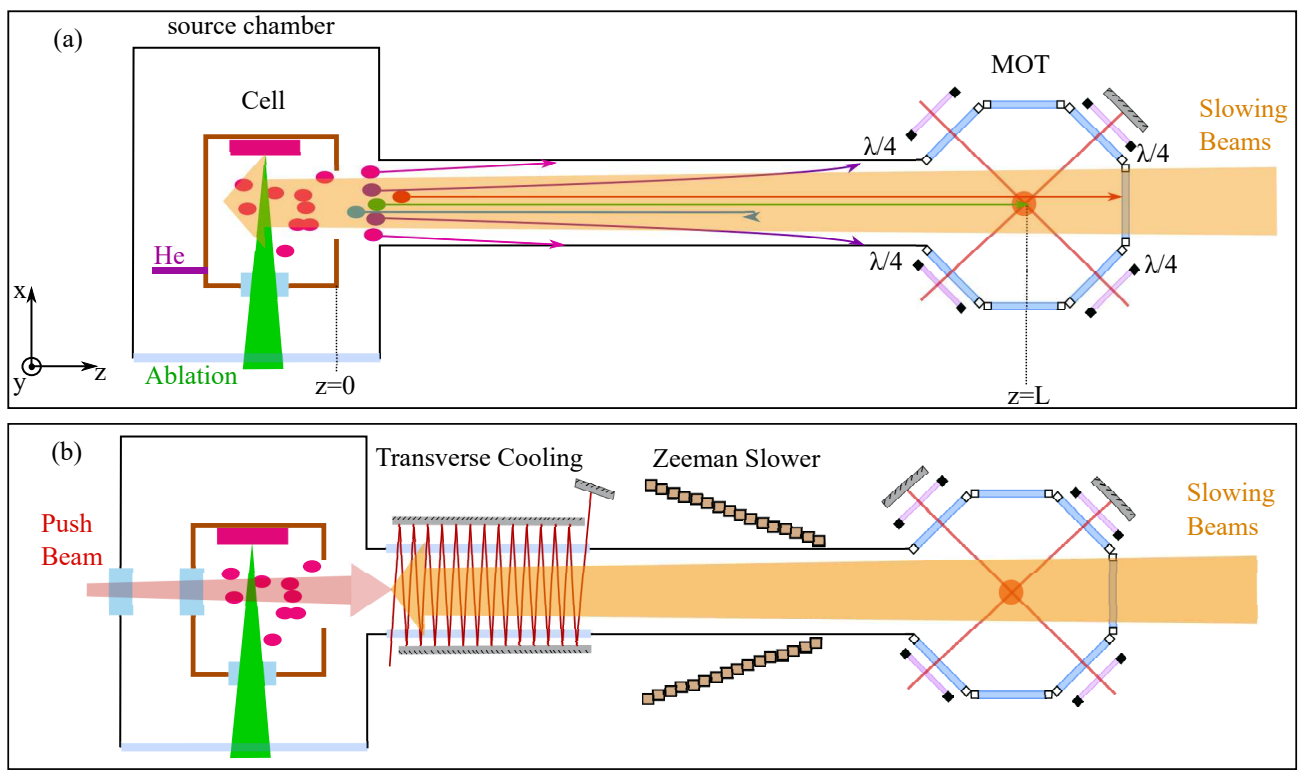}%
		\caption{(a) Typical experimental setup for a molecule MOT.  Molecules generated by ablation of a target are cooled by introducing He buffer gas into a cell at temperature $T_{cell}\le 4$\,K.  Upon extraction from the cell ($z=0$), they are subsequently slowed by a counter-propagating laser beam.  Most molecules have transverse velocities too high to be captured (pink and purple, with the latter representing molecules that would have reached the MOT in absence of slowing); referred to as `pluming'.  Of the molecules that do have sufficiently small transverse velocities, some are overslowed (blue), some are too fast to be slowed sufficiently before reaching the MOT at $z=L$ (red), and some are slowed such that $0\le v_{z}\le v_{cap}$, where $v_{cap}$ is the capture velocity (green).  Only the latter can be trapped in the MOT.  (b) Illustration of proposals discussed in Sec.~\ref{sec:slowingImprovements} for mitigating pluming (transverse cooling) and overslowing (push beam via windows added on the back of the cell and source chamber and/or Zeeman slower).  \label{fig:moleculeExperimentSchematic}}
	\end{figure}

	All experiments involving direct laser-cooling and trapping of molecules into MOTs thus far follow the same general approach, illustrated in Fig.~\ref{fig:moleculeExperimentSchematic}.  First, a slow ($\langle v_{z,0}\rangle\sim 100$\,m/s) beam of molecules is generated using a cryogenic buffer gas beam (CBGB) source~\cite{bsd2011,hld2012}.  To capture into a MOT, molecules from this beam must be slowed to $v_{z}<v_{cap}$, where $v_{cap}$ is the MOT capture velocity.  To do this, the CBGB is slowed by a counter-propagating laser beam.  Sufficiently slowed molecules are then captured by a MOT.  We discuss slowing in greater detail in Sec.~\ref{sec:slowingImprovements}.  
	
	 Most sources produce $\sim10^{11}$ molcules/pulse in the $N=1$ state~\cite{bsd2011,hld2012}; in other words, the best molecular MOTs to date only trap a small fraction ($\sim10^{-5}$) of the potential molecules.  In Sec.~\ref{sec:slowingImprovements}, we discuss ways in which this can be improved.

	\section{Simulation Techniques}
	\label{sec:simTechniques}
	\subsection{Optical Bloch Equations}
	
	In order to simulate the interaction between the applied laser fields and the molecule, we mostly follow the approaches illustrated in~\cite{dta2016,dta2018}.  We solve for the evolution of the density matrix, $\rho$, (expressed using the `hyperfine basis' $|F,J,m_{F}\rangle$) using the master equation:
	
	\begin{equation}
		\frac{\partial \rho}{\partial t}=\frac{1}{i\hbar}\left[H,\rho\right] - \left[\sum_{p=-1}^{1}\frac{\Gamma}{2}(S_{p}^{\dagger}S_{p}\rho-S_{p}\rho S_{p}^{\dagger}+h.c)\right],
		\label{eq:master}
	\end{equation}

	\noindent where 
	
	\begin{eqnarray}
	\nonumber \fl	H=\sum_{F,J,m_{F}} E_{F,J}|F,J,m_{F}\rangle\langle F,J,m_{F}|+\sum_{F',J',m_{F}'}E_{F',J'}|F',J',m_{F}'\rangle\langle F',J',m'_{F}|\\
	+\left(-\hat{\mathbf{d}}\cdot \hat{\mathbf{E}}(\bm{r},t)-\hat{\mathbf{\mu}}\cdot \hat{\mathbf{B}}(\bm{r},t)\right).
		\label{eq:ham}
	\end{eqnarray}

	\noindent Here $\hat{\mathbf{d}}$ is the electric dipole operator, $\hat{\mathbf{E}}$ is the total electric field applied by all lasers in the system, $\hat{\mathbf{\mu}}$ is the magnetic dipole operator, $\hat{\mathbf{B}}$ the magnetic field, subscript $p$ indicates light polarization in spherical vector components, and $S_{p}\propto\sum_{F,J,m,F'}\left(\langle F',J',m_{F}'|T_{p}^{(1)}(\mathbf{d})|F,J,M_{F}\rangle\right)|F,J,m_{F}\rangle\langle F',J',m'=m+p|$, where the first term is the matrix element of the dipole operator.  The decomposition of Eq.~\ref{eq:master} into separate equations for the time evolution of the matrix elements $\rho_{mn}$ constitutes the Optical Bloch Equations~\cite{gas1979,uwc1989}.

	When only the first term on the right side of Eq.~\ref{eq:master} is included, the resulting Liouville-von Neumann equation describes the evolution of the density matrix $\rho$ due to Hamiltonian $H$, while ignoring dissipation via spontaneous emission (which is described by the second term of Eq.~\ref{eq:master}).  The first two terms of $H$ (Eq.~\ref{eq:ham}) correspond to the energy levels of the hyperfine manifolds $E_{F,J}$ and $E_{F',J'}$ (see Fig.~\ref{fig:levelStructure}(a)), respectively.  
	
	\subsubsection{$-\hat{\mathbf{d}}\cdot \hat{\mathbf{E}}(\mathbf{r})$:}
	\label{subsubsec:de}
	
	To decompose this term, we followed the procedure described in~\cite{brown}.  Ultimately, we find (in normalized units):
	
	\begin{equation}
		\fl\frac{-\hat{\mathbf{d}}\cdot \hat{\mathbf{E}}(\mathbf{r})}{\hbar\Gamma} = -\sum_{j,p}\sqrt{\frac{s_{j}}{8}}\exp\left[-i\omega_{j} t - i\beta_{j}\sin(\Omega_{j}t)\right] C_{p}\tilde{E}_{p,j}(\omega_{rf},\mathbf{r}) + h.c.
		\label{eq:dETerm}
	\end{equation}
	
	\noindent Here, subscript $j$ refers to each laser frequency applied to the system, and subscript $p$ refers to the spherical components of the local polarization vector.  
	
	The saturation parameter, $s_{j}$, is defined as $s_{j}=\sqrt{I_{j}/I_{sat}}$, where $I_{j}$ is the peak intensity of the laser beam, which is assumed to have a Gaussian profile, and $I_{sat}=\hbar c\Gamma k^{3}/12\pi$ is the standard definition for `saturation intensity' for a transition with wavenumber $k$ and linewidth $\Gamma=\frac{4}{3\hbar}\frac{1}{4\pi\epsilon_{0}}k^{3}d^{2}$~\cite{mvd}.  Each laser $j$ is associated with either the X$\rightarrow$A or X$\rightarrow$B electronic transition, and the term $\omega_{j}$ represents the frequency of laser $j$ after either $\omega_{XA}$ or $\omega_{XB}$ (see Fig.~\ref{fig:levelStructure}(a)), respectively, is subtracted, as we are working in the interaction picture~\cite{mvd}.  States $\textrm{A}^{2}\Pi_{1/2}$ and $\textrm{B}^{2}\Sigma$ are energetically far enough away that cross-talk from lasers addressing the different electronic transitions can be ignored.  The terms $\beta_{j}$ (modulation index) and $\Omega_{j}$ (modulation frequency) describe phase modulation used, e.g., for spectral broadening.  The term $C_{p}$ represents a `coupling matrix' such that $\langle F',m'|C_{p}|F,m\rangle$ is the dimensionless matrix element of the electric dipole matrix operator (e.g., these are `Clebsch-Gordan'-like terms).  The values of $C_{p}$ for the system depicted in Fig.~\ref{fig:levelStructure}A, including $J$-mixing, are derived in ~\ref{sec:CpAndUDeriv}.  Finally, $\tilde{E}_{p,j}(\omega_{rf},\mathbf{r})$ refers to the normalized component of the electric field with polarization $p$ at position $\textbf{r}$ provided by laser $j$ (including the effect of the finite beam waist), and the $\omega_{rf}$ term indicates that, for rfMOTs, the polarization flips circularity whenever the magnetic field sign changes. 
	
	The exact form of $\tilde{E}_{p,j}(\omega_{rf},\mathbf{r})$ depends on the application being simulated.  In this paper, we simulate:\begin{itemize}
		\item 2D Transverse cooling with and without simultaneous application of a 1D slowing laser (see Sec.~\ref{subsec:trans})
		\item 1D Slowing (see Secs.~\ref{subsec:zeeSlow} and~\ref{subsec:slowPush})
		\item dc and rf 3D MOTs (Secs.~4-6).
		
	\end{itemize}
	
	\noindent In ~\ref{sec:EpDeriv}, we derive the form of $\tilde{E}_{p,j}(\omega_{rf},\mathbf{r})$ for each of these cases.
	
	\subsubsection{$-\hat{\mathbf{\mu}}\cdot \hat{\mathbf{B}}$:}
	\label{subsubsec:uB}
	
	This term is handled in two different ways, depending on whether the Zeeman energy $H_{z}\sim\mu_{B}B$ is small relative to the typical hyperfine energy splitting $H_{HF}\sim E_{F1,J1}-E_{F2,J2}$ (Fig.~\ref{fig:levelStructure}A).
	
	If $H_{z}\ll H_{HF}$, we assume that the magnetic field does not mix different $|F,J\rangle$ states significantly and that the Zeeman shifts are all linear in the hyperfine basis. We write $-\hat{\mathbf{\mu}}\cdot \hat{\mathbf{B}}=\mu_{B}\left(\sum_{F,J}g_{F,J}\mathbf{B}\cdot\mathbf{F}+\sum_{F',J'}g_{F',J'}\mathbf{B}\cdot\mathbf{F'}\right)$.  This matches the treatment in~\cite{dta2018}.  We make this approximation for the transverse cooling calculation in Sec.~\ref{subsec:trans}, the `pushed white-light slower' described in Sec.~\ref{subsec:slowPush}, and the 3D MOTs of SrF and CaF described in Secs.~\ref{sec:bichromTrapping} and~\ref{sec:blueMOT}.  In this case~\cite{dta2018},
	
	\begin{equation}
	\fl	-\hat{\mathbf{\mu}}\cdot \hat{\mathbf{B}} = g_{F,J}\mu_{B}\sum_{q=-1}^{1}(-1)^{q}B_{q}(\omega_{rf},\mathbf{r})T_{-q}^{1}(F) + g_{F',J'}\mu_{B}\sum_{q=-1}^{1}(-1)^{q}B_{q}(\omega_{rf},\mathbf{r})T_{-q}^{1}(F'),
	\end{equation}
	
	\noindent where $B_{q}(\omega_{rf},\mathbf{r})$ is the inner-product of $\mathbf{B}(\omega_{rf},\mathbf{r})$ with spherical unit basis vector $\hat{n}_{q}$ and we allow for the possibility of $B$ to vary with $\textbf{r}$ (e.g. for the anti-Helmoltz coil configuration used in magneto-optical trapping).  We use the Wigner-Eckart theorem to express matrix elements of $T^{1}_{-q}$~\cite{brown}:
	
	\begin{equation}
		\fl\langle F,J,m_{F,1}|T_{-q}^{1}(F)|F,J,m_{F,2}\rangle= (-1)^{F-m_{F,1}}\sqrt{F(F+1)(2F+1)}\left( \begin{array}{ccc}
			F & 1 & F \\
			-m_{F,1} & -q & m_{F,2}\end{array} \right).
	\end{equation}
	
	If $H_{z}\ge H_{HF}$, this approximation is invalid.  In this case, our approach is to first express $-\hat{\mathbf{\mu}}\cdot \hat{\mathbf{B}}$ in the `Zeeman basis' $|m_{S},m_{N},m_{I}\rangle$, where $m_{N}$, $m_{I}$ and $m_{S}$ are the magnetic quantum numbers of the molecular rotation, nuclear spin, and electronic spin, respectively.  We then use a unitary transformation to convert $H_{z}$ to the `hyperfine basis' $|F,J,m_{F}\rangle$, in which all other terms in Eq.~\ref{eq:master} are expressed.  
	
	As an example, consider a magnetic field along the $\hat{z}$ direction.   Making the approximation $g_{s}=2$ and neglecting the rotational and nuclear magnetic moments, we find $-\hat{\mathbf{\mu}}\cdot \hat{\mathbf{B}}=H_{z}^{m_{s},m_{I},m_{N}}= \sum_{m_{N},m_{I},m_{S}}2\mu_{B}Bm_{s}|m_{S},m_{N},m_{I}\rangle\langle m_{S},m_{N},m_{I}|$ (e.g., the Hamiltonian is diagonal).  This is converted to the hyperfine basis $H_{z}^{F,J,m_{F}}=U^{\dagger}H_{z}^{m_{S},m_{N},m_{I}}U$, where $U_{ab}=\langle m_{S,a},m_{N,a},m_{I,a}|F_{b},J_{b},m_{F,b}\rangle$.  The explicit form for $U$, including the effect of $J$-mixing, is derived in \ref{sec:CpAndUDeriv}.
	
	The magnetic field term is treated in this way for the simulations of Type-II Zeeman slowers described in Sec.~\ref{subsec:zeeSlow}, and for simulations of MOTs of MgF, SrOH, and CaOH described in Sec.~\ref{sec:ZeeMOT}, all of which have at least one pair of hyperfine manifolds within the $\textrm{X}^{2}\Sigma$ state with minimal ($\le0.4\Gamma$) hyperfine splitting.
	
	\subsubsection{Spontaneous Emission:}
	The final term of Eq.~\ref{eq:master} handles the effect of spontaneous emission.  This ultimately reduces to~\cite{dta2018}:
	
	\begin{eqnarray}
		\fl \frac{d\rho_{mn}}{dt}_{sp}= -\left[\sum_{p=-1}^{1}\frac{\Gamma}{2}(S_{p}^{\dagger}S_{p}\rho-S_{p}\rho S_{p}^{\dagger}+h.c)\right]\nonumber, \textrm{where}\\
		\frac{d\rho_{mn}}{dt}_{sp}= -\rho_{mn} (\textrm{m,n both correspond to excited states}), \nonumber\\
		\frac{d\rho_{mn}}{dt}_{sp}= -\frac{1}{2}\rho_{mn} (\textrm{either m is ground and n excited, or vice versa}), \textrm{and} \nonumber\\
		\frac{d\rho_{mn}}{dt}_{sp}= \sum_{p}C_{eff,p}\rho_{eff}C^{\dagger}_{eff,p} (\textrm{m,n both correspond to ground states}). \label{eq:spont} 
	\end{eqnarray}
	
	\noindent Subscript $sp$ indicates that this is the contribution of the spontaneous decay to the evolution of $\rho$.  Here $C_{eff,p,nm}=C_{p,nm}\exp\left[i\omega_{nm}t\right]$ if $n$ is an excited state and $m$ is a ground state (0 otherwise), with $C_{p}$ defined in Eq.~\ref{eq:dETerm}, and $\rho_{eff,mn}=\rho_{mn}$ if $m,n$ both correspond to excited states (0 otherwise).
	
	\subsection{Determining Forces}
	Now that all terms in Eq.~\ref{eq:master} are determined, we solve for the evolution of $\rho$ given a starting velocity $\mathbf{v}$ and position $\mathbf{r}$ of a molecule.  We use the \textit{Julia} programming language~\cite{julia}, a compiled language with built-in implementation of the openBLAS linear algebra libraries that makes it both fast and easy to use for this application.  For computational convenience, we round all frequencies to the nearest integer multiple of a common, low frequency $\omega_{r}$.  The value of $\omega_{r}$ is itself chosen to be an integer fraction of the unit frequency $\Gamma$: $\omega_{r}=\Gamma/N_{H}$.  Similarly, we round all speeds, $v$, to an integer multiple of $\omega_{r}/k_{XA}=(\Gamma/k_{XA})/N_{H}$, where $k_{XA}$ is the wavenumber of the X$\rightarrow$A transition and $\Gamma/k_{XA}$ is the unit speed.  This approach guarantees that the Hamiltonian is periodic, with period $\tau=2\pi/\omega_{R}$~\cite{ytb2016,dta2018}.  In practice, we use $N_{H}=100$ (10) when $v\le 0.5\Gamma/k_{XA}$ ($>0.5\Gamma/k_{XA}$).  The system is evolved for a sufficiently long time, $t_{tr}$, such that transients related to the initial conditions dissipate.  At this point, the expectation values of molecular operators evolve with the same periodicity as the Hamiltonian (e.g. force $\langle F(t)\rangle=\langle F(t+\tau)\rangle$)~\cite{ytb2016,dta2018}.  Thus, once $t_{tr}$ is reached, we evolve this system for one additional period and calculate $\langle F(\bm{r},\bm{v},t)\rangle$, as in~\cite{dta2018}.  Using the Heisenberg picture time derivative, one finds:
	
	\begin{equation}
	\fl F(\mathbf{r},\mathbf{v},t)=\frac{dp_{i}}{dt}=-\frac{\partial}{\partial r_{i}}H=\sum_{j,p}\sqrt{\frac{s_{j}}{8}}\exp\left[i\omega_{j} t + i\beta_{j}\sin(\Omega_{j}t)\right] C_{p}\frac{\partial\tilde{E}_{p,j}(\omega_{rf},\mathbf{r})}{\partial r_{i}}. 
	\end{equation}

	\noindent Here, subscript $i$ refers to the Cartesian component of the force.  The force averaged over the density matrix is then:
	
	\begin{equation}
		\langle F_{i}(\mathbf{r},\mathbf{v},t) \rangle=\textrm{Tr}\left[\rho(t)\frac{\partial p_{i}}{\partial t}\right]
	\end{equation}

	Next, the ensemble-averaged force is averaged over the period:
	
	\begin{equation}
		\langle F_{i}(\mathbf{r},\mathbf{v})\rangle=\frac{1}{\tau}\int_{t_{tr}}^{t_{tr}+\tau} \langle F_{i}(t) \rangle dt.
	\end{equation}

	\noindent  Finally, to average over different initial starting conditions within the polarization and intensity gradient created by the light field, we average $\langle F_{i}(\mathbf{r},\mathbf{v})\rangle$ over a minimum of 50 trajectories with randomized positions within the cube defined by corners $(kr_{x},kr_{y},kr_{z})$ and $(kr_{x}+2\pi,kr_{y}+2\pi,kr_{z}+2\pi)$.

	\section{Improvements to redMOT capture velocity with two-color trapping}
	\label{sec:bichromTrapping}
	
	All molecular redMOTs to date, whether rf-redMOTs~\cite{nmd2016,aad2017} or dual-frequency dc-redMOTs~\cite{twt2017,aad2017,cdy2018,dwy2020}, drive only the X$^{2}\Sigma\rightarrow$ A$^{2}\Pi_{1/2}$ transition.  Here we consider possible advantages to also using the X$^{2}\Sigma\rightarrow$B$^{2}\Sigma$ transition for improved MOT performance~\cite{tst2015,xko2021}.  We focus here on SrF; analogous simulations for CaF are discussed in~\ref{sec:CaFMOT}.  
	
	Unlike the A$^{2}\Pi_{1/2}$ state, the B$^{2}\Sigma$ state has a substantial $g$-factor as well as resolvable hyperfine structure (for SrF, $E_{B\Sigma,1'}-E_{B\Sigma,0'}\sim 2\Gamma$).  As was discussed in~\cite{dta2016}, the larger $g$-factor presents an opportunity, as substantial trapping could be achieved without a dual-frequency approach.  However, the resolvable hyperfine structure presents a complication.  Consider transitions from the $F=1$ states of X$^{2}\Sigma$ to $F'=0,1$ states of B$^{2}\Sigma$.  A laser that is detuned by $\Delta\sim -\Gamma$ relative to the $|\textrm{X}^{2}\Sigma,F=1\rangle\rightarrow|\textrm{B}^{2}\Sigma,F'=1\rangle$ transition is also detuned by $\Delta\sim +\Gamma$ relative to the $|\textrm{X}^{2}\Sigma,F=1\rangle\rightarrow|\textrm{B}^{2}\Sigma,F'=0\rangle$ transition; the latter has an adverse effect on trapping.  
	
\begin{figure*}[h!]
		\centering
		\includegraphics[scale=1]{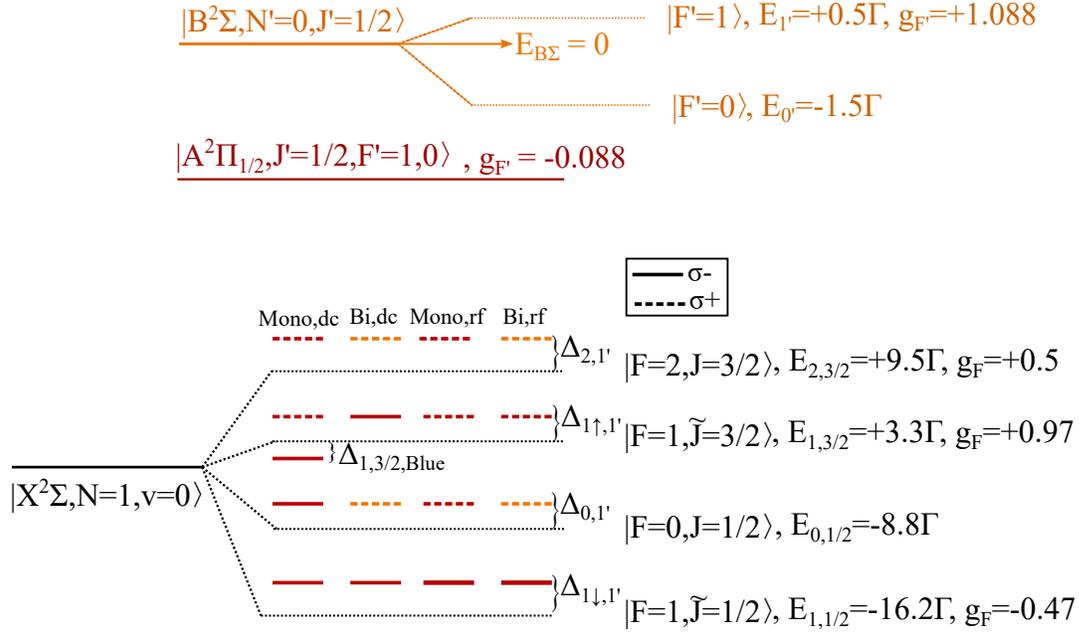}
		\caption{Level diagram for SrF.  The redMOT configurations described in Table~\ref{tb:rateEqParams} are also indicated.  The `Mono dc' configuration uses dual frequency trapping via an additional laser tuned to the blue of $|1,3/2\rangle$.  Colored lines in the ground state manifold indicate that an applied laser frequency would connect such a line to the appropriate excited state level (red for X$\rightarrow$A and orange for X$\rightarrow$B).  $\Delta_{F,F'}$ (Table~\ref{tb:rateEqParams}) refers to the magnitude of the detuning of this laser frequency from resonance.  Here, unless otherwise noted, all lasers are red-detuned, so values of $\Delta_{F,F'}$ are negative.  Throughout the paper, light that is red (blue) detuned of a transition involving a state in the X$^{2}\Sigma$ hyperfine manifold will be indicated by lines above (below) that manifold.  Energies $E_{F,J}$ and $E_{F'}$ are defined relative to the `hyperfine-free' energy of a given state (as shown for the B$^{2}\Sigma$ state).
		\label{fig:SrFLevelDiagram}}
	\end{figure*}

	Here, we propose a novel two-color molecular redMOT configuration that uses \textit{both} electronically excited states.  We excite both $F=1$ states in the X($N=1$) manifold to A$^{2}\Pi_{1/2}$, while the $F=2$ and $F=0$ states are coupled to B$^{2}\Sigma$ (see Fig.~\ref{fig:SrFLevelDiagram}).  This avoids the complication described in the previous paragraph, since neither $F=2$ nor $F=0$ can couple to $F'=0$.  
	
	\begin{table}[ht]
		
		\begin{tabular}{|l|l|l|l|l|}
			\hline
			Label & Transition & $s_{j}$ & $\Delta_{F,F'}(\Gamma)$&  $\hat{p}$ \\
			\hline
			\multirow{5}{4em}{Mono dc} & \multirow{5}{4em}{$X\rightarrow A$ (all)} & 10 & $\Delta_{1\downarrow\!,1'}=-1.1$ &  $\sigma^{-}$ \\
			&  & 20 & $\Delta_{0,1'}=-2.3$ &  $\sigma^{-}$ \\
			&  & \textbf{8.7} & $\bm{\Delta_{1\uparrow\!,1'}=-1.2}$ &  $\bm{\sigma^{+}}$ \\
			&  & 31.3 & $\Delta_{2,1'}=-0.9$ &  $\sigma^{+}$ \\
			&  & \textbf{10} & $\bm{\Delta_{1\uparrow\!,1'}=+1}$ &  $\bm{\sigma^{-}}$ \\
			\hline
			\multirow{4}{4em}{Bi dc} & $X\rightarrow A$ & 20 & $\Delta_{1\downarrow\!,1'}=-2$ &  $\sigma^{-}$ \\
			&  $X\rightarrow B$ & 20 & $\Delta_{0,1'}=-2$ &  $\sigma^{+}$ \\
			&  $\bm{X\rightarrow A}$ & \textbf{20} & $\bm{\Delta_{1\uparrow\!,1'}=-4.5}$, $\bm{\Delta_{2,1'}=+1.8}$ &  $\bm{\sigma^{-}}$ \\
			&  $\bm{X\rightarrow B}$ & \textbf{20} & $\bm{\Delta_{2,1'}=-2}$ &  $\bm{\sigma^{+}}$ \\
			\hline
			\multirow{4}{5em}{Bi dc,\\Optimized (Sec. 4.1.1)} & $X\rightarrow A$ & 30 & $\Delta_{1\downarrow\!,1'}=-1$ &  $\sigma^{-}$ \\
			&  $X\rightarrow B$ & 10 & $\Delta_{0,1'}=-1$ &  $\sigma^{+}$ \\
			&  $\bm{X\rightarrow A}$ & \textbf{30} & $\bm{\Delta_{1\uparrow\!,1'}=-5}$, $\bm{\Delta_{2,1'}=+1.3}$ &  $\bm{\sigma^{-}}$ \\
			&  $\bm{X\rightarrow B}$ & \textbf{45} & $\bm{\Delta_{2,1'}=-3.5}$ &  $\bm{\sigma^{+}}$ \\
			\hline
			\multirow{4}{4em}{Mono rf} &  \multirow{4}{4em}{$X\rightarrow A$ (all)} & 20 & $\Delta_{1\downarrow\!,1'}=-2$ &  $\sigma^{-}$ \\
			&  & 20 & $\Delta_{0,1'}=-2$ &  $\sigma^{-}$ \\
			&  & 20 & $\Delta_{1\uparrow\!,1'}=-2$ &  $\sigma^{+}$ \\
			&  & 20 & $\Delta_{2,1'}=-2$ &  $\sigma^{+}$ \\
			\hline
			\multirow{4}{4em}{Bi rf} & $X\rightarrow A$ & 20 & $\Delta_{1\downarrow\!,1'}=-2$ &  $\sigma^{-}$ \\
			&  $X\rightarrow B$ & 20 & $\Delta_{0,1'}=-2$ &  $\sigma^{+}$ \\
			&  $X\rightarrow A$ & 20 & $\Delta_{1\uparrow\!,1'}=-2$ &  $\sigma^{+}$ \\
			&  $X\rightarrow B$ & 20 & $\Delta_{2,1'}=-2$ &  $\sigma^{+}$ \\
			\hline

		\end{tabular}
		
		\caption{Parameters used for SrF redMOT simulations, whose results are shown in Fig.~\ref{fig:phaseExample}.  The $\Delta_{F,F'}$ column shows the detuning of each laser relative to a specific transition between hyperfine manifolds $F\rightarrow F'$.  If a laser frequency is within $|5\Gamma|$ of resonance with multiple $F\rightarrow F'$ transitions,  we list both such transitions  (e.g., the third laser in the `Bi dc' configuration in Table~\ref{tb:rateEqParams} is -4.5$\Gamma$ with respect to the X$\rightarrow$A $|F=\!1\!\uparrow\rangle\rightarrow |F'=1\rangle$ transition and also is $+1.8\Gamma$ with respect to the $|F=2\rangle\rightarrow |F'=1\rangle$ transition, see also Fig.~\ref{fig:SrFLevelDiagram}).  \textbf{Bold font} indicates a pair of lasers that participate in a quasi-dual-frequency scheme (e.g. have opposite polarizations and detunings with respect to a hyperfine manifold, and are both within $|3.5\Gamma|$ of resonance with that manifold).  For example,the last two lasers in `Bi dc' are both within $3\Gamma$ of resonance with a transition from the $F=2$ state.   For these MOT simulations, $\bm{B}=b_{0}\cos(\omega_{rf}t)(x\hat{x}+z\hat{z}-2y\hat{y})$.  For rf, $b_{0}=12.5$\,G/cm and $\omega_{rf}=\Gamma/5$.  For dc, $b_{0}=8.8$\,cm and $\omega_{rf}=0$.  The $\hat{p}$ column gives the polarization of each laser beam propagating in the +z direction, when $b_{0}\cos(\omega_{rf}t)$ is positive.  \label{tb:rateEqParams}}
	\end{table}
	
	We compare these proposed two-color redMOTs to the `one-color' X$\rightarrow$A MOTs that have been demonstrated experimentally thus far, for both rf and dc configurations.  All configurations have labels indicating whether they are one-color (`mono') or two-color (`bi'), and whether the corresponding redMOT is rf or dc; see Table~\ref{tb:rateEqParams}.  The laser parameters used in each case are shown in Table~\ref{tb:rateEqParams}.  Four additional, unlisted, vibrational repumping lasers are used in all simulations, set to resonance with transitions between the four hyperfine levels of $|\textrm{X}^{2}\Sigma,v=1\rangle$ and $|\textrm{A}^{2}\Pi_{1/2},v'=0\rangle$, all with $s_{j}=20$.  Each laser has a `detuning' $\Delta_{F,F'}$ relative to the resonant frequency coupling hyperfine manifolds $F$ and $F'=1$, see Fig.~\ref{fig:SrFLevelDiagram}.  Here and throughout, $F\!=\!1\!\uparrow$ refers to the $|F=1,\tilde{J}=3/2\rangle$ manifold and $F\!=\!1\!\downarrow$ refers to the $|F=1,\tilde{J}=1/2\rangle$ manifold.

	\subsection{Measuring capture velocity, temperature, and $\sigma$}
	\label{subsec:SrFRed}

\begin{figure}
		\centering
		\includegraphics{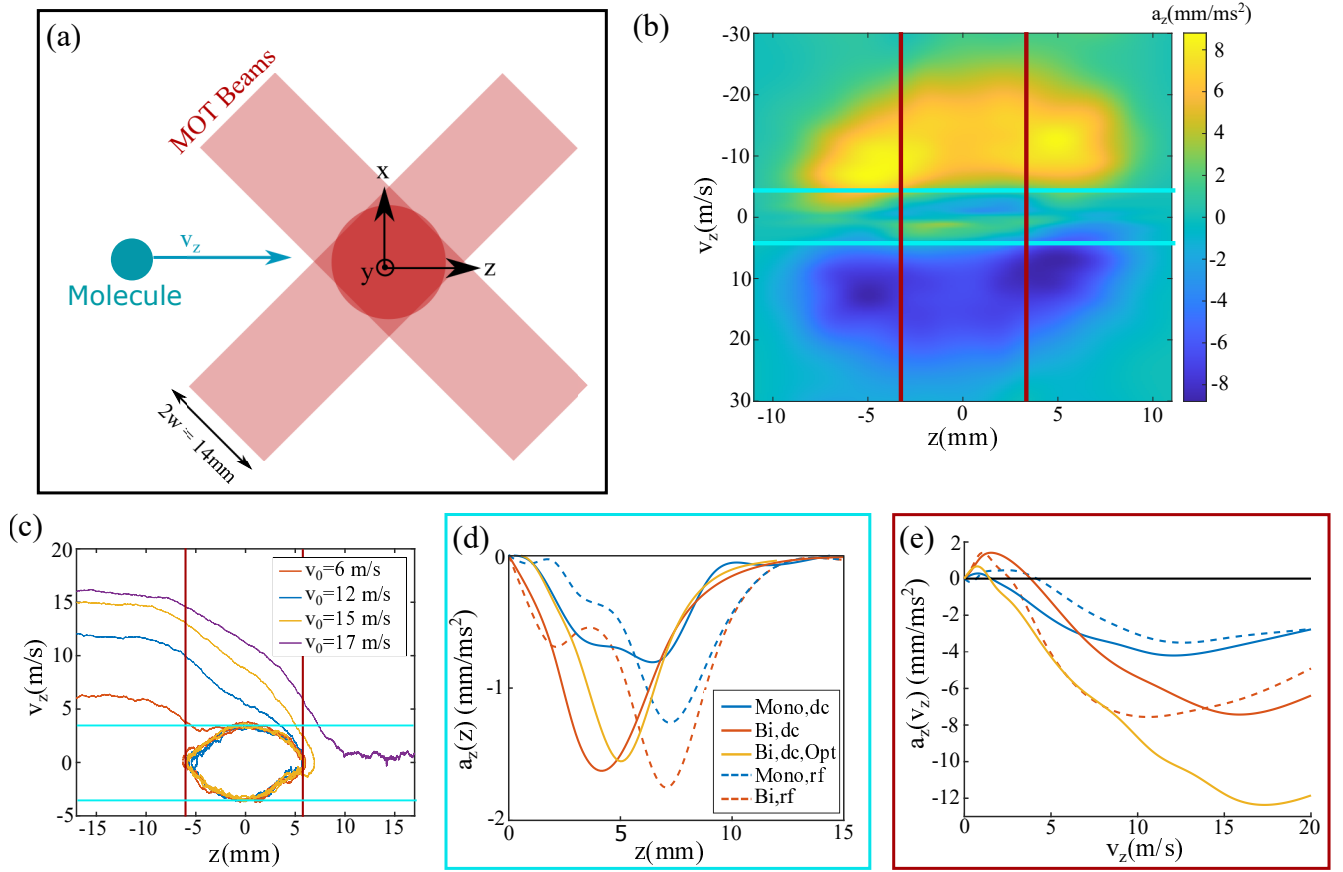}%
		\caption{(a) Geometry for capture into a redMOT.  Slowed molecules enter the MOT region along $z$.  (b) Plot of $a_{z}(v_{z},z)$ from simulation of the `Bi-rf' configuration in Table~\ref{tb:rateEqParams}.  (c) Trajectory simulations, used to determine $v_{cap}$, $T$, and $\sigma$ as described in text.  Molecules with $v_{0}>v_{cap}$ escape.  Here, $v_{cap}=15.7$\,m/s, $T=46$\,mK, and $\sigma=4.3$\,mm, see Table~\ref{tb:capVelTempAndSig}.  The sub-Doppler heating here results in the circular trajectory of the particle in $(z,v_{z})$ space.  (d) and (e): Plots of $a_{z}(z)$ and $a_{z}(v_{z})$ by averaging over $|v_{z}|\le v_{z,max}$ and $|z|\le z_{max}$, respectively (see cyan and dark red lines, respectively, in (b-c)).  Here we show results for all cases in Table~\ref{tb:rateEqParams}.  Sub-Doppler heating can clearly be seen in (e) as the reversal of the sign of $a_{z}$ at low $v_{z}$. \label{fig:phaseExample}}
	\end{figure}

	For all MOT simulations in this work, we set $\bm{v}\parallel\hat{z}$ and $\bm{r}\parallel\hat{z}$ (e.g. to study the capture of molecules entering the 3D MOT region after slowing, see Fig.~\ref{fig:phaseExample}(a)).  We calculate the acceleration $a_{z}(z,v_{z})$ due to the atom-light interaction felt by the molecule along this axis of motion as a function of displacement $z$ and velocity $v_{z}$.  A typical plot of $a_{z}(z,v_{z})$ is shown in Fig.~\ref{fig:phaseExample}(b).
	
	Using $a_{z}(z,v_{z})$, the particle trajectory for a choice of initial position and velocity can be determined.  To measure the redMOT capture velocity $v_{cap}$, we start with a position at the `beginning' of the MOT region (in this paper, taken as $z=-17$\,mm from the MOT center, determined by the trapping laser beam $1/e^{2}$-intensity radius of $w_{MOT}=7$\,mm).  We then vary the starting velocity up until the molecule `escapes' from the trap (here defined as reaching $z=+17$\,mm), see Fig.~\ref{fig:phaseExample}(c).  We note here that this will only measure $v_{cap}$ for a molecule that travels directly along the slowing axis; in general, $v_{cap}$ will be reduced as the displacement from the slowing axis increases (and thus the molecule begins to `miss' the high intensity regions of the MOT lasers).
	
	For molecules that are captured, an additional trajectory time of 100\,ms is used to obtain convergence for $\sigma=\sqrt{\langle z^{2}\rangle}$ and $v_{T}=\sqrt{\langle v_{z}^{2}\rangle}$, the rms displacement and velocity of trapped particles, respectively.  Temperature is given by $T=mv_{T}^{2}/k_{B}$.  During this trajectory, we add the effect of random photon kicks due to spontaneous emission; the probability of a kick occuring during a trajectory evolution timestep $t_{s}$ is $p_{k}=P_{exc}(z,v_{z})\Gamma t_{s}$, where $P_{exc}(z,v_{z})$ is the total excited state population and $t_{s}\ll \Gamma^{-1}$.  A kick occurs whenever a random number $r<p_{k}$, where $0\le r\le 1$.
	
	We also show plots of `spatial deceleration' and `velocity deceleration', defined by $a_{z}(z)=\frac{1}{2v_{z,max}}\int_{-v_{z,max}}^{v_{z,max}}a_{z}(z,v_{z})dv_{z}$ and $a_{z}(v_{z})=\frac{1}{2z_{max}}\int_{-z_{max}}^{z_{max}}a_{z}(z,v_{z})dz$, respectively, where $z_{max}$ and $v_{z,max}$ are the maximum displacement and velocity of a trapped particle in $(z,v_{z})$ space (Fig.~\ref{fig:phaseExample}(d-e)). 

	In Table~\ref{tb:capVelTempAndSig}, we list the values of $v_{cap}$, $T$, and $\sigma$ found for the redMOT configurations in Table~\ref{tb:rateEqParams}.  In~\ref{sec:CaFMOT}, we show that similar laser parameters give good trapping for two-color rf and dc-redMOTs of CaF as well, demonstrating that this approach is generalizable.
	
	\begin{table}[ht]
		
		\begin{tabular}{|l|l|l|l|}
			\multicolumn{4}{c}{SrF redMOT configurations} \\
			\hline
			Label & $v_{cap}$ & $T$ (mK) & $\sigma$ (mm)\\
			\hline
			Mono,dc & 8.6 & 23 & 4.8\\
			Bi, dc & 9.8 & 57 & 6.2 \\
			Mono,rf & 9.4 & 67 & 5.2 \\
			Bi, rf & 15.7 & 46 & 4.3\\
			Bi, dc, optimized (Sec. 3.1.1) & 12.5 & 14 & 2.7 \\
			\hline
			\multicolumn{4}{c}{CaF redMOT configurations} \\
			\hline
			Label & $v_{cap}$ & $T$ (mK) & $\sigma$ (mm)\\
			\hline
			Mono,dc & 12.2 & 50 & 7.4\\
			Bi, dc & 17.8 & 36 & 5.5 \\
			Mono,rf & 14.9 & 50 & 5.8 \\
			Bi, rf & 20.2 & 39 & 4.8\\
			\hline
		\end{tabular}
		
		\caption{Capture velocity $v_{cap}$, temperature $T$, and rms width $\sigma$ determined from simulations (including effect of $v=1$ repumping), for SrF and CaF in various redMOT configurations.  In general, capture velocities are higher for CaF, primarily due to its lower mass.  We observe that rf-redMOTs are generally more effective than dc-redMOTs at capturing molecules, as has been observed experimentally~\cite{aad2017}. Further, two-color MOTs (here with label `Bi') are generally better than one-color (`Mono') MOTs.  Finally, we find that slight changes to the choices for detunings and the intensities of each laser can have substantial effects on all three performance parameters (compare the different two-color DC MOTs in SrF, see Table~\ref{tb:rateEqParams} and Sec. 4.3).  The parameters used for CaF are shown in \ref{sec:CaFMOT}. \label{tb:capVelTempAndSig}}
	\end{table}

	\subsubsection{Optimizing the two-color dc-redMOT capture velocity}
	\label{subsec:OBEOptRed}
	
	The sub-Doppler heating described in Refs.~\cite{dta2016,dta2018,weh1994} and observed in Fig.~\ref{fig:phaseExample}(e) limits both the capture velocity of the MOT and how low the temperature can reach.  Thus, we varied the choices of frequency and the intensity addressing each hyperfine transition, with an eye on keeping the overall intensity realistically achievable in experiments.  Ultimately, we found a set of values that dramatically reduces (but does not completely eliminate) the effect of sub-Doppler heating for the two-color dc-redMOT.  These are shown in Table~\ref{tb:rateEqParams} (labeled as Bi dc, Optimized), and the results are displayed in Fig.~\ref{fig:phaseExample} and Table~\ref{tb:capVelTempAndSig}.  For the optimized case, the sub-Doppler heating is less severe while the Doppler cooling is more effective (Fig.~\ref{fig:phaseExample}(e)), leading to substantially lower temperatures.   The $a_{z}(z)$ curves for the two cases, however, are remarkably similar (Fig.~\ref{fig:phaseExample}(d)), so the lower temperature corresponds to a more compact cloud of trapped molecules.

	\subsection{MOT compression}
	
	After the molecules are captured in the one-color redMOTs used to date, it is common to increase the phase space density of the trapped cloud by reducing the laser intensities and increasing the magnetic field~\cite{nmd2016,dwy2020,aad2017,vhd2022,twt2017}.  Lowering the intensity reduces the scattering rate and thus the `random-walk' heating, and it also reduces the magnitude of the sub-Doppler force~\cite{tar2015}.  These both act to lower the temperature at the cost of reducing the overall magnitude of the trapping forces.  The larger magnetic field increases the spatial gradient of the trapping force, thus compressing the molecules  
	
\begin{figure*}[h!]
		\centering
		\includegraphics[scale=1]{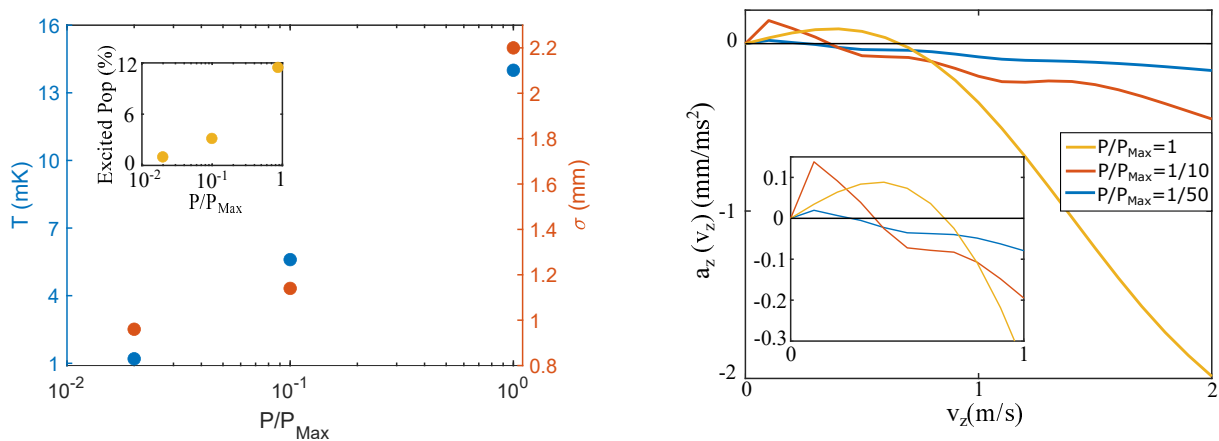}
		\caption{Compression of the optimized two-color redMOT.  Left: $T$ and $\sigma$ vs laser power.  Inset: Population in excited states vs laser power.  Right: $a_{z}(v_{z})$ for different laser powers.  Inset: Zoom in to $v_{z}\le1$\,m/s.  As the power is lowered, the velocity at which the sign of the damping force reverses is reduced.}
		\label{fig:motCompression}
	\end{figure*}

	To determine whether the two-color dc-redMOT behaves similarly, we varied the laser powers for the `Bi, dc optimized' case (Tables~\ref{tb:rateEqParams} and \ref{tb:capVelTempAndSig}).  For simulations at lower power, we also increase $b_{0}$ from 8.8\,G/cm to 20\,G/cm.  Results are shown in Fig.~\ref{fig:motCompression}.   We indeed see that the excited state population (and thus the scattering heating rate) decreases with power, as does the range of velocities over which sub-Doppler forces dominate.  As a consequence, both $T$ and $\sigma$ also decrease with power, similar to what has been observed both in experiments~\cite{nmd2016,aad2017,twt2017} and in previous simulations of one-color MOTs~\cite{dta2018}. 
	
	It is possible that improvements to compression could be made by changing other parameters such as the laser detuning, or by reducing the power of different laser frequencies by different factors, during the ramp.  Further optimization of the compression is outside of the scope of this work.  Nevertheless, the main point is clear: the temperature and size of the two-color redMOT can be reduced to reach values similar to those achieved in one-color compressed redMOTs ($T\sim 1$\,mK and $\sigma\sim 1$\,mm), using essentially the same protocol.

	\section{BlueMOTs: Improvements to trap density and temperature}
	\label{sec:blueMOT}
	
	Sub-Doppler heating limits how low $T$ and $\sigma$ one can achieve in either one-color or two-color molecular redMOTs.  Similar behavior was observed in atomic Type-II redMOTs~\cite{jst2018}, and in previous one-color molecular redMOT simulations~\cite{dta2018}.  Logically, if sub-Doppler heating is present in a redMOT, then one should be able to achieve sub-Doppler cooling in a blueMOT.  Indeed, with this motivation, a blueMOT has been demonstrated for a Type-II trap of Rb atoms~\cite{jdt2018}.  However, a blue-MOT has yet to be demonstrated, or simulated, in molecular systems.  (A recent proposal described how to `engineer' a sub-Doppler force in MOTs where red-detuned light is still primarily responsible for the spatial confinement~\cite{xko2022}.)
	
	In this section, we show results of simulations of a two-color dc-blueMOT, where sub-Doppler cooling \textit{and} spatial confinement are provided simultaneously by the blue-detuned light.  As in Sec.~\ref{sec:bichromTrapping}, here we drive X$\rightarrow$A transitions on the two X$^{2}\Sigma,F=1$ states, and X$\rightarrow$B transitions on the $F=0$ and $F=2$ states, this time with all $\Delta_{F,F'}>0$.  Then, as in Sec.~\ref{subsec:OBEOptRed}, we optimized the choices of intensity and detuning.  This time, however, instead of \textit{minimizing} the effect of \textit{red}-detuned sub-Doppler \textit{heating}, we \textit{maximize} the effect of \textit{blue}-detuned sub-Doppler \textit{cooling}.  
	
	\begin{table}[ht]
		
		\begin{tabular}{|l|l|l|l|}
			\multicolumn{4}{c}{SrF two-color blueMOT configuration (Fig.~\ref{fig:blueMOT})} \\
			\hline
			Transition & $s_{j}$ & $\Delta_{F,F'}(\Gamma)$&  $\hat{p}$ \\
			\hline
			$X\rightarrow A$ & 12 & $\Delta_{1\downarrow\!,1'}=+2$ &  $\sigma^{+}$ \\
			$X\rightarrow B$ & 4 & $\Delta_{0,1'}=+1$ &  $\sigma^{-}$ \\
			$\bm{X\rightarrow A}$ & \textbf{4} & $\bm{\Delta_{1\uparrow\!,1'}=+1}$ &  $\bm{\sigma^{-}}$ \\
			$\bm{X\rightarrow B}$ & \textbf{16} & $\bm{\Delta_{2,1'}=+3}$, $\bm{\Delta_{1\uparrow\!,1'}=-3.3}$ &  $\bm{\sigma^{+}}$ \\
			\hline
			
		\end{tabular}
		
		\caption{Optimized laser parameters used in simulations of a two-color dc-blueMOT of SrF.  Lasers participating in a quasi-dual-frequency scheme (both addressing $F=1\!\uparrow$) are bolded.  For this simulation, we set $b_{0}=25$\,G/cm. \label{tb:blueSrFParams}}
	\end{table}
	
	\begin{figure*}[h!]
		\centering
		\includegraphics[scale=1]{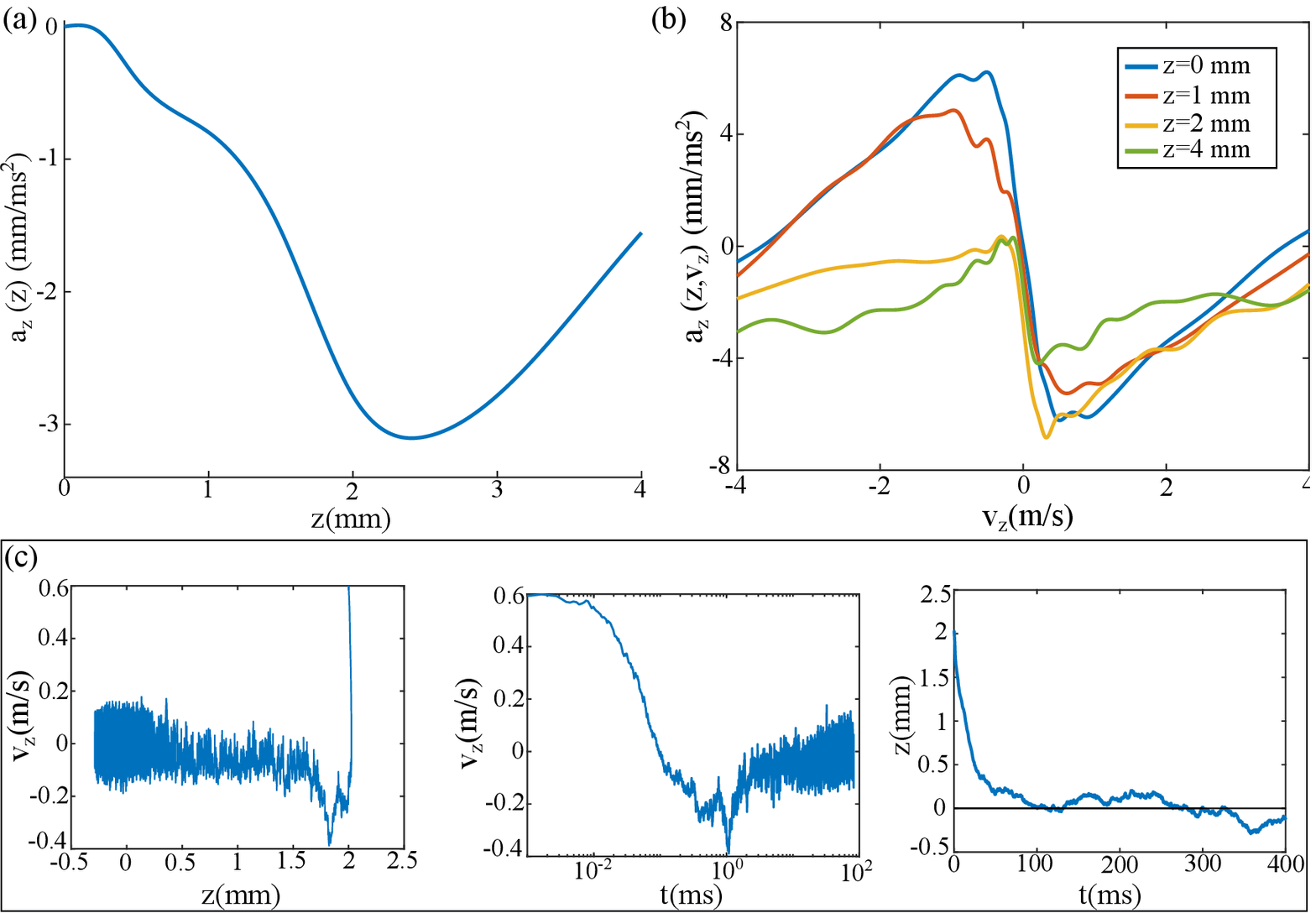}
		\caption{Two-color dc-blueMOT behavior for the optimized parameters in Table~\ref{tb:blueSrFParams}.  (a) $a_{z}(z)$ for $a_{z}(z,v_{z})$ integrated over $|v_{z}|\le0.1$\,m/s.  (b): $a_{z}(v_{z},z)$ for various $z$.  Note the effect of the spatial restoring force here: as $z$ increases, the whole curve is shifted towards negative $a_{z}$.  The sub-Doppler cooling effect (signified by the sharp slope around $v_{z}=0$) is present even out to $z=4$\,mm (where $B=10$\,G), demonstrating that it is remarkably robust with respect to magnetic field.  (c): Trajectory of a particle (including random photon scattering) with $v_{z,0}=2v_{T,Red}$ (0.6\,m/s) and $z_{0}=2\sigma_{Red}$ (2\,mm) (see text).  The velocity evolves rapidly to a value slightly below zero, indicating motion towards the MOT center (middle panel) while the position converges to the center of the trap on a timescale of $\sim 50$\,ms (right panel).  In the left panel we plot $v(z)$ for the trajectory, and we see here that, unlike in the molecule redMOTs, there is no `merry-go-round' effect.}
		\label{fig:blueMOT}
	\end{figure*}
	
	We indeed find that sub-Doppler cooling can be achieved simultaneously with trapping.  In Fig.~\ref{fig:blueMOT}, we show some key features of this new type of molecule MOT for the set of laser parameters listed in Table~\ref{tb:blueSrFParams}, which we found to be a good choice for SrF.  In Fig.~\ref{fig:blueMOT}(a), we plot $a_{z}(z,v_{z})$ integrated over $|v_{z}|\le 0.1$\,m/s.  The confining forces are strong, with magnitudes comparable to the redMOTs (Fig.~\ref{fig:phaseExample}), and are effective out to $z>4$\,mm.
	
	We also observe that the slope of $a_{z}(v_{z})$ at low velocities is quite sharp, which leads to exceptional cooling (Fig.~\ref{fig:blueMOT}(b)).  In addition, the velocity damping is robust out to $z=4$\,mm ($B=10$\,G) in this system.  This is a major difference between Type-II and Type-I sub-Doppler forces: for Type-I systems, sub-Doppler cooling is only effective for near zero $B$~\cite{wdp1992,dta2016} (typically, $B\le 0.05\hbar\Gamma/\mu_{B}$, corresponding to $B\le0.25$\,G for SrF) while for Type-II systems, sub-Doppler cooling is robust out to at least $B\sim \hbar\Gamma/\mu_{B}$ ($\sim$5\,G for SrF), as seen in~\cite{dta2016}.  Here, we see that it is actually effective out to at least 10\,G.  
	
	This blueMOT would be a poor choice for capturing from the CBGB, as restoring forces are only achieved for low velocities, $v_{z}\lesssim 3$\,m/s.  However, it can capture nearly all molecules from a compressed redMOT with $T\sim 1$\,mK ($v_{T,Red}=\sqrt{k_{B}T/m}\sim0.3$\,m/s) and $\sigma_{Red}\sim 1$\,mm.  We demonstrate this by plotting the trajectory of a particle with $z_{0}=2$\,mm and $v_{z}=2v_{T,Red}=0.6$\,m/s: since this particle is captured, we can say that all particles within 2 standard deviations of the mean velocity and position should be retained when switching from a compressed two-color dc-redMOT to the two-color dc-blueMOT (see Fig.~\ref{fig:blueMOT}(c)).
	
	In addition, once the simulated trajectory in Fig.~\ref{fig:blueMOT}(c) stabilizes, we record the time evolution of $z$ and $v_{z}$ due to the combination of random photon kicks and the MOT forces; the long-time rms values are then used to determine $\sigma$ and $T$ respectively.  We find $T=25\,\mu$K and $\sigma=0.13$\,mm.
	
	This two-color dc-blueMOT stage, when added after the two-color dc-redMOT compression stage, would provide very favorable conditions for loading into an ODT.  Since the blueMOT reduces $\sigma$ from 1\,mm to $\sim 100\,\mu$m, the density would be increased by $\sim 10^{3}$ relative to that used in current experiments~\cite{aad2017,dwy2020,lhc2022,twt2017,smd2016}.  Typically, experiments load an ODT by turning off the compressed MOT, then turning on $\Lambda$-enhanced gray molasses cooling~\cite{cad2018,ljd2021,wbd2021,lhc2022,hvd2022} along with the ODT for loading.  With this protocol, the ODT capture fraction is roughly proportional to the number of molecules originally within the ODT beam diameter (typically 100$\,\mu$m or less); using a blueMOT should enable near unit efficiency for molecules from the MOT being captured in the ODT, compared to the current state of the art of $<5\%$, which heretofore has been the case~\cite{ljd2021}.  
	
	Because the blueMOT molecule temperature is already much lower than typical ODT trap depths ($T_{D}\lesssim 600\,\mu$K) used for loading molecules~~\cite{cad2018,aad2018,ljd2021,wbd2021,lhc2022,vhd2022}, it may also enable direct ODT loading from the confined gas of molecules, rather than from an untrapped, expanding, $\Lambda$-cooled gas, as is done currently~\cite{cad2018,aad2018,ljd2021,wbd2021,lhc2022,vhd2022}.  Direct loading should enhance the time during which molecules can be loaded, and thus the eventual loading fraction from the MOT to the ODT, relative to the case of loading from a molasses-cooled, but expanding, cloud.  Simulations of ODT loading, however, are beyond the scope of this paper.

	\section{Simulations of MOTs for molecules with minimal hyperfine splitting}
	\label{sec:ZeeMOT}
	
	The MOT simulations in the previous sections were done using the approximation $H_{z}\ll H_{HF}$ (e.g., assumes $F,J,m_{F}$ are `good' quantum numbers).  This is the case for small displacements from the MOT center for SrF and CaF, as the smallest relevant hyperfine splitting is $V_{HF,B}\approx2\Gamma$ (SrF B$^{2}\Sigma$ state), while $H_{z}\approx0.2B\Gamma$ (with $B$ in Gauss) for both molecules.  
	
	This assumption no longer holds for large displacements from the MOT center, or for molecules with smaller energy differences between hyperfine manifolds such as SrOH, CaOH, and MgF.  Solving the OBEs for simulations of MOTs for these molecules requires treating the $-\bm{\mu\cdot B}$ term first in the $|m_{s},m_{I},m_{J}\rangle$ basis, in and then converting into the $|F,J,m_{F}\rangle$ basis which all other components of the Hamiltonian are expressed in (this procedure was described in Sec.~\ref{subsubsec:uB}).  In this section, we perform simulations using the more general approach.
	
\begin{figure}
		\includegraphics{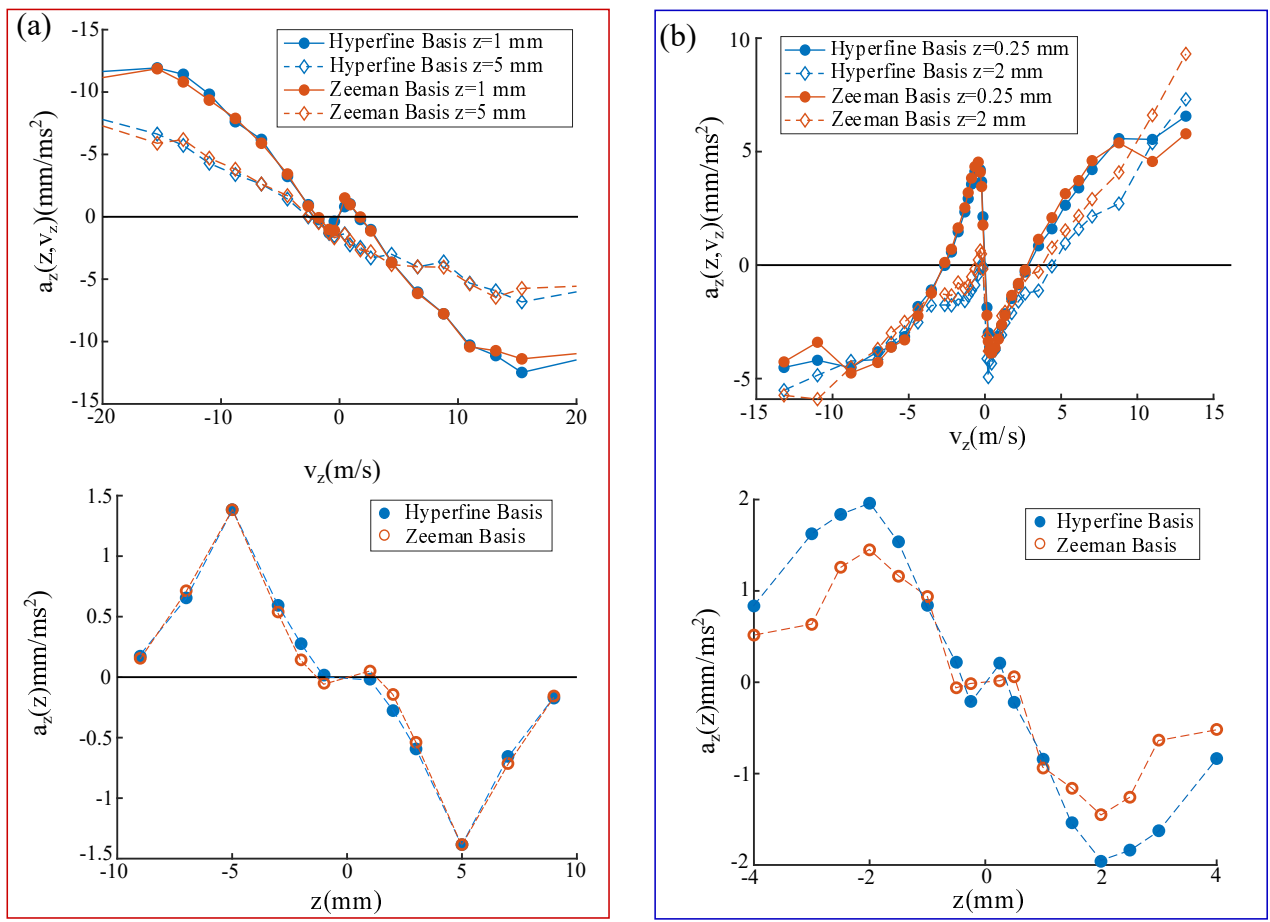}%
		\caption{Comparison between simulations where Zeeman-induced mixing between hyperfine states is ignored (Hyperfine) and where it is accounted for (Zeeman) for (a) the optimal two-color redMOT (see Table~\ref{tb:rateEqParams}) and (b) the two-color blueMOT (Table~\ref{tb:blueSrFParams}), in SrF.    \label{fig:ZeemanVsHyperfine}}
	\end{figure}

	To verify the code for the generalized case, we checked to ensure that a simulation using this approach for $-\bm{\mu\cdot B}$ gives the same results as one using the `Hyperfine' approach for SrF.  Since SrF has well resolved hyperfine structure in the X$^{2}\Sigma$ and B$^{2}\Sigma$ state, both approaches should yield similar results.  The results of the comparison are shown in Fig.~\ref{fig:ZeemanVsHyperfine}.  Indeed, they match quite well, with a slight divergence between the two arising for $|z|\gtrsim 2$\,mm ($|B|\gtrsim 5$\,G) in the blueMOT.  As $B$ increases, some divergence is expected, as this is where the assumption $H_{z}\ll H_{HF}$ begins to break down.
	
	\subsection{SrOH and CaOH}
	\label{subsec:OH}
	
	We next applied the generalized code to molecules with minimal hyperfine structure. The alkaline earth hydroxides CaOH and SrOH have both been laser-cooled~\cite{kbd2017,bvd2020}; additionally, CaOH has been trapped in an rf-redMOT~\cite{vhd2022} and subsequently optically trapped~\cite{hvd2022}.  Here, we perform simulations of effective \textit{dc}-redMOTs and \textit{dc}-blueMOTs of both molecules.  
	
	Here, we restrict ourselves to one-color X$\rightarrow$A MOTs, since decays from the B$^{2}\Sigma$ state populate more vibrational modes (including bending modes) than decays from A$^{2}\Pi_{1/2}$, in these hydroxides~\cite{lld2022}.  Further, we do \textit{not} include the effect of a $v=1$ repumping laser.  Repumping for these molecules has typically been done through the B$^{2}\Sigma$ state \footnote{The Franck-Condon factors for the B$^{2}\Sigma$ state are less diagonal in these hydroxides than in SrF and CaF, so repumping through B$^{2}\Sigma$ is possible without needing an unrealistic amount of laser power (as in~\cite{vhd2022}).}.  This breaks the $\Lambda$ system between the $|\textrm{X}^{2}\Sigma,v=0\rangle$ and $|\textrm{X}^{2}\Sigma,v=1\rangle$ states, mitigating the deleterious effect of the a $v=1$ repumper.  Hence, simulations that ignore the repumper should still give accurate results for forces in these systems.
	
	\begin{figure}
		\includegraphics{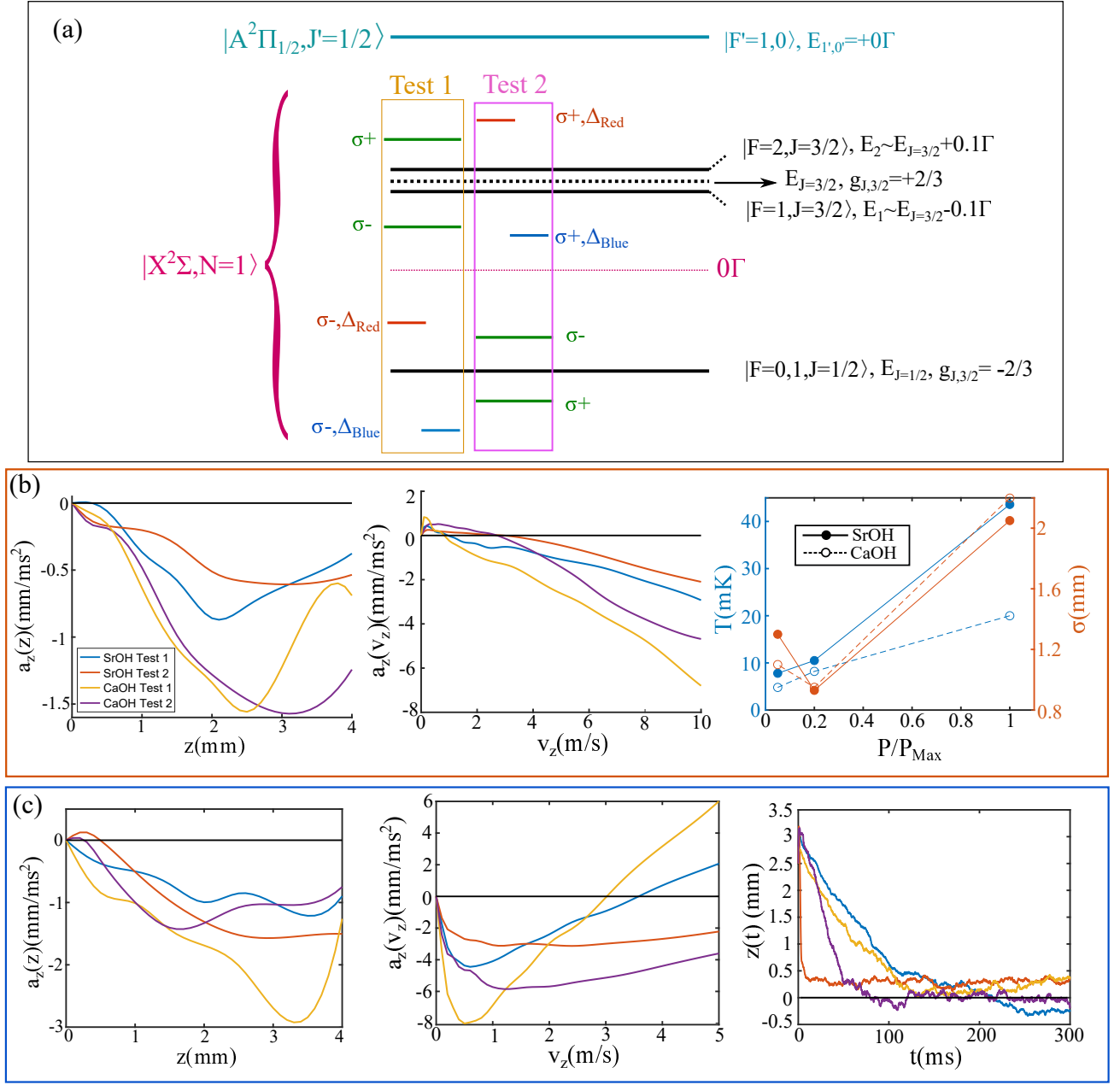}%
		\caption{(a): MOT configurations for SrOH \& CaOH, where hyperfine levels of the same $J$ are nearly degenerate.  Spatial confinement is generated by a dual-frequency mechanism (green), either on $|J=3/2\rangle$ (Test 1) or $|J=1/2\rangle$ (Test 2).  An additional laser set to \textit{either} red or blue of the other state provides Doppler cooling or sub-Doppler cooling, respectively.  Results of simulations for redMOTs (Table~\ref{tb:redMOTHydroxideParameters}) are shown in (b).  Left and middle panel: $a_{z}(z)$ and $a_{z}(v_{z})$, respectively, for the compressed redMOT (Table~\ref{tb:redMOTHydroxideParameters}).  Note that, for Test 2, the sub-Doppler force is effective for $v\lesssim 3$\,m/s, which leads to lack of confinement.  Right panel: $T$ and $\sigma$ vs. $P/P_{max}$ for Test 1.  As in SrF, reducing power results in lower $\sigma$ and $T$. (c): BlueMOT (Table~\ref{tb:blueMOTHydroxideParameters}).  Trapping and sub-Doppler cooling are achieved for all configurations.  The right panel shows $z(t)$ for a molecule with $z_{0}=2\sigma_{red}$ and $v_{z,0}=2v_{T,red}$; this demonstrates how quickly the blueMOT compresses the cloud of molecules.   \label{fig:OH}}
	\end{figure}
	
	\begin{table}[ht]
		
		\begin{tabular}{|l|l|l|l|l|l|l|}
			\multicolumn{7}{c}{Hydroxide redMOT configurations(Fig.~\ref{fig:OH}(b))} \\
			\hline
			\multicolumn{4}{|c|}{Laser Parameters} & \multicolumn{3}{c|}{Results} \\
			\hline
			Label & $s_{j,max}$ & $\Delta_{J}(\Gamma$)&  $\hat{p}$ & $v_{cap}(s_{max})$& $T(s_{max}/20)$& $\sigma(s_{max}/20)$\\
			\hline
			\multirow{4}{4em}{Test 1} & \textbf{20} & $\bm{\Delta_{3/2}=-1.5}$ &  $\bm{\sigma^{+}}$ & \multicolumn{3}{c|}{SrOH} \\\cline{5-7}
			&  \textbf{20} & $\bm{\Delta_{3/2}=+1.5}$ &  $\bm{\sigma^{-}}$ & 17.0\,m/s  & 7.8\,mK  & 1.3\,mm \\\cline{5-7}
			&  \multirow{2}{2em}{20} & \multirow{2}{5em}{$\Delta_{1/2}=-2$} &  \multirow{2}{2em}{$\sigma^{-}$} & \multicolumn{3}{c|}{CaOH}  \\\cline{5-7}
			& & & & 23.3\,m/s & 4.2\,mK & 1.1\,mm \\\cline{5-7}
			\hline
			
			\multirow{4}{4em}{Test 2} & \textbf{30} & $\bm{\Delta_{1/2}=-1.5}$ &  $\bm{\sigma^{-}}$ & \multicolumn{3}{c|}{SrOH} \\\cline{5-7}
			&  \textbf{10} & $\bm{\Delta_{1/2}=+1}$ &  $\bm{\sigma^{+}}$ & 10.4\,m/s  & ---  & --- \\\cline{5-7}
			&  \multirow{2}{2em}{20} & \multirow{2}{5em}{$\Delta_{3/2}=-2$} &  \multirow{2}{2em}{$\sigma^{+}$} & \multicolumn{3}{c|}{CaOH}  \\\cline{5-7}
			& & & & 17.8\,m/s & --- & --- \\\cline{5-7}
			\hline
			
		\end{tabular}
		
		\caption{Left: Parameters used for hydroxide redMOT simulations (Fig.~\ref{fig:OH}(b)).  Detunings $\Delta_{J}$ are indexed relative to the indicated $J$ (see Fig.~\ref{fig:OH}(a)).  Bold font indicates levels that participate in a dual-frequency scheme.  Right: Results.  First column indicates $v_{cap}$ for the `capture redMOT' ($s_{j}=s_{j,max}$ for all lasers, $b_{0}=12.5$\,G/cm).  In the next two columns, $T$ and $\sigma$ for the low power `compressed redMOT' ($s_{j}=s_{j,max}/20$ for all lasers, $b_{0}=25$\,G/cm) are reported. For Test 2, the molecules were no longer confined in the low-power MOT, so $T$ and $\sigma$ are only listed for Test 1.  \label{tb:redMOTHydroxideParameters}}
	\end{table}
	
	\begin{table}[ht]
		
		\begin{tabular}{|l|l|l|l|l|l|l|}
			\multicolumn{7}{c}{Hydroxide blueMOT configurations (Fig.~\ref{fig:OH}(c))} \\
			\hline
			\multicolumn{4}{|c|}{Laser Parameters} & \multicolumn{3}{c|}{Results} \\
			\hline
			Label &  $s_{j}$ & $\Delta_{J}(\Gamma)$&  $\hat{p}$ & $t_{D}$& $T$& $\sigma$\\
			\hline
			\multirow{4}{4em}{Test 1} & \textbf{4} & $\bm{\Delta_{3/2}=-1}$ &  $\bm{\sigma^{+}}$ & \multicolumn{3}{c|}{SrOH} \\\cline{5-7}
			& \textbf{4} & $\bm{\Delta_{3/2}=+1}$ &  $\bm{\sigma^{-}}$ & 150\,ms  & 55$\,\mu$K  & 0.27\,mm \\\cline{5-7}
			& \multirow{2}{2em}{20} & \multirow{2}{5em}{$\Delta_{1/2}=+4$} &  \multirow{2}{2em}{$\sigma^{-}$} & \multicolumn{3}{c|}{CaOH}  \\\cline{5-7}
			& & & & 100\,ms & 70$\,\mu$K & 0.24\,mm \\\cline{5-7}
			\hline
			
			\multirow{4}{4em}{Test 2} & \textbf{6} & $\bm{\Delta_{1/2}=-1.5}$ &  $\bm{\sigma^{-}}$ & \multicolumn{3}{c|}{SrOH} \\\cline{5-7}
			&  \textbf{2} & $\bm{\Delta_{1/2}=+1}$ &  $\bm{\sigma^{+}}$ & 10\,ms  & 53$\,\mu$K  & 0.31\,mm \\\cline{5-7}
			&  \multirow{2}{2em}{16} & \multirow{2}{5em}{$\Delta_{3/2}=+3$} &  \multirow{2}{2em}{$\sigma^{+}$} & \multicolumn{3}{c|}{CaOH}  \\\cline{5-7}
			& & & & 50\,ms & $90\,\mu$K & 0.09\,mm \\\cline{5-7}
			\hline
			
		\end{tabular}
		
		\caption{Parameters used for hydroxide blueMOT simulation (Fig.~\ref{fig:OH}(c)).  Detunings $\Delta_{J}$ are indexed relative to the indicated $J$ (see Fig.~\ref{fig:OH}(a)).  Bold font indicates levels that participate in a dual-frequency scheme.  Here, $t_{D}$ is the timescale of spatial compression, see Fig.~\ref{fig:OH}(c) right panel. \label{tb:blueMOTHydroxideParameters}}
	\end{table}
	
	Both hydroxide molecules considered have very similar hyperfine structures, illustrated in Fig.~\ref{fig:OH}(a).  The spin-rotation interaction splits states with $J=3/2$ and $J=1/2$.  However, because of the small hyperfine interactions, there is minimal J-Mixing and only small splitting between the two hyperfine levels that share the same $J$.  Our setup uses a dual-frequency approach on one of the two $J$ levels, when this is $J=3/2$ ($J=1/2$), we refer to the simulation as Test 1 (2).  For both cases, we use either red-detuned (for Doppler cooling in a redMOT) or blue-detuned (for sub-Doppler cooling in a blueMOT) light on the other transition.  Basically, one $J$ level can be thought of as the `trapping' level and the other as the `cooling' level.  A similar approach, with red-detuning on the `cooling' level, was demonstrated to work experimentally for YO, which has a pair of ground states in the X$^{2}\Sigma$ state that are nearly degenerate~\cite{dwy2020}.  Since the $g_{J}$-factors have different signs for $J=3/2$ versus $J=1/2$~\cite{bvd2020}, the signs of the blue and red polarizations in the dual-frequency approach are reversed between Test 1 and Test 2.  The laser parameters used to obtain the results in Fig.~\ref{fig:OH} are shown in Tables~\ref{tb:redMOTHydroxideParameters}-~\ref{tb:blueMOTHydroxideParameters}.
	
	For the redMOTs, we performed simulations both for a high power `capture redMOT' ($s_{j}=s_{j,max}$ for each laser with $B_{Grad}=12.5$\,G/cm), and a lower power `compressed redMOT' (see Table~\ref{tb:redMOTHydroxideParameters}).  Generally Test 1 yielded better results for redMOTs of both molecules; stronger trapping and cooling forces are observed, leading to higher capture velocities in the capture redMOT and lower temperatures $T$ and cloud size $\sigma$ in the compressed redMOT.  In fact, for Test 2, the molecules are not confined in the compressed MOT at all.  This is because the sub-Doppler heating in Test 2 is effective for $v\lesssim 3$\,m/s (compare to $\sim 1$\,m/s for Test 1); molecules at this speed cannot be recaptured by the spatial confining forces (Fig.~\ref{fig:OH}(b)).
	
	There is very little difference between the results for the two molecules; the larger forces for CaOH primarily result from its lower mass.  Even at low power, the temperatures are still quite high, with $T\sim 5$\,mK ($v_{T}\sim 0.7$\,m/s) for Test 1, again primarily resulting from the sub-Doppler heating.  No effort was made here to find more optimal values of MOT control parameters for these molecules.
	
	In the blueMOT, `Test 2' yielded generally better results for both molecules.  Though all test cases ultimately demonstrated molecular cooling and confinement (see Table~\ref{tb:blueMOTHydroxideParameters}), in `Test 2' we generally found more robust velocity damping and much faster timescales ($t_{D}$) for spatial compression, where $t_{D}$ is defined to be the earliest time for which $z<\sigma$ (Fig.~\ref{fig:OH}(c)).  This is demonstrated by monitoring the trajectory $z(t)$ (evolving under $a_{z}(v_{z},z)$ plus random photon kicks, as in Sec.~\ref{sec:blueMOT}) for a molecule with initial position $r_{0}=2$\,mm and initial velocity $v_{z,0}=2.5$\,m/s (corresponding to at least 95\% of molecules from a compressed `Test 1' redMOT).  For all tests, $\sigma\sim 200\,\mu$m and $T\sim 70\,\mu$K (Table~\ref{tb:blueMOTHydroxideParameters}).
	
	The results here indicate that it should be possible to capture both CaOH and SrOH in a dc-redMOT.  Although the achievable temperature in the compressed redMOT is still somewhat high, nearly all molecules from a `Test 1' redMOT can be recaptured when switching to either a `Test 1' or `Test 2' blueMOT.  Recently, an rf-redMOT of CaOH at a similar gradient to that simulated here was demonstrated to yield a molecular cloud size of $800\,\mu$m~\cite{vhd2022}; 1\% of the molecules in this MOT could then be transferred into an ODT~\cite{hvd2022}.  Implementing a subsequent blueMOT, thus shrinking the cloud size to $\sim 200\,\mu$m, should increase the density by a factor of $\sim 50$, and thus dramatically improve ODT loading fraction, just as discussed in Sec.~\ref{sec:blueMOT} for the case of SrF.

	\subsection{MgF}
	
\begin{figure}
		\includegraphics{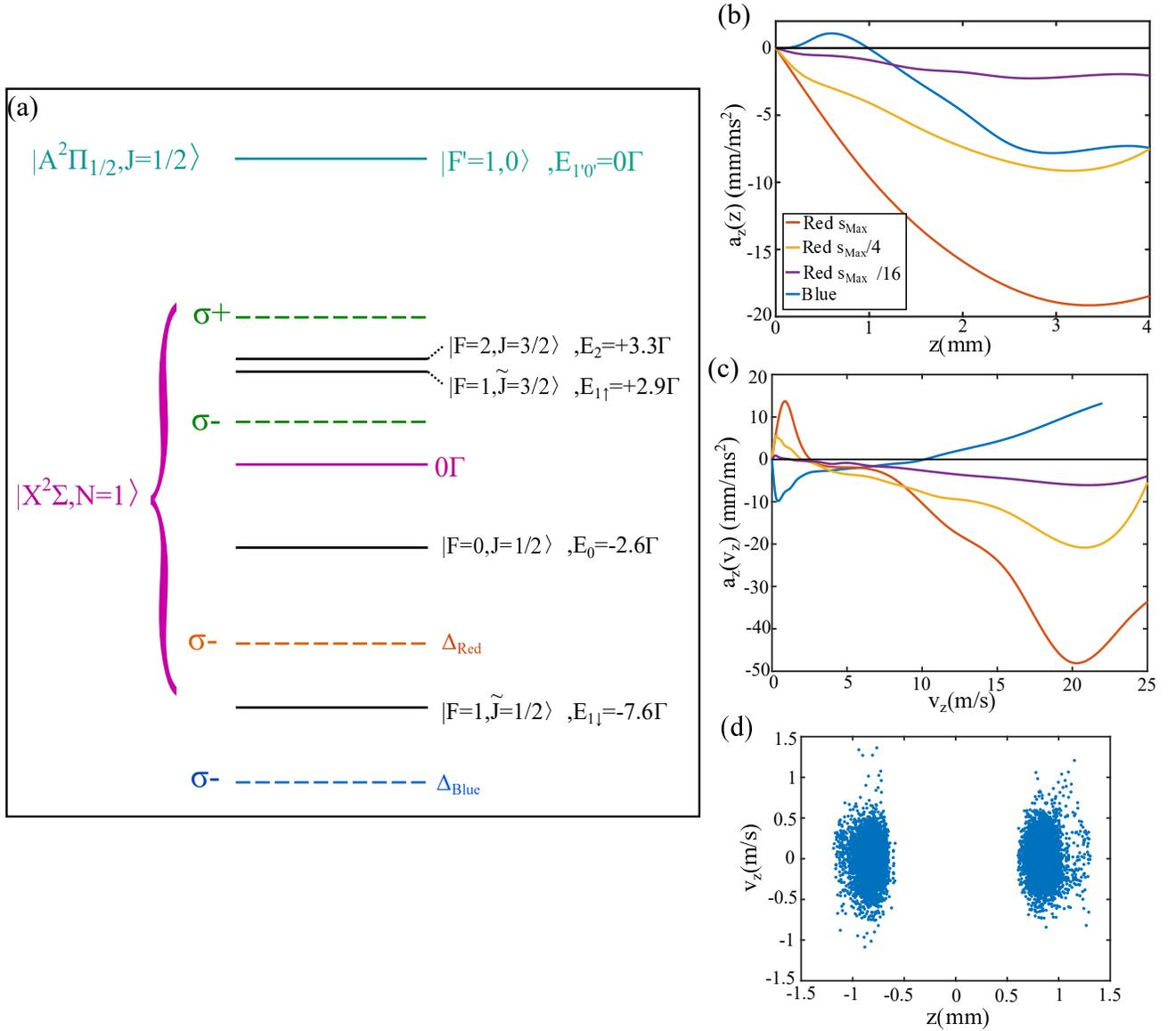}%
		\caption{(a): Scheme for laser cooling and trapping of MgF.  Lasers 1 and 2 (green dashed) are used in all trapping schemes to give a dual frequency mechanism on these states (green dashed lines).  Velocity damping (via detuning either red for Doppler cooling or blue for Sub-Doppler cooling) is accomplished by an additional laser detuned from the $|F=1,\tilde{J}=1/2\rangle$ state.  (b): Simulation results for $a_{z}(z)$ for the cases indicated in Tables~\ref{tb:redMOTMgFParameters} and~\ref{tb:blueMOTMgFParameters}.  Clearly trapping is achieved in all configurations. (c): Results for $a_{z}(v_{z})$.  Again, note the impacts of Doppler and sub-Doppler forces.  (d): Phase-space diagram for blueMOT trajectories with 100 different starting values $z_{0}$ and $v_{z,0}$ derived from the low power red-MOT (where $\sigma=1.24$\,mm and $v_{T}=1.3$\,m/s.)  Molecules collect around two stable points in $z$, as expected from the $a_{z}(z)$ curve for the blueMOT.  \label{fig:MgF}}
	\end{figure}

	The approach described in the previous section should be generalizeable to other molecules with small hyperfine splittings.  One such molecule is MgF.  Its level diagram is illustrated in Fig.~\ref{fig:MgF}(a).  MgF has received much interest recently as a good candidate for molecular cooling and trapping~\cite{kgj2015,xyy2016,xxy2019,dwt2022}, largely because its light mass, large scattering rate ($\Gamma_{XA}/2\pi=22$\,MHz), and high wavenumber/short wavelength of the cycling transition ($\lambda_{XA}=359$\,nm) all enable large decelerating forces.  For example, $a_{max}\propto \hbar k_{XA}\Gamma/m$ is 15x higher for MgF than SrF.  Hence, experiments with MgF should be able to achieve both higher MOT capture velocities and shorter slowing lengths than prior experiments.  
	
	Compared to the molecules discussed in previous sections, MgF has much higher saturation intensity ($I_{sat,MgF}=62$\,mW/cm$^{2}$).  Moreover, less laser power is available at the ultraviolet X$\rightarrow$A wavelength for MgF than for the visible wavelengths used for those other molecules.  Here we restrict the total laser power to a realistic value of 600\,mW, and thus a total saturation parameter (summed over all lasers) of $\sum_{j} s_{j,max}=12$. 
	
	\begin{table}[ht]
		
		\begin{tabular}{|l|l|l|l|l|l|l|}
			\multicolumn{7}{c}{MgF redMOT configuration(Fig.~\ref{fig:MgF})} \\
			\hline
			\multicolumn{4}{|c|}{Laser Parameters} & \multicolumn{3}{c|}{Results} \\
			\hline
			$J$ & $s_{j,max}$ & $\Delta_{HF}/\Gamma$&  $\hat{p}$ & $v_{cap}(s_{max})$& $T(s_{max}/16)$& $\sigma(s_{max}/16)$\\
			\hline
			\textbf{3/2} & \textbf{4} & \textbf{-1.5} &  $\bm{\sigma^{+}}$ & \multirow{3}{2em}{29.0\,m/s} & \multirow{3}{2em}{9\,mK} & \multirow{3}{2em}{1.24\,mm} \\
			\textbf{3/2} & \textbf{4} & \textbf{+1.5} &  $\bm{\sigma^{-}}$ & & &\\
			1/2 ($F=1$) & 4 & -2 & $\sigma^{-}$  & & & \\
			\hline

		\end{tabular}
		
		\caption{Left: Parameters used for MgF redMOT simulation.  Detunings $\Delta_{HF}$ are indexed relative to the indicated $J$ (see Fig.~\ref{fig:MgF}(a)).  Bold font indicates levels that participate in a dual-frequency scheme.  Right: Results.  First column indicates $v_{cap}$ for the `capture redMOT' ($s_{j}=s_{j,max}$ for all lasers, $b_{0}=25$\,G/cm).  In the next two columns, $T$ and $\sigma$ for the low power `compressed redMOT' ($s_{j}=s_{j,max}/16$ for all lasers, $b_{0}=50$\,G/cm) are reported.  In Fig.~\ref{fig:MgF}, we also plot results for $s_{j}=s_{j,max}/4$ and $b_{0}=50$\,G/cm.   \label{tb:redMOTMgFParameters}}
	\end{table}
	
	\begin{table}[ht]
		
		\begin{tabular}{|l|l|l|l|l|l|l|}
			\multicolumn{7}{c}{MgF blueMOT configuration, $b_{0}=50$\,G/cm (Fig.~\ref{fig:MgF})} \\
			\hline
			\multicolumn{4}{|c|}{Laser Parameters} & \multicolumn{3}{c|}{Results} \\
			\hline
			$J$ & $s_{j}$ & $\Delta_{HF}/\Gamma$&  $\hat{p}$ & $t_{D}$& $T$& $\sigma$\\
			\hline
			\textbf{3/2} & \textbf{1} & \textbf{-1.5} &  $\bm{\sigma^{+}}$ & \multirow{3}{3em}{10\,ms} & \multirow{3}{3em}{270$\,\mu$K} & \multirow{3}{5em}{0.84\,mm} \\
			\textbf{3/2} & \textbf{1} & \textbf{+1.5} &  $\bm{\sigma^{-}}$ & & &\\
			1/2 ($F=1$) & 4 & +4 & $\sigma^{-}$  & & & \\
			\hline

		\end{tabular}
		
		\caption{Parameters used for MgF blueMOT simulation.  Detunings $\Delta_{HF}$ are indexed relative to the indicated $J$ (see Fig.~\ref{fig:MgF}(a)).  Bold font indicates levels that participate in a dual-frequency scheme.   \label{tb:blueMOTMgFParameters}}
	\end{table}
	
	The parameters used to generate the redMOT and blueMOT $a_{z}(z)$ and $a_{z}(v_{z})$ curves in Fig.~\ref{fig:MgF} are shown in Tables~\ref{tb:redMOTMgFParameters} and~\ref{tb:blueMOTMgFParameters}, as are the results.  Generally, we follow the same approach as `Test 1' in Sec.~\ref{subsec:OH}, where a dual-frequency mechanism on the $J=3/2$ manifold is used for trapping and light set to either red (Doppler cooling) or blue (sub-Doppler cooling) of $|F=1,\tilde{J}=1/2\rangle$ is used for velocity damping.  We do not directly address $|F=0,J=1/2\rangle$, but instead let it be excited by the blue dual-frequency light that addresses $J=3/2$, which is only $-3\Gamma$ detuned from $|F=0,J=1/2\rangle$.
	
	For the redMOT, we observed, as expected, a higher $v_{cap}$ than in any molecule tested previously.  However, we also observed that sub-Doppler heating forces are much higher here, resulting in high temperatures ($T=9$\,mK, $v_{T}=1.3$\,m/s) even in the `compressed redMOT'.
	
	Similarly, in the blueMOT we observe strong sub-Doppler cooling forces, as well as strong confinement for $|z|>1$\,mm.  The sub-Doppler force is effective out to $v_{z}=8.5$\,m/s, more than sufficient for capturing molecules from the low-power redMOT, even with its high temperature.  Unfortunately, we also observe a reversal in the sign of $a_{z}(z)$ for $|z|\lesssim 1\textrm{mm}$; similar `Sisyphus like' forces in $a_{z}(z)$ have been observed in previous work on simulations of Type-II MOTs~\cite{dta2016}.  This corresponds to there being multiple `stable points' to which molecules can be attracted. Fig.~\ref{fig:MgF}(d) shows a phase-space plot showing long-time trajectories for molecules with initial velocity and position chosen from the distribution in the `compressed redMOT' (Table~\ref{tb:redMOTMgFParameters}).  As expected from the form of $a_{z}(z)$, we observe two stable regions centered at $z=\pm 920\,\mu$m.  Calculating $\sigma=\sqrt{\langle z^{2}\rangle}$ and $T=k_{B}\langle v_{z}^{2}\rangle/m$ from the phase plot, we find $\sigma = 0.84$\,mm and $T=270\,\mu$K, though of course the distribution of $z$ values is far from gaussian.  It may be possible to mitigate this issue by choosing a different set of laser parameters, but we have not attempted such optimization.

	\section{Grand summary of MOT simulations}

	In the last three sections, we discussed a number of ways to improve molecular MOTs.  First, we demonstrated that two-color redMOTs, where different hyperfine manifolds within the X$^{2}\Sigma$ state are excited to two distinct electronic excited states, can lead to increases in capture velocity for both rf and dc configurations (Sec.~\ref{sec:bichromTrapping}).  It has been observed that, for white light slowers, the number of slowed molecules that reach the MOT region increases rapidly with their velocity~\cite{bsd2012}, so the number of captured molecules likely increases as a large power of the capture velocity (at least $\propto v_{cap}^{2}$ and possibly faster).  Hence, we expect this to lead to significant gains in the number of molecules that can be captured.  Finally, we found that compression via lowering the laser power and increasing the field gradient in the optimized two-color redMOT leads to similar reductions in $T$ and $\sigma$ as have been observed in one-color MOTs~\cite{nmd2016,aad2017,dwy2020,twt2017}.
	
	However, the temperatures in all molecular redMOTs observed to date experimentally~\cite{nmd2016,twt2017,aad2017,dwy2020,vhd2022,lhc2022}, and in both our and others' simulations~\cite{dta2018}, are still much higher than $T_{D}$, the Doppler temperature.  This is because of the sub-Doppler heating characteristic of Type-II transitions.  Here, we introduced two-color blueMOTs that yield both sub-Doppler cooling and trapping forces strong enough to recapture all molecules from a compressed redMOT of SrF.  This reduces the spatial extent of the cloud by a factor of $\sim10$, and the temperature by a factor of $\sim 40$, relative to the compressed redMOT.  We believe that this would present a much more favorable starting condition for the subsequent loading of an ODT. 
	
	Finally, we also presented the results of simulations for one-color MOTs of the molecules SrOH, CaOH, and MgF, where, due to small hyperfine energy splittings in the X$^{2}\Sigma$ state, a slightly modified approach is required.  We demonstrated that confinement in can be achieved in these systems using a dual-frequency trapping scheme applied to a pair of hyperfine levels with small splitting, similar to what has been demonstrated in redMOTs for YO~\cite{dwy2020}.  We also showed that redMOT compression and blueMOT capture and cooling can be effective in these species.

	\section{Simulation results II: Methods for improved slowing of a Cryogenic Buffer Gas Beam}
	\label{sec:slowingImprovements}

	Thus far, we have focused on techniques for improving the density $n$ and capture velocity $v_{cap}$ of magneto-optical traps.  Here, we turn our focus to improving the total flux of capturable molecules, i.e., those that reach the MOT capture region (here defined to be $r_{end}\le w_{MOT}$, where $r_{end}$ is the displacement from the $z$-axis when the molecule reaches the end of the slower (Fig.~\ref{fig:moleculeExperimentSchematic}); we take $w_{MOT}=7$\,mm) with $v_{z}<v_{cap}$.  In Sec.~\ref{subsec:expSetup}, we discussed the typical experimental setup used for direct molecular laser-cooling and trapping experiments.  Typical CBGBs have approximately Gaussian longitudinal ($\sigma_{v,z}=25$\,m/s, with $\langle v_{z0}\rangle\sim140$\,m/s) and transverse ($\sigma_{v\perp}=25$\,m/s) velocity distributions~\cite{bsd2011,hld2012}.  Although a CBGB can contain $\sim 10^{11}$ molecules in the $N=1$ state per pulse~\cite{bsd2011}, thus far, only up to $10^{6}$ have ever been captured in a MOT~\cite{aad2017}.  The major inefficiency contributing to this loss is non-ideal behavior of the slowing force at low velocities.   
	
	Ideally, the molecules would be decelerated to $v_{cap}$, and no further; in other words, we would like the slowing curve $a_{z}(v_{z})$ to fall off as sharply as possible for $v\le v_{cap}$.  If the cut-off is more gradual, then molecules with small enough transverse velocity to reach the MOT capture region for $v_{z}=v_{cap}$, but large enough to `miss' for some lower $v_{z}$, will be lost as they continue to be slowed passed the necessary $v_{cap}$ (represented by purple arrows in Fig.~\ref{fig:moleculeExperimentSchematic}(b)).  The more gradual the cut-off, the more molecules are lost due to this `pluming' process.  In extreme cases, even molecules with zero transverse velocity can `miss' the capture region due to slowing to $v_{z}<0$; we refer to this phenomenon, which can occur when $a_{z}(v_{z}\rightarrow 0)<0$, as `overslowing' (blue arrows in Fig.~\ref{fig:moleculeExperimentSchematic}(b)).  
	
	In experiments conducted to date, molecules from the CBGB are slowed either through `white light' slowing (WLS)~\cite{bsd2012}, where the laser is spectrally broadened to cover all Doppler shifts from $kv_{cap}$ up to $k\langle v_{z0}\rangle$ (as well as the full hyperfine spectrum of the molecule, $E_{2,3/2}-E_{1,1/2}$, see Fig.~\ref{fig:levelStructure}); or through chirped slowing~\cite{twt2017NJP}, where the detuning of the laser is shifted dynamically during the slowing process to compensate for the changing Doppler shift as the beam is slowed.  The discussion for the rest of this section consists of comparisons to, and modifications from, WLS on the X$\rightarrow$B transition.  In Fig.~\ref{fig:zeeman}, we show $a(v_{z})$ for WLS.  Unfortunately, the curve has a very gradual cut-off; reduction of the deceleration from its maximum value $a_{max}$ to $a_{max}/2$ occurs over a range in $v_{z}$ of $\Delta v_{cut}\sim$20\,m/s.  Moreover, $a_{z}(v_{z}=0)$ is negative and still rather large.  Per the previous paragraph, this leads to pluming and overslowing.  \footnote{We have also simulated chirped light slowing~\cite{twt2017NJP}, (not shown in Fig.~\ref{fig:zeeman}) and found it to have a similarly gradual cut-off.}
	
	Here, we consider two approaches to sharpening the low velocity cut-off of $a_{z}(v_{z})$, while still maintaining the ability to slow molecules from $\langle v_{z0}\rangle$ down to $v_{cap}$ by the time they reach $z=L$ (Fig.~\ref{fig:moleculeExperimentSchematic}(a)).  In Sec.~\ref{subsec:zeeSlow} we propose a novel version of a Type-II Zeeman slower, and in Sec~\ref{subsec:slowPush}, we discuss using a `push' beam that \textit{co-propagates} with the molecular beam, to accumulate molecules with $v_{z}\approx v_{cap}$ and eliminate overslowing.  Pluming can be reduced by transverse cooling, which we discuss in Sec.~\ref{subsec:trans}.  Finally, in Sec.~\ref{subsec:expGain} we discuss the expected increases in the flux of slowed molecules that result from implementing these techniques.

	\subsection{Zeeman Slowing To Increase Molecule Capture Efficiency}
	\label{subsec:zeeSlow}

	In this section, we show that the deceleration curve has a much sharper cut-off for Zeeman slowers than for the white light slowers that have typically been used in molecular cooling experiments~\cite{bsd2012,twt2017NJP}.  We also demonstrate a novel, two-color, molecular Zeeman slower. 
	
\begin{figure}
		\centering
		\includegraphics{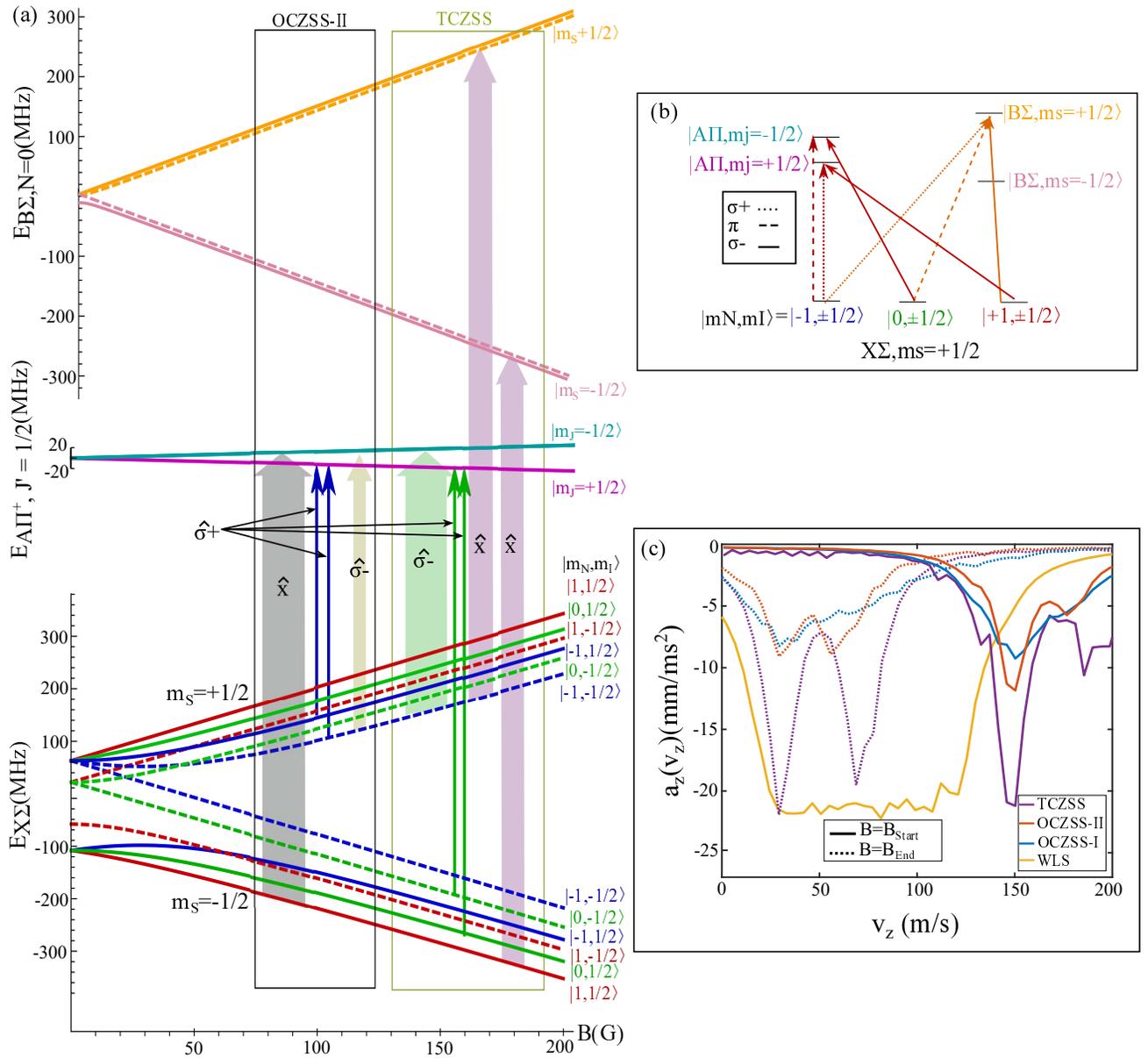}%
		\fl \caption{(a): Schematic illustrating slowing schemes OCZSS-II and TCZSS.  Slower states are addressed with narrow $\sigma^{+}$ light (solid arrows).  All other light is spectrally broadened to address repump states (spectral width indicated by arrow thickness), as described in text.   In the TCZSS, $m_{N}=\pm 1$ states are coupled to B$^{2}\Sigma$ with $\hat{x}$ polarized light broadened to cover the Doppler width and hyperfine splitting, similar to in a `white light' slower.  The same laser addresses both $m_{s}$ branches, since the $B$-field dependence for energies in the $m_{s}=\pm 1/2$ branches of B$^{2}\Sigma$ is similar to that of the X$^{2}\Sigma$ state.  Throughout, dashes indicate states with $m_{I}=-1/2$. (b): Illustration of polarizations that can couple X$^{2}\Sigma$ to the two excited states.  (c): Slowing curves for SrF, using the parameters in Table~\ref{tb:zeemanParams}.  Note that the Zeeman slowing curves drop off at low $v_{z}$ more sharply than the WLS curve.   \label{fig:zeeman}}
	\end{figure}

	There have been two prior proposals published for implementing a molecular (type-II) longitudinal Zeeman slower ($\bm{B}$ field along the molecular beam axis $\hat{z}$), where Zeeman shifts are engineered to cancel Doppler shifts as molecules are slowed (just as in an atomic Zeeman slower~\cite{pme1982}).  This method results in both slowing and cooling of the longitudinal velocity~\cite{pko2018,lby2019}.  In these prior proposals, lasers are used to excite only the X$^{2}\Sigma\rightarrow$A$^{2}\Pi_{1/2}$ transition, and thus we refer to them as `one-color Zeeman slowing schemes' (OCZSS).  A sufficiently large field is applied such that $\mu_{B}B\gg V_{HF}$, resulting in a decoupling of the electronic spin, nuclear spin, and rotational angular momentum~\cite{pko2018}.  This splits the energy spectrum in the X$^{2}\Sigma$ state into two branches, corresponding to $m_{s}=\pm 1/2$, with energies $E_{z}=\pm g_{S}m_{S}\mu_{B}B$ (ignoring nuclear and rotational $g$ factors), see Fig.~\ref{fig:zeeman}(a).  Throughout, we use the approximation $g_{S}\approx 2$.  The energy spectrum in the A$^{2}\Pi_{1/2}$ (B$^{2}\Sigma$) state splits into two branches with $E_{z}\propto \pm g^{\prime}m_{J}\mu_{B}B$ (with $g^{\prime} = g^{\prime}_{J,A\Pi}$ or $g^{\prime}_{J,B\Sigma}$, respectively).  For molecules considered in this paper, $-0.2<g^{\prime}_{J,A\Pi}< 0$ and $2< g^{\prime}_{J,B\Sigma}< 2.2$ (see \ref{sec:CpAndUDeriv}).  
	
	The polarizations required to couple $|$X$^{2}\Sigma,N=1\rangle$ states to $|$A$^{2}\Pi_{1/2},J'=1/2\rangle$ and $|$B$^{2}\Sigma,N'=0\rangle$, in the strong magnetic field, are illustrated in Fig.~\ref{fig:zeeman}(b).  Since the X$^{2}\Sigma$ and B$^{2}\Sigma$ states are both well described by Hund's case (b)~\cite{brown}, the polarization dependence is intuitive: nominally, only states of the same $m_{s}$ and $m_{I}$ are coupled, and states of $m_{N}$ and $m_{N}'=m_{N}+q$ are coupled by light with polarization having spherical vector projection $q$.  The polarization dependence for coupling to the A$^{2}\Pi_{1/2}$ state is less intuitive, due to it being well described by Hund's case (a).  Because a Hund's case (a) eigenstate is a superposition of case (b) eigenstates~\cite{brown}, states with $m_{s}=1/2$ can couple to either of the $m_{J}'=\pm1/2$ manifolds. 
	
	As discussed in Sec.~\ref{sec:level}, slowing molecules without a Zeeman slower requires broadening the spectrum of a laser counter-propagating to the molecular beam source in order to cover both the molecular Doppler shift ($k(\langle v_{z,0}\rangle-v_{cap})/2\pi\sim200$\,MHz) and the frequency difference between $|2,3/2\rangle$ and $|1,1/2\rangle$ (typically $\sim 200$\,MHz) resulting from hyperfine and spin-rotation splitting (Fig.~\ref{fig:levelStructure}).  In both prior molecular Zeeman slower proposals, the Zeeman and Doppler shifts are designed to cancel along the $m_{s}=1/2$ branch of the X state, and thus we refer to $m_{s}=1/2$ as the `slower branch'~\cite{pko2018,lby2019}.  On the other hand, in the $m_{s}=-1/2$ branch, the Doppler shift and Zeeman shift `anti compensate', and thus an additional laser must be applied.  This laser must be spectrally broadened to cover all three shifts (Zeeman, Doppler, and hyperfine structure).  Here, we refer to the $m_{s}=-1/2$ branch as the `repump' branch.
	
	The two prior proposals differ in how they handle the hyperfine structure in the slower branch.  In~\cite{pko2018}, hereafter referred to as OCZSS-I, $\hat{y}$ linearly polarized light (a linear combination of $\sigma^{\pm}$ polarizations) is broadened to cover the whole hyperfine structure.  In~\cite{lby2019} (OCZSS-II), separate $\sigma^{-}$ and $\sigma^{+}$ polarized lasers are used; the $\sigma^{-}$ beam is broadened to cover transitions from the $|m_{S}=1/2,m_{N}=0,1,m_{I}=\pm 1/2\rangle$ states, while the $\sigma^{+}$ laser has two narrow and discrete components that address the $|m_{S}=1/2,m_{N}=-1,m_{I}=\pm 1/2\rangle$ states.  Thus, unless the Zeeman and Doppler shifts exactly cancel (which occurs at some `design velocity' $v$ for a given field), molecules become 'stuck' in the `slower states' $|\psi_{\pm 1/2}\rangle =|m_{S}=1/2,m_{N}=-1,m_{I}=\pm 1/2\rangle$.  This was originally proposed for use with BaF~\cite{lby2019}.  In both slowers, $\hat{x}$ polarized light (another orthogonal linear combination of $\sigma^{\pm}$) is broadened to cover all three frequency shifts in the `repump' branch.  These proposals are illustrated in Fig.~\ref{fig:zeeman}(a).   
	
	Here, we present a new approach to Zeeman slowing, hereafter referred as a two-color Zeeman slower scheme (TCZSS), that can be used for any molecule with a usable $\textrm{B}^{2}\Sigma$ state (i.e., one with favorable Franck-Condon factors), including CaF and SrF.  This excitation pattern for the TCZSS is illustrated in Fig.~\ref{fig:zeeman}(a).  For the molecules considered here, the \textit{relative} Zeeman shift between states of the same spin in B$^{2}\Sigma$ and X$^{2}\Sigma$ is negligible, since $g^{\prime}_{J,B\Sigma}-g_{S}<0.1$ (~\ref{sec:CpAndUDeriv}).  Thus, $\Delta m_{s}=0$ transitions here are insensitive to the magnetic field, making them a poor choice for providing the magnetic field-selective transitions necessary for Doppler compensation.  However, this same feature makes them a natural choice for repumping (see Fig.~\ref{fig:zeeman}(a)).  Here, the Zeeman sensitivity is provided by narrow $\sigma^{+}$ lasers that couple the `slower states' $|\psi_{\pm 1/2}\rangle=|m_{S}=-1/2,m_{N}=0,m_{I}=\pm 1/2\rangle$ states to $\textrm{A}^{2}\Pi_{1/2}$.  The $|m_{S}=+1/2,m_{N}=0,m_{I}=\pm 1/2\rangle$ states in the `repump' branch are repumped by coupling to $\textrm{A}^{2}\Pi_{1/2}$ with a broadened $\sigma^{-}$ laser; for this `repump' branch, the spectral breadth must be sufficient to cover the combined Zeeman and Doppler shifts.  Finally, all other states (i.e., all those with $m_{N}\neq 1$) are repumped by a single $\hat{x}$ polarized laser driving the X$\rightarrow$B transition.  Since this transition is Zeeman insensitive, it only needs to be spectrally broadened to cover the Doppler shift and the hyperfine splitting in the X$^{2}\Sigma$ manifold.
	
	By introducing an additional excited electronic state, the scattering rate, $R_{s}=\Gamma p_{Exc}$ (where $p_{Exc}$ is the fraction of molecules in the excited state), and thus the slowing acceleration ($a_{z}\propto R_{s}$), increases.  In the approximation $\Gamma_{XB}\approx \Gamma_{XA} = \Gamma$ (true for all molecules considered here), the maximum scattering rate is $R_{s}\approx\Gamma\times n_{E}/(n_{E}+n_{G})$~\cite{mvd}.  Here, $n_{G}=24$ is the total number of `ground' states (including substructure of $v=1$ as well as $v=0$), and $n_{E}$ is the total number of `excited' states, with either $n_{E}=4$ (if only exciting to A$^{2}\Pi_{1/2}$) or $n_{E}=8$ (when exciting to both states as in this new proposal).  Hence, driving both electronic transitions can increase the slowing force by a factor of nearly 2.
	
	In Fig.~\ref{fig:zeeman}(c) we plot $a_{z}(v_{z})$ at two points along a SrF Zeeman slower for this new TCZSS, as well as the OCZSS's of Refs.~\cite{pko2018,lby2019}.  The parameters used in these simulations are shown in Table~\ref{tb:zeemanParams}.  All detunings $\Delta_{j}$ are indexed relative to the hyperfine-free resonance frequencies ($\omega_{XA}$ and $\omega_{XB}$ in Fig.~\ref{fig:levelStructure}).  As expected, we observe a significantly stronger force for the TCZSS.  We also compare Zeeman slowing to WLS in Fig.~\ref{fig:zeeman}(c).   Here, the WLS excites the X$\rightarrow$B transition while vibrationally pumping on the X$\rightarrow$A transition (see Table 8), avoiding the $\Lambda$ structure that slows scattering in any one-color scheme.  Though the peak magnitudes of the forces are similar, the $a_{z}(v_{z})$ curve for the TCZSS falls off much more sharply for velocities away from resonance.  This significantly mitigates losses from pluming and overslowing.
	
	\begin{table}[ht]
		
	\begin{tabular}{|l|l|l|l|l|l|l|}
		\multicolumn{7}{c}{OCZSS-I~\cite{pko2018}, $B_{Start}=340$\,G ($v=150$\,m/s) $B_{End}=465$\,G ($v=30$\,m/s)} \\
		\hline
		X$^{2}\Sigma$ states addressed & Transition & $s_{j}$ & $\Delta_{j}/\Gamma$& $\beta_{j}$ (rad) & $\Omega_{j}/\Gamma$ & $\hat{p}$ \\
		\hline
		$m_{s}=+1/2$ (all) & $X,v=0\rightarrow A$ & 40 & -113 &  9 & 1 & $\hat{y}$\\
		\hline
		$m_{s}=-1/2$ & $X,v=0\rightarrow A$ & 360 & +82 &  45 & 1 & $\hat{x}$\\
		\hline
		$m_{s}=+1/2$ ($v=1$ repump) & $X,v=1\rightarrow A$ & 180 & -113 &  20 & 1 & $\hat{y}$\\
		\hline
		$m_{s}=-1/2$ ($v=1$ repump) & $X,v=1\rightarrow A$ & 450 & +82 &  60 & 1 & $\hat{x}$\\
		\hline
		
		\multicolumn{7}{c}{ }\\
		\multicolumn{7}{c}{OCZSS-II~\cite{lby2019}, $B_{Start}=340$\,G ($v=150$\,m/s) $B_{End}=465$\,G ($v=30$\,m/s)} \\
		\hline
		X$^{2}\Sigma$ states addressed & Transition & $s_{j}$ & $\Delta_{j}/\Gamma$& $\beta_{j}$ (rad) & $\Omega_{j}/\Gamma$ & $\hat{p}$ \\
		\hline
		$m_{s}=+1/2$, $|m_{N}=-1,m_{I}=1/2\rangle$ & $X,v=0\rightarrow A$ & 2 & -111.3 &  0 & 0 & $\sigma^{+}$\\
		\hline
		$m_{s}=+1/2$, $|-1,-1/2\rangle$& $X,v=0\rightarrow A$ & 2 & -103.8 &  0 & 0 & $\sigma^{+}$\\
		\hline
		$m_{s}=+1/2$ (remaining states) & $X,v=0\rightarrow A$ & 40 & -109 &  12 & 1 & $\sigma^{-}$\\
		\hline
		$m_{s}=-1/2$ & $X,v=0\rightarrow A$ & 360 & +82 &  45 & 1 & $\hat{x}$\\
		\hline
		$m_{s}=+1/2$ ($v=1$ repump) & $X,v=1\rightarrow A$ & 180 & -113 &  20 & 1 & $\hat{y}$\\
		\hline
		$m_{s}=-1/2$ ($v=1$ repump) & $X,v=1\rightarrow A$ & 450 & +82 &  60 & 1 & $\hat{x}$\\
		\hline
		
		\multicolumn{7}{c}{ }\\
		\multicolumn{7}{c}{TCZSS, $B_{Start}=478$\,G ($v=150$\,m/s) $B_{End}=335$\,G ($v=30$\,m/s)} \\
		\hline
		X$^{2}\Sigma$ states addressed & Transition & $s_{j}$ & $\Delta_{j}/\Gamma$& $\beta_{j}$ (rad) & $\Omega_{j}/\Gamma$ & $\hat{p}$ \\
		\hline
		$m_{s}=-1/2$, $|0,1/2\rangle$& $X,v=0\rightarrow A$ & 1 & +63.2 &  0 & 0 & $\sigma^{+}$\\
		\hline
		$m_{s}=-1/2$, $|0,-1/2\rangle$ & $X,v=0\rightarrow A$ & 1 & +53.8 &  0 & 0 & $\sigma^{+}$\\
		\hline
		$m_{s}=\pm1/2$ (all $m_{N}=\pm1$ states) & $X,v=0\rightarrow B$ & 450 & -21.9 &  60 & 1 & $\hat{x}$\\
		\hline
		$m_{s}=+1/2$ ($m_{N}=0$) & $X,v=0\rightarrow A$ & 225 & -91.8 &  60 & 1 & $\sigma^{-}$\\
		\hline
		$m_{s}=-1/2$ ($v=1$ repump) & $X,v=1\rightarrow A$ & 180 & +69 &  20 & 1 & $\hat{y}$\\
		\hline
		$m_{s}=+1/2$ ($v=1$ repump) & $X,v=1\rightarrow A$ & 450 & -102.4 &  60 & 1 & $\hat{x}$\\
		\hline
		
		\multicolumn{7}{c}{ }\\
		\multicolumn{7}{c}{WLS, $\bm{B}=B_{0}(\hat{x}+\hat{y})/\sqrt{2}$ where $B_{0}$=5\,G} \\
		\hline
		X$^{2}\Sigma$ states addressed & Transition & $s_{j}$ & $\Delta_{j}/\Gamma$& $\beta_{j}$ (rad) & $\Omega_{j}/\Gamma$ & $\hat{p}$ \\
		\hline
		All & $X\rightarrow B$ & 450 & -25.6 &  44 & 0.6 & $\hat{x}$\\
		\hline

	\end{tabular}

	\caption{Parameters used for simulations of molecular beam slowers for SrF shown in Fig.~\ref{fig:zeeman}(c).  The velocities listed in the title correspond to the (primary) peak of the deceleration curves observed in Fig.~\ref{fig:zeeman}.  These parameters were chosen to work well for a SrF beam source with a longitudinal velocity distribution described by $\langle v_{z,0}\rangle\sim 140$\,m/s and $\sigma_{vz}=25$\,m/s.  The values of $B_{End}$ and $B_{Start}$ could be optimized for any given experiment, to slow the largest fraction of molecules to $v\le v_{cap}$.  For the WLS, a field transverse to the slowing beam set at a 45$^{\circ}$ with respect to the polarization is used to remix Zeeman dark states.   \label{tb:zeemanParams}}
\end{table}
	
	We also find that the $a_{z}(v_{z})$ curves for TCZSS and OCZSS-II are narrower than for OCZSS-I.  This is because, in the latter, addressing all 6 states in the slowing branch with a single polarization makes it more likely for molecules with velocities outside of the desired velocity (at a given field) to be off-resonantly repumped (since all 6 frequency components can address the all 6 substates).  In the other schemes, the designated slowing states can, \textit{in principle}, only be addressed by the two narrowband lasers.  \footnote{Due to residual hyperfine mixing, there is still some likelihood in both TCZSS and OCZSS-II for one of the two `slower states' (in both cases, $|\psi_{1/2}\rangle$) to be repumped.  In TCZSS, the X$\rightarrow$B light drives this repumping, which could not happen without hyperfine mixing since the slower states have $m_{N}=0$ and thus require $\hat{\pi}$ polarization (Fig.~\ref{fig:zeeman}(b)). In OCZSS-II, this is driven by by the moderately broadened $\sigma^{-}$ light; again, without hyperfine mixing this could not happen, as only $\sigma^{+}$ and $\pi$ light can address the slower levels, which have $m_{N}=-1$ (Fig.~\ref{fig:zeeman}(b)).  This hyperfine-induced coupling is responsible for the `second' peak observed in the deceleration curves for these slowers (Fig.~\ref{fig:zeeman}(c)), which arises when the laser resonant with $|\psi_{1/2}\rangle$ at the design velocity $v_{1}$ for a given field is Doppler-shifted to resonance with the other slower state $|\psi_{-1/2}\rangle$, for molecules at a different velocity $v_{2}$.  As a result, at $v_{2}$ as well as $v_{1}$, both $|\psi_{-1/2}\rangle$ and $|\psi_{1/2}\rangle$ can cycle photons.  The peak in scattering rate for $v_{2}$ is lower and broader than for $v_{1}$.  To avoid overslowing, the slower should be designed such that the sharper peak is at lower velocity.  \textit{This} is why we use `slower states' states in the $m_{s}=-1/2$ manifold for the TCZSS.}
	
	In general, the optimal configuration for a molecular Zeeman slower will depend on the species considered.   In \ref{sec:ZeeCaFSrOH}, we discuss the results of Zeeman slower simulations for CaF and SrOH, and demonstrate that our TCZSS usefully generalizes to other species.

	\subsection{`Pushed' White Light slowing to increase molecule capture efficiency}
	\label{subsec:slowPush}
	
\begin{figure*}[h!]
		\centering
		\includegraphics[scale=1]{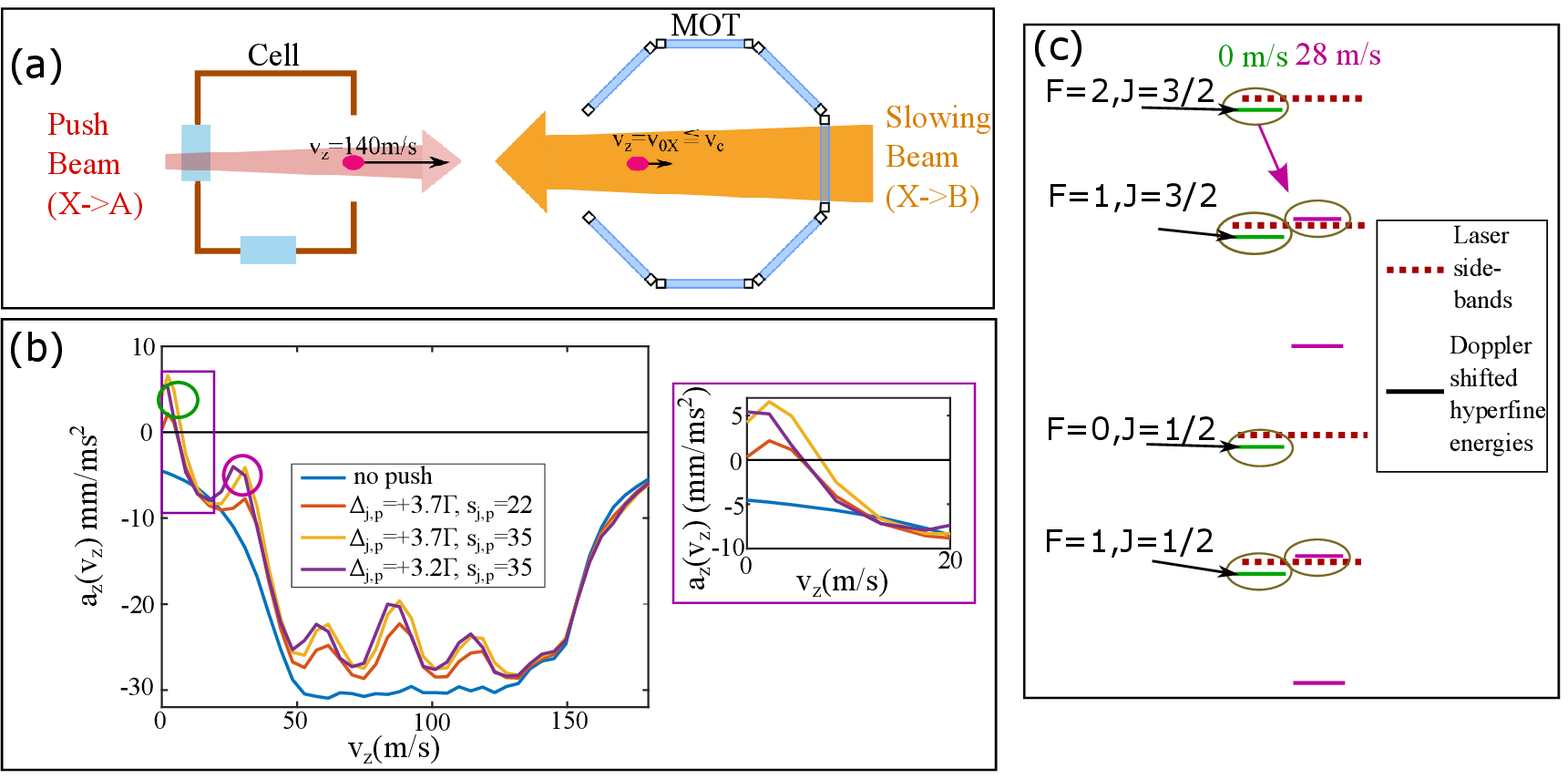}
		\caption{(a) Scheme for `pushed' white light slowing.  In addition to a X$\rightarrow$B `white light' slower beam (orange), we apply a near-resonant, counter-propagating, X$\rightarrow$A `push' beam (red).  The velocity of the molecule (magenta) is reduced to $v_{z}=v_{0X}$ by the time it reaches the MOT.  (b): Results from simulations of pushed white-light slowing of SrF.  For all cases, the slower parameters are \{$s_{j,s}$, $\Delta_{j,s}/\Gamma$, $\beta_{j,s}$(rad), $\Omega_{j,s}/\Gamma$\}=\{280, -22.4, 44, 0.6\}.  Adding the push adds a small positive acceleration near $v_{z}=0$ (green circle) and hence creates a zero crossing in the total deceleration curve at a finite positive velocity $v_{0X}$.  The value of $v_{0X}$ can be tuned by adjusting the saturation parameter ($s_{j,p}$) and detuning ($\Delta_{j,p}$) of the push laser (inset). Structure in the deceleration curve (e.g., additional peak indicated with a magenta circle) results from molecular hyperfine structure.  (c): Effect of Doppler shift on push laser.  We phase-modulate the push beam with $\beta_{j,p}=2.5$ and $\Omega_{j,p}/\Gamma = 6.5$; this results in a beam with sidebands resonant with each hyperfine state for atoms with velocity near 0\,m/s (green).  However, at some nonzero velocities, such as 28\,m/s for the case of SrF (magenta), some hyperfine states again become resonant with laser sidebands (circles).  This causes the `bumps' in the deceleration curves shown in (b). }
		\label{fig:whiteLightPush}
	\end{figure*}

	Engineering a sharp cut-off in the $a_{z}(v_{z})$ curve can also be accomplished by adding a push beam that counter-propagates with the slowing laser, and is near resonant for $v_{z}=0$ molecules (Fig.~\ref{fig:whiteLightPush}(a)).  The principle is similar that of the `2D-plus' MOT, where additional beams along the atomic beam axis are used to guide the longitudinal velocity to a small but non-zero value; this serves to guide atoms to a 3D-MOT with velocities low enough to be captured~\cite{dsw1998}.  
	
	Adding a push beam enables a tunable zero-crossing at some velocity $v_{0X}$ in the $a(v_{z})$ curve.  The intensity ($s_{j,p}$) and detuning ($\Delta_{j,p}$) of the push beam can be adjusted to tune the value of $v_{0X}$ such that $0\le v_{0X}\lesssim v_{cap}$.  This should completely eliminate any over-slowing, as molecules will `pile-up' at $v_{0X}$, and also significantly mitigate losses due to pluming.
	
	In Fig.~\ref{fig:whiteLightPush}(b), we display results of simulations of pushed white light slowing of SrF.   As anticipated, it is indeed possible to tune the position of the zero crossing by adjusting the frequency and intensity of the push beam.  Adding the push beam does slightly reduce the effectiveness of the white light slower, particularly at certain velocities where some sidebands of the push beam are Doppler shifted to resonance with hyperfine levels other than the ones they are designed to address at $v_{z}\sim 0$ (Fig.~\ref{fig:whiteLightPush}(c)).  However, this only reduces the maximum deceleration of the WLS by $\sim 20$\%, and this effect can be mitigated by only turning on the push beam towards the end of the slowing time.

	\subsection{Transverse Cooling To Reduce Molecular Pluming}
	\label{subsec:trans}
	
	Recently, two dimensional gray-molasses cooling was demonstrated to decrease the transverse temperature of a CBGB of YbF from 25\,mK to $T<220\,\mu$K\cite{alt2021}, using the X$^{2}\Sigma\rightarrow \textrm{A}^{2}\Pi_{1/2}$ transition.  In that experiment, however, simultaneous transverse cooling \textit{and} longitduinal slowing was not considered; this likely will be required for transverse cooling to aid in MOT loading (see Fig.~\ref{fig:moleculeExperimentSchematic}(b)).  Since gray-molasses cooling relies on cycling between `bright' and `dark' states~\cite{weh1994,dta2016}, it is possible that introduction of a longitudinal slowing laser may disrupt this mechanism, as it can out-couple molecules from the dark-states of the transverse cooling.
	
\begin{figure}
		\centering
		\includegraphics{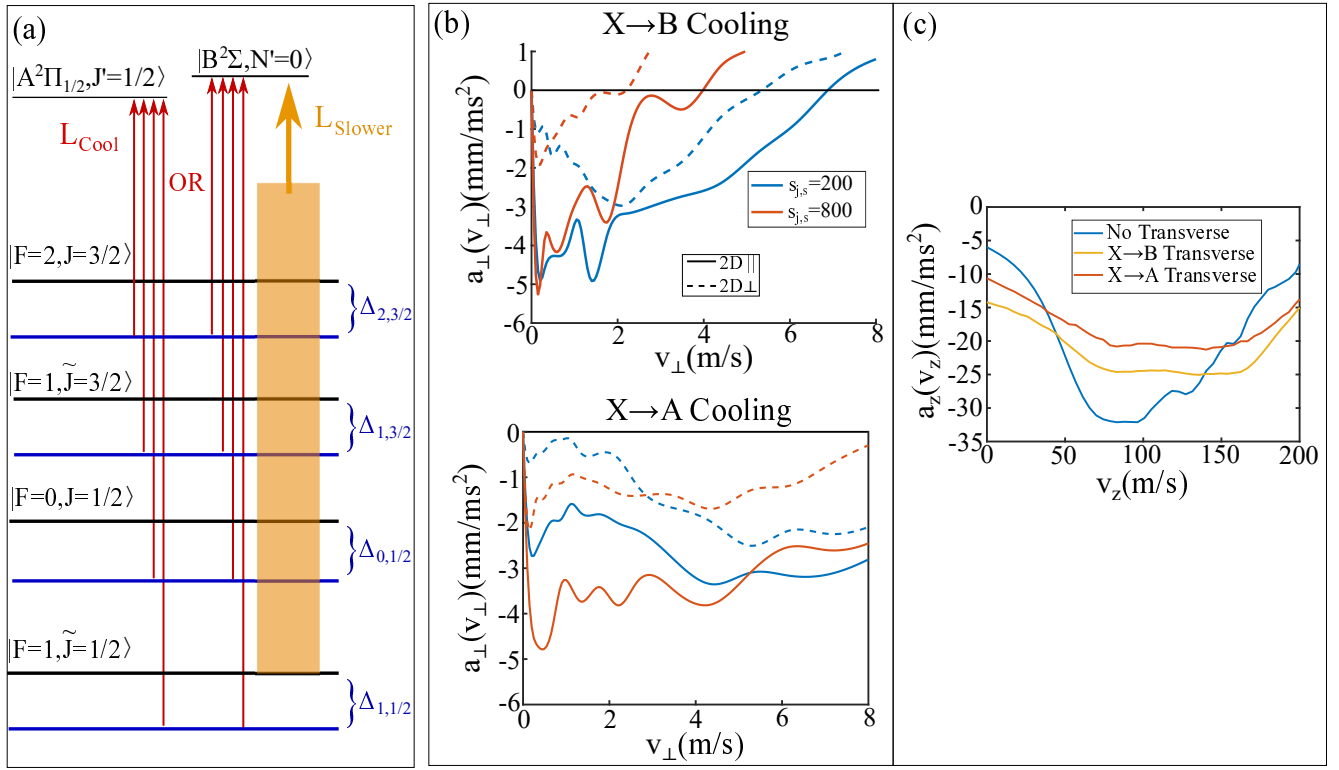}%
		\caption{(a): Level diagram with transverse cooling and longitudinal slowing.  For the simulations done here, blue-detuned transverse cooling, addressing all hyperfine levels in the $|\textrm{X}^{2}\Sigma,N=1\rangle$ state, is applied on either the $\textrm{X}\rightarrow \textrm{A}$ or $\textrm{X}\rightarrow \textrm{B}$ transition.  White-light slowing is applied on the $\textrm{X}\rightarrow \textrm{B}$ transition, with spectral breadth indicated by the shaded orange area.  (b): Results for the case described in the text.  The transverse cooling is robust to and, for some configurations, enhanced by, the presence of slowing light.  (c): Slowing curves $a_{z}(v_{z})$ with and without the transverse cooling light. \label{fig:trans}}
	\end{figure}

	Here, we consider simultaneous transverse cooling (testing both X$^{2}\Sigma\rightarrow$A$^{2}\Pi_{1/2}$ and X$^{2}\Sigma\rightarrow$ B$^{2}\Sigma$ transitions) and WLS (for the simulations here, the white light slower parameters are \{$\Delta_{j,s}/\Gamma$, $\beta_{s}$(rad), $\Omega_{s}/\Gamma$\}=\{-25.4, 44, 0.6\}) for SrF.  As in~\cite{alt2021}, we consider 2D$_{\parallel}$ (with both transverse cooling lasers polarized along $\hat{z}$) and 2D$_{\perp}$ (lasers cross polarized, one along $\hat{x}$ and the other along $\hat{z}$) configurations.  Here, we assume lasers address all 4 hyperfine components at a common detuning $\Delta_{T}=+5\Gamma$ (Fig.~\ref{fig:trans}(a)), each with intensity $s_{j,T}=20$ (finite laser width is not considered for the transverse cooling simulations.).  All transverse cooling simulations are done at zero $B$-field.
	
	 We observe that adding the longitudinal slowing does not completely preclude transverse cooling (Fig.~\ref{fig:trans}(b)).  In fact, for X$\rightarrow$A cooling, increasing the slowing power (saturation parameter $s_{j,s}$) actually enhances the  transverse cooling forces for $v_{\perp}\lesssim 3$\,m/s.  \footnote{We speculate that in `standard' gray molasses (e.g. without the addition of the longitudinal slower), molecules addressed by this transition spend a sub-optimally long time in dark-states~\cite{dta2016}.}
	
	Another interesting feature is that the 2D$_{\parallel}$ configuration works at all, even though we have simulated with zero $B$-field.  Typically, with $B=0$, molecules in this configuration would be immediately pumped into $|F=2,m_{F}=\pm 2\rangle$ states, which are not coupled to the excited state by the $\hat{z}$ polarized transverse cooling light. In~\cite{alt2021}, it was shown that a $B$-field of 1\,G was sufficient to cycle molecules out of the dark states.  Here, instead, the longitudinal slowing light, with polarization $\hat{x}$, provides the remixing. 
	
	Finally, we examined the effect of transverse cooling on the longitudinal slowing curve (here assumed to be from WLS, with $s_{j,s}=800$ and all other parameters the same as those used in Fig.~\ref{fig:whiteLightPush}), see Fig.~\ref{fig:trans}(c).  We find that adding the transverse light causes the maximum deceleration to diminish by $\sim 30\%$.  The transverse light also appears to broaden the range of $v_{z}$ over which longitudinal slowing is effective.  In absence of transverse cooling, the `edges' of the slowing curve are determined by when the red (blue) portion of the broadened slowing light is Doppler shifted out of resonance with $|F=2,J=3/2\rangle$ ($|F=1,J=1/2\rangle$).  When transverse light is overlapped with the slowing light, the former can out-couple population from those states, allowing for further scattering from the WLS beam (similar to the mechanism discussed above).  As discussed earlier, a gradual cut-off of $a_{z}(v_{z})$ near $v_{z}=0$ makes it more likely for molecules to be lost by pluming or overslowing.  However, since transverse cooling will primarily be applied towards the beginning of the slowing region (Fig.~\ref{fig:moleculeExperimentSchematic}(b)), this should not be an issue.  
	
	\subsection{Expected gains in trappable molecular flux from these improvements}
	\label{subsec:expGain}
	
	In order to determine which of these changes can yield the most benefit to MOT loading, we next consider the expected gain in trappable molecular flux.  Throughout, we assume a slowing length of $L=1$\,m (see Fig.~\ref{fig:moleculeExperimentSchematic}).  For cases where the transverse slower is implemented, we assume that it is applied from $z=10$\,cm to $z=20$\,cm ($z=0$ at the cell aperture).
	
	In all cases, we determine $a_{z}(z,r,v_{z})$ and $a_{\perp}(z,v_{\perp})$, where $r$ is the transverse displacement from the $z$ axis and $a_{\perp}=\bm{a}\cdot\bm{\hat{r}}$.  The dependence on $r$ comes from the finite width of the slowing beam.  The dependence on $z$ comes from focusing the slowing beams; we choose the $1/e^{2}$-radius to be 2.5\,mm at the cell and 7.5\,mm at the MOT, as in typical experiments~\cite{smd2016}. For the Zeeman slower, there is an additional dependence on $z$ due to the changing $B$-field.  Here, we assume the field has a functional form $B(z) = B_{start}+(B_{end}-B_{start})(z/L)^{2}$.  For cases where a push beam is implemented, we take it to have the same spatial profile as the slowing beams.  
	
	We assume that molecules emanate from the cell aperture with initially uncorrelated $r$ and $v_{\perp}$ (due to in-cell collisions with He), and molecular origins are spread evenly over the area of the aperture of the cell (here assumed to be 3\,mm in diameter).  We assume mean longitudinal velocity $\langle v_{z}\rangle = 140$\,m/s, longitudinal velocity spread $\sigma_{vz}=25$\,m/s, and transverse spread $\sigma_{v\perp}=25$\,m/s~\cite{bsd2011}.  We take the MOT laser beam waists to be $w_{MOT}=7$\,mm.
	
	We then determine whether a molecule with a given set of initial values $v_{z0}$, $v_{\perp,0}$, and $r_{0}$, evolving under $a_{z}$ and $a_{r}$, (i) makes it to $z=L$ without being turned around (i.e. overslowed), (ii) is within the MOT capture volume (i.e., when it reaches $z=L$, it has transverse displacement $r_{end}$ such that $r_{end}\le w_{MOT}$), and (iii) has final velocity $v_{z,End}\le v_{cap}$ when $z=L$.  If all conditions are met, then the molecule is taken to be capturable. 
	
	\begin{figure}
		\centering
		\includegraphics{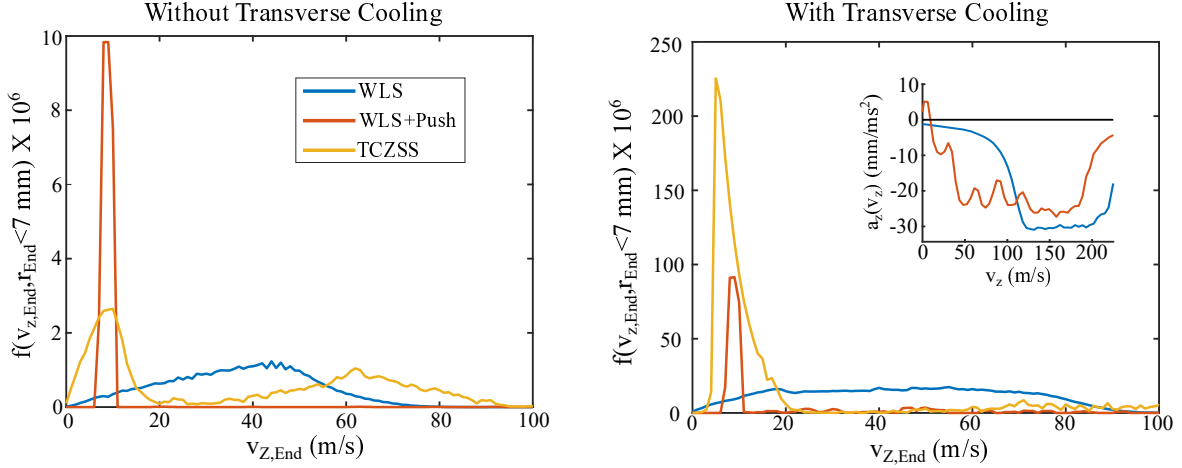}%
		\caption{Longitudinal velocity distribution of molecules that are within the capture radius when the reach the end of the slower, for various slowing configurations (here, those that optimize $f_{cap}(10\,\textrm{m/s})$).  We adjust the WLS slower detuning to seperately optimize $f_{cap}$ with and without the push beam; the resulting $a_{z}(v_{z},z=L,r=0)$ for WLS (blue) and WLS+Push (red) are shown in an inset (parameters used for TCZSS are the same as in Table~\ref{tb:zeemanParams} and Fig.~\ref{fig:zeeman}(c)).  The difference in the $a_{z}(v_{z})$ curve is what causes the absence of molecules at $v_{z}\gtrsim 10$\,m/s for the WLS+Push configuration. \label{fig:endDistro}}
	\end{figure}
	
	\begin{table}[ht]
		
		\begin{tabular}{|l|l|l|l|l|}
			\multicolumn{5}{c}{Fraction of capturable molecules $f_{cap}(v_{cap})$ ($\times 10^{6}$) for various slowing configurations} \\
			\hline
		    Slowing configuration & $f_{cap}(5\,\textrm{m/s})$ & $f_{cap}(10\,\textrm{m/s})$ & $f_{cap}(15\,\textrm{m/s})$ & $f_{cap}(20\,\textrm{m/s})$  \\
			\hline
			WLS & 0.44 & 1.8 & 4.0 & 6.8\\
			WLS + Push & 26 & 34 & 42 & 49\\
			TCZSS & 5 & 17 & 31 & 44\\
			WLS + Transverse Cooling & 15 & 61 & 120 & 200\\
			WLS + Push + Transverse Cooling & 130 & 270 & 560 & 1200\\
			TCZSS + Transverse Cooling & 400 & 1000 & 1300 & 1400\\
			\hline
			
		\end{tabular}
		
		\caption{Fraction of molecules that are capturable (as defined in main text) for various slowing configurations. \label{tb:fracCaptured}}
	\end{table}

	 In Table~\ref{tb:fracCaptured}, we list the fraction of capturable molecules that meet these conditions for various slowing configurations and values of $v_{cap}$, and in Fig.~\ref{fig:endDistro}, we plot the longitudinal velocity distribution $f(v_{z,End})$ of the subset of molecules that are within the capture volume when they reach of the slowing region, when the slowers are optimized to maximize the capturable fraction for $v_{cap}=10$\,m/s.  By either adding a push beam to a WLS, or switching to a TCZSS, gains of $\gtrsim 10$ in capturable fraction $f_{cap}(v_{cap})$ for $v_{cap}\sim 10$\,m/s can be achieved. Similar gains can be achieved by adding a transverse cooling stage to a WLS.  Implementing both transverse cooling and one of the `improved' slowing schemes leads to gains as large as $\gtrsim 100$ in MOT number, according to the simulations. 
	 
	 As noted in Sec.~\ref{subsec:SrFRed}, $v_{cap}$ will depend on $r_{end}$.  We have simplified this by just setting the capture probability to 0 for $r_{end}>w_{MOT}$, and to 1 if $r_{end}<w_{MOT}$ and $v_{z}<v_{cap}$, and thus the values of $f_{cap}(v_{cap})$ in Table~\ref{tb:fracCaptured} are to be taken only as an approximate measure.  Since this approximation was applied for all slowing configurations, we expect that the relative differences will be generally accurate.
	
	\section{Conclusion}

	Here, we have proposed, and used simulations to validate, a number of new techniques which should lead to improvements in the number, density and temperature of trapped molecular clouds produced by direct laser cooling and trapping.  Specifically, the number of molecules loaded into a MOT can potentially be increased by a factor of $\sim 100$ or more via improvements to both the MOT capture velocity and the flux of slowed molecules and, furthermore, the volume of the trapped cloud can be compressed by a factor of $\sim 100$ by implementing a blueMOT, which should increase the fraction of molecules loaded from the MOT into an ODT from $\sim 5$\%~\cite{ljd2021} to near unity.  These improvements, when all combined, have the potential to increase the number of molecules loaded into an ODT by a factor of $\sim 10^{3}$ or more.  
	
	Many laboratories are now seeking to achieve quantum degeneracy using directly laser cooled and trapped molecules.  Most such experiments are currently limited by the low molecule numbers and densities that can be initially loaded into an ODT, since this limits how quickly evaporative cooling can take place.  Enhancing the initial density would allow for evaporation to occur on a timescale faster than the many loss-rates in the system caused, e.g. by inelastic collisions~\cite{jhi2011,iju2010,cad2020}, phase noise-induced transitions from the microwave shielding~\cite{abd2021,sbl2022} used to avoid said inelastic collisions~\cite{lqu2018,khu2018}, and black-body radiation induced vibrational transitions~\cite{vdu2007}.  Achieving quantum degeneracy in directly cooled molecules will increase the chemical diversity of quantum degenerate molecular systems dramatically; benefits include the additional spin degree of freedom afforded by the unpaired spin in $^{2}\Sigma$ molecules~\cite{mbz2006}, the ability to study a more diverse range of quantum chemical reactions~\cite{ibk2020}, the potential for quantum degenerate gases of triatomic~\cite{vhd2022,hvd2022} or even polyatomic~\cite{adk2020,mld2022} laser coolable molecules, and advantages for precision measurements~\cite{lah2018,ald2020,khu2017,nleEDM2018}.
	
	\ack
	
	This work was supported by AFOSR.  We are grateful to Varun Jorapur, Qian Wang, and Geoffrey Zheng for helpful advice and discussion.
	
	\appendix
	
	\section{Derivation of the terms $C_{p}$ and $U$ for $^{2}\Sigma$ ground state molecules with $J$-Mixing included}
	\label{sec:CpAndUDeriv}
	
	In this section, we derive the coupling matrix $C_{p}$ and the unitary matrix $U$ for converting between Hamiltonians expressed in the Zeeman-basis $|m_{S},m_{N},m_{I}\rangle$ and the Hyperfine-basis $|F,J,m_{F}\rangle$, with $J$-mixing included.  The levels $|F=1,J=3/2\rangle$ and $|F=1,J=1/2\rangle$ are mixed by the hyperfine interaction, which has a non-diagonal hamiltonian in the $|F,J,m_{F}\rangle$ basis~\cite{swh1996}.  Thus, we express them as:
	
	\begin{eqnarray}
			\nonumber|F=1,\tilde{J}=3/2\rangle = a|F=1,J=3/2\rangle+b|F=1,J=1/2\rangle\\
			|F=1,\tilde{J}=1/2\rangle = -b|F=1,J=3/2\rangle+a|F=1,J=1/2\rangle
	\end{eqnarray}
	
	\noindent where the tilde indicates that it is a mixed state.  In Table~\ref{tb:aAndBParams} we list the values of $a$ and $b$ for all of the molecules discussed in the paper.
	
	\begin{table}[ht]
		
		\begin{tabular}{|l|l|l|}
			\hline
			Molecule & $a$ & $b$\\
			\hline
			SrF & 0.888 & 0.460\\
			CaF & 0.772 & 0.635\\
			MgF & 0.699 & 0.715\\
			SrOH \& CaOH & 1 & 0\\ 
			\hline

		\end{tabular}
		
		\caption{$J$-Mixing parameters for the molecules discussed in the main text.  The mixing is negligible in for the cases with hydroxide ligands due to the minimal hyperfine interaction. \label{tb:aAndBParams}}
	\end{table}
	
	Throughout this section, the ordering of the X$^{2}\Sigma$ hyperfine states is given in Table~\ref{tb:XStateOrder} and the ordering of both the A$^{2}\Pi_{1/2}$ and B$^{2}\Sigma$ hyperfine states are given in Table~\ref{tb:ABStateOrder}
	
	\begin{table}[ht]
		
		\begin{tabular}{|l|l|}
			$m$& $|F,\tilde{J},m_{F}\rangle$\\
			\hline
			1 & $|1,1/2,-1\rangle$\\
			\hline
			2 & $|1,1/2,0\rangle$\\
			\hline
			3 & $|1,1/2,+1\rangle$\\
			\hline
			4 & $|0,1/2,0\rangle$\\
			\hline
			5 & $|1,3/2,-1\rangle$\\
			\hline
			6 & $|1,3/2,0\rangle$\\
			\hline
			7 & $|1,3/2,+1\rangle$\\
			\hline
			8 & $|2,3/2,-2\rangle$\\
			\hline
			9 & $|2,3/2,-1\rangle$\\
			\hline
			10 & $|2,3/2,0\rangle$\\
			\hline
			11 & $|2,3/2,1\rangle$\\
			\hline
			12 & $|2,3/2,2\rangle$\\
			\hline			
		\end{tabular}
		
		\caption{Order in which X$^{2}\Sigma$ hyperfine states appear in matrix expressions \label{tb:XStateOrder}}
	\end{table}
	
		\begin{table}[ht]
		
		\begin{tabular}{|l|l|}
			$n$& $|F',J',m_{F}'\rangle$\\
			\hline
			1 & $|1,1/2,-1\rangle$\\
			\hline
			2 & $|1,1/2,0\rangle$\\
			\hline
			3 & $|1,1/2,+1\rangle$\\
			\hline
			4 & $|0,1/2,0\rangle$\\	
			\hline
		\end{tabular}
		
		\caption{Order in which A$^{2}\Pi_{1/2}$ and B$^{2}\Sigma$ hyperfine states appear in matrix expressions \label{tb:ABStateOrder}}
	\end{table}
	
	\subsection{$C_{p}$ for $|$X$^{2}\Sigma, N=1\rangle\rightarrow |$B$^{2}\Sigma, N'=0\rangle$ transitions}
	
	For this transition, both states are well described by Hund's case (b), in which the set of quantum numbers that describes the molecular state is $|\Lambda=0,S=1/2,I=1/2,N,F,J,m_{F}\rangle$~\cite{brown}.  Our task is then to determine the matrix elements $C_{p,mn}=\langle F,N,J,m_{F}|T^{1}_{p}(\bm{d})|F',N',J',m_{F}'\rangle$, where $\bm{d}$ is the dipole-moment operator, $p$ is the polarization, $m$ refers to the set of quantum numbers \{$F,J,m_{F},N=1$\} in the ground state and $n$ to the excited state quantum numbers \{$F',J'=1/2,m_{F}',N'=0$\}.  Using the Wigner-Eckhart and spectator~\cite{brown} theorems, we ultimately obtain:
	
	\begin{equation}
		\langle F,N,J,m_{F}|T^{1}_{p}(\bm{d})|F',N',J',m_{F}'\rangle=c_{1}c_{2}c_{3}c_{4}
	\end{equation}
	
	\noindent where
	
	\begin{eqnarray}
			\nonumber c_{1} = (-1)^{F-m_{F}}\left(\begin{array}{ccc}F&1&F'\\-m_{F}&-p&m_{F}'\end{array}\right)\\
			c_{2} = (-1)^{F'+J+1+I}\sqrt{(2F+1)(2F'+1)}\left\{\begin{array}{ccc}J'&F'&I\\F&J&1\end{array}\right\}\\
			\nonumber c_{3} = (-1)^{J'+N+1+S}\sqrt{(2J+1)(2J'+1)}\left\{\begin{array}{ccc}N'&J'&S\\J&N&1\end{array}\right\}\\
			\nonumber c_{4} = (-1)^{N}\sqrt{(2N+1)(2N'+1)}\left(\begin{array}{ccc}N&1&N'\\0&0&0\end{array}\right).
	\end{eqnarray}
	
	With this, we can determine the $C_{p,mn}$ matrix elements (rows correspond to X state, columns to B state) for the  $|$X$^{2}\Sigma, N=1\rangle\rightarrow |$B$^{2}\Sigma, N'=0\rangle$ transition.  
	
	\begin{equation}
		C_{p=-1} = \left[\begin{array}{cccc}0&0&0&0\\\frac{a}{3}-\frac{b}{3\sqrt{2}}&0&0&0\\0&\frac{a}{3}-\frac{b}{3\sqrt{2}}&0&-\frac{a}{3}-\frac{b\sqrt{2}}{3}\\\frac{1}{3}&0&0&0\\0&0&0&0\\\frac{a}{3\sqrt{2}}+\frac{b}{3}&0&0&0\\0&\frac{a}{3\sqrt{2}}+\frac{b}{3}&0&\frac{a\sqrt{2}}{3}-\frac{b}{3}\\0&0&0&0\\0&0&0&0\\\frac{1}{3\sqrt{2}}&0&0&0\\0&\frac{1}{\sqrt{6}}&0&0\\0&0&\frac{1}{\sqrt{3}}&0\end{array}\right]\\
	\end{equation}

	\begin{equation}
		C_{p=0} = \left[\begin{array}{cccc}\frac{a}{3}-\frac{b}{3\sqrt{2}}&0&0&0\\0&0&0&-\frac{a}{3}-\frac{b\sqrt{2}}{3}\\0&0&-\frac{a}{3}+\frac{b}{3\sqrt{2}}&0\\0&-\frac{1}{3}&0&0\\\frac{a}{3\sqrt{2}}+\frac{b}{3}&0&0&0\\0&0&0&\frac{a\sqrt{2}}{3}-\frac{b}{3}\\0&0&-\frac{a}{3\sqrt{2}}-\frac{b}{3}&0\\0&0&0&0\\\frac{1}{\sqrt{6}}&0&0&0\\0&\frac{\sqrt{2}}{3}&0&0\\0&0&\frac{1}{\sqrt{6}}&0\\0&0&0&0\end{array}\right]\\
	\end{equation}

	\begin{equation}
		C_{p=+1} = \left[\begin{array}{cccc}0&-\frac{a}{3}+\frac{b}{3\sqrt{2}}&0&-\frac{a}{3}-\frac{b\sqrt{2}}{3}\\0&0&-\frac{a}{3}+\frac{b}{3\sqrt{2}}&0\\0&0&0&0\\0&0&\frac{1}{3}&0\\0&-\frac{a}{3\sqrt{2}}-\frac{b}{3}&0&\frac{a\sqrt{2}}{3}-\frac{b}{3}\\0&0&-\frac{a}{3\sqrt{2}}-\frac{b}{3}&0\\0&0&0&0\\\frac{1}{\sqrt{3}}&0&0&0\\0&\frac{1}{\sqrt{6}}&0&0\\0&0&\frac{1}{3\sqrt{2}}&0\\0&0&0&0\\0&0&0&0\end{array}\right]\\
	\end{equation}
	
	\subsection{$C_{p}$ for $|$X$^{2}\Sigma, N=1\rangle\rightarrow |$A$^{2}\Pi_{1/2}, J'=1/2\rangle$ transitions}
	
	Unlike the X and B states, the A$^{2}\Pi_{1/2}$ state is well described by Hund's case (a), in which the set of good quantum numbers is $|\Lambda,S,\Sigma,\Omega,I,F,J,m_{F}\rangle$.  To determine the matrix elements between the case (b) X state and the case (a) A State, we follow the procedure outlined in~\cite{wkt2008}.  First, we express the X$^{2}\Sigma$ states in the case (a) basis~\cite{brown}:
	
	\begin{equation}
		\fl|\Lambda;N,S,J\rangle = \sum_{\Omega=-1/2}^{1/2}\sum_{\Sigma=-1/2}^{1/2}(-1)^{J+\Omega}\sqrt{2N+1}\left(\begin{array}{ccc}S&N&J\\ \Sigma&\Lambda&-\Omega\end{array}\right)|\Lambda,S,\Sigma,\Omega,J\rangle
		\label{eq:XInCaseA}
	\end{equation}

	We also must express the A$^{2}\Pi_{1/2}^{+}$ state as a sum of states $|\Lambda=+1,\Sigma=-1/2,\Omega=+1/2\rangle$ and $|\Lambda=-1,\Sigma=+1/2,\Omega=-1/2\rangle$.  For the positive parity basis~\cite{brown}:
	
	\begin{equation}
		\fl|\textrm{A}^{2}\Pi_{1/2}^{+}\rangle = \frac{1}{\sqrt{2}}\left(|\Lambda=+1,\Sigma=-1/2,\Omega=+1/2\rangle+|-1,+1/2,-1/2\rangle\right)
		\label{eq:caseAParity}
	\end{equation}
	
	Finally, we use the Wigner-Eckhart and spectator theorems to decompose the terms $C_{p,mn}=\langle \Lambda,S,\Sigma,\Omega,I,F,J,m_{F}|T^{1}_{p}(\bm{d})|\Lambda',S,\Sigma',\Omega',I,F',J',m_{F}'\rangle$, similarly to~\cite{wkt2008,brown}:
	
	\begin{equation}
		\fl\langle \Lambda,S,\Sigma,\Omega,I,F,J,m_{F}|T^{1}_{p}(\bm{d})|\Lambda',S,\Sigma',\Omega',I,F',J',m_{F}'\rangle=c_{1}c_{2}c_{3},
	\end{equation}
	
	\noindent where
	
	\begin{eqnarray}
		\nonumber c_{1} = (-1)^{F-m_{F}}\left(\begin{array}{ccc}F&1&F'\\-m_{F}&-p&m_{F}'\end{array}\right)\textrm{;}\\
		\label{eq:xToADecompose} c_{2} = (-1)^{F'+J+1+I}\sqrt{(2F+1)(2F'+1)}\left\{\begin{array}{ccc}J'&F'&I\\F&J&1\end{array}\right\}\textrm{; and}\\
		\nonumber c_{3} = \sum_{q=-1}^{1}(-1)^{J-\Omega}\left(\begin{array}{ccc}J&1&J'\\-\Omega&q&\Omega'\end{array}\right).
	\end{eqnarray}

	Applying Eq.~\ref{eq:xToADecompose} to the X$\rightarrow$A transition, with X expressed in the case (a) basis through Eq.~\ref{eq:XInCaseA} and A decomposed into the positive parity basis case (Eq.~\ref{eq:caseAParity}), we find (rows correspond to X state, columns to A state):
	
	\begin{equation}
		C_{p=-1} = \left[\begin{array}{cccc}0&0&0&0\\\frac{a\sqrt{2}}{3}+\frac{b}{6}&0&0&0\\0&\frac{a\sqrt{2}}{3}+\frac{b}{6}&0&-\frac{a\sqrt{2}}{3}+\frac{b}{3}\\\frac{\sqrt{2}}{3}&0&0&0\\0&0&0&0\\-\frac{a}{6}+\frac{b\sqrt{2}}{3}&0&0&0\\0&-\frac{a}{6}+\frac{b\sqrt{2}}{3}&0&-\frac{a}{3}-\frac{b\sqrt{2}}{3}\\0&0&0&0\\0&0&0&0\\-\frac{1}{6}&0&0&0\\0&-\frac{1}{2\sqrt{3}}&0&0\\0&0&-\frac{1}{\sqrt{6}}&0\end{array}\right]\\
	\end{equation}
	
	\begin{equation}
		C_{p=0} = \left[\begin{array}{cccc}\frac{a\sqrt{2}}{3}+\frac{b}{6}&0&0&0\\0&0&0&-\frac{a\sqrt{2}}{3}+\frac{b}{3}\\0&0&-\frac{a\sqrt{2}}{3}-\frac{b}{6}&0\\0&-\frac{\sqrt{2}}{3}&0&0\\-\frac{a}{6}+\frac{b\sqrt{2}}{3}&0&0&0\\0&0&0&-\frac{a}{3}-\frac{b\sqrt{2}}{3}\\0&0&\frac{a}{6}-\frac{b\sqrt{2}}{3}&0\\0&0&0&0\\-\frac{1}{2\sqrt{3}}&0&0&0\\0&-\frac{1}{3}&0&0\\0&0&-\frac{1}{2\sqrt{3}}&0\\0&0&0&0\end{array}\right]\\
	\end{equation}
	
	\begin{equation}
		C_{p=+1} = \left[\begin{array}{cccc}0&-\frac{a\sqrt{2}}{3}-\frac{b}{6}&0&-\frac{a\sqrt{2}}{3}+\frac{b}{3}\\0&0&-\frac{a\sqrt{2}}{3}-\frac{b}{6}&0\\0&0&0&0\\0&0&\frac{\sqrt{2}}{3}&0\\0&\frac{a}{6}-\frac{b\sqrt{2}}{3}&0&-\frac{a}{3}-\frac{b\sqrt{2}}{3}\\0&0&\frac{a}{6}-\frac{b\sqrt{2}}{3}&0\\0&0&0&0\\-\frac{1}{\sqrt{6}}&0&0&0\\0&-\frac{1}{2\sqrt{3}}&0&0\\0&0&-\frac{1}{6}&0\\0&0&0&0\\0&0&0&0\end{array}\right]\\
	\end{equation}
	
	\subsection{Derivation of $U$, the unitary matrix used to convert from the Zeeman basis to the Hyperfine basis}
	
	As described in Sec.~\ref{subsubsec:uB}, for simulations in which the zeeman term $H_{z}\propto \mu_{B}B> H_{HF}$, where $H_{HF}$ is the typical energy scale for hyperfine splitting, the approximation $\langle-\bm{\hat{\mu}\cdot \hat{B}}\rangle=\mu_{B}\bm{B\cdot F}$ is not valid.  This applies to all Zeeman slower simulations and all simulations involving SrOH, CaOH, and MgF, all of which have very small hyperfine splitting in the X$^{2}\Sigma$ level.  For these cases it is convenient to express the $-\bm{\hat{\mu}\cdot \hat{B}}$ term in the $|m_{s},m_{N},m_{I}\rangle$ basis.  
	
	The elements for the unitary matrix that converts from a basis $|\psi\rangle$ to another basis $|\phi\rangle$ can be expressed as $U_{mn}=\langle\psi_{m}|\phi_{n}\rangle$.  Converting the Hamiltonian as expressed in basis $|\psi\rangle$ ($H^{\psi}$) to basis $|\phi\rangle$ is accomplished via $H^{\phi}=U^{\dagger}H^{\psi}U$.
	
	In this case, $|\psi\rangle = |m_{N},m_{S},m_{I}\rangle$ and $|\phi\rangle = |F,J,m_{F}\rangle$.  The terms in $U_{mn}$ are then just the Clebsch-Gordan coefficients that arise from first decomposing $F$ into $J+I$ and then $J$ into $N+S$:
	
	\begin{eqnarray}
		\nonumber\fl\langle N,m_{N},S,m_{S},I,m_{I}|N,S,I,F,J,m_{F}\rangle = (-1)^{-J+I-m_{F}}\sqrt{2F+1}\left(\begin{array}{ccc}J&I&F\\m_{J}=m_{S}+m_{N}&m_{I}&-m_{F}\end{array}\right)\\
		\times (-1)^{-N+S-m_{S}-m_{I}}\sqrt{2J+1}\left(\begin{array}{ccc}N&S&J\\m_{N}&m_{S}&-m_{J}=-m_{N}-m_{S}\end{array}\right)
	\end{eqnarray}

	Throughout, this section, the ordering of the $|m_{S},m_{N},m_{I}\rangle$ states in X$^{2}\Sigma$ is given by Table~\ref{tb:XStateZeeOrder}.
	
	\begin{table}[ht]
		
		\begin{tabular}{|l|l|}
			$m$& $|m_{S},m_{N},m_{I}\rangle$\\
			\hline
			1 & $|-1/2,-1,-1/2\rangle$\\
			\hline
			2 & $|-1/2,-1,+1/2\rangle$\\
			\hline
			3 & $|-1/2,0,-1/2\rangle$\\
			\hline
			4 & $|-1/2,0,+1/2\rangle$\\
			\hline
			5 & $|-1/2,-+1,-1/2\rangle$\\
			\hline
			6 & $|-1/2,+1,+1/2\rangle$\\
			\hline
			7 & $|+1/2,-1,-1/2\rangle$\\
			\hline
			8 & $|+1/2,-1,+1/2\rangle$\\
			\hline
			9 & $|+1/2,0,-1/2\rangle$\\
			\hline
			10 & $|+1/2,0,+1/2\rangle$\\
			\hline
			11 & $|+1/2,+1,-1/2\rangle$\\
			\hline
			12 & $|+1/2,+1,+1/2\rangle$\\
			\hline			
		\end{tabular}
		
		\caption{Order in which X$^{2}\Sigma$ Zeeman states appear in matrix expressions \label{tb:XStateZeeOrder}}
	\end{table}
	
	With the field-free $J$-mixing between states of the same $F$ but different $J$ included, $U$ can be expressed by:
	
	\begin{equation}
		\fl \resizebox{1\hsize}{!}{$U_{^{2}X\Sigma}=\left[\begin{array}{cccccccccccc}0&0&0&0&0&0&0&1&0&0&0&0\\\frac{b\sqrt{3}}{2}&0&0&0&\frac{-a\sqrt{3}}{2}&0&0&0&\frac{1}{2}&0&0&0\\\frac{a}{\sqrt{3}}-\frac{b}{\sqrt{6}}&0&0&0&\frac{b}{\sqrt{3}}&0&0&0&\frac{\sqrt{2}}{2}&0&0&0\\0&\frac{a}{\sqrt{6}}&0&-\frac{1}{\sqrt{6}}&0&\frac{b}{\sqrt{6}}-\frac{a}{\sqrt{3}}&0&0&0&\frac{1}{\sqrt{3}}&0&0\\0&\frac{a}{\sqrt{3}}&0&\frac{1}{\sqrt{3}}&0&\frac{b}{\sqrt{3}}+\frac{a}{\sqrt{6}}&0&0&0&\frac{1}{\sqrt{6}}&0&0\\0&0&\frac{\sqrt{2}a}{\sqrt{3}}+\frac{b}{\sqrt{12}}&0&0&0&\frac{\sqrt{2}b}{\sqrt{3}}-\frac{a}{\sqrt{12}}&0&0&0&\frac{1}{2}&0\\-\frac{\sqrt{2}a}{\sqrt{3}}-\frac{b}{\sqrt{12}}&0&0&0&-\frac{b\sqrt{2}}{\sqrt{3}}+\frac{a}{\sqrt{12}}&0&0&0&\frac{1}{2}&0&0&0\\0&-\frac{a}{\sqrt{3}}+\frac{b}{\sqrt{6}}&0&\frac{1}{\sqrt{3}}&0&-\frac{b}{\sqrt{3}}-\frac{a}{\sqrt{6}}&0&0&0&\frac{1}{\sqrt{6}}&0&0\\0&-\frac{a}{\sqrt{6}}-\frac{b}{\sqrt{3}}&0&-\frac{1}{\sqrt{6}}&0&-\frac{b}{\sqrt{6}}+\frac{a}{\sqrt{3}}&0&0&0&\frac{1}{\sqrt{3}}&0&0\\0&0&-\frac{a}{\sqrt{3}}+\frac{b}{\sqrt{6}}&0&0&0&-\frac{b}{\sqrt{3}}-\frac{a}{\sqrt{6}}&0&0&0&\frac{\sqrt{2}}{2}&0\\0&0&-\frac{b\sqrt{3}}{2}&0&0&0&\frac{a\sqrt{3}}{2}&0&0&0&\frac{1}{2}&0\\0&0&0&0&0&0&0&0&0&0&0&1\end{array}\right]$}
	\end{equation}

	\noindent where rows correspond to the Zeeman basis and columns to the hyperfine basis.

	The Zeeman term is treated slightly differently for B$^{2}\Sigma$ and A$^{2}\Pi_{1/2}$.  In these cases, the Zeeman term is expressed in $|m_{J},m_{I}\rangle$.  For example, for $\bm{B}=B\hat{z}$, this gives:
	
	\begin{equation}
		\fl-\bm{\hat{\mu}\cdot \hat{B}}=H_{z}^{m_{J},m_{I}}=\sum_{m_{J},m_{I}}g'_{J}\mu_{B}Bm_{J}|m_{J},m_{I}\rangle\langle m_{J},m_{I}|
	\end{equation}
	
	\noindent The values of $g'_{J}$ in A$^{2}\Pi_{1/2}$ and B$^{2}\Sigma$ for the molecules discussed in this paper are shown in Table~\ref{tb:gFExcited}.  The non-zero $g_{J}$-factor in the A state (and the corresponding departure from $g_{J}=g_{s}\approx 2$ for the B State) arises primarily from mixing between the A and B states by rotational and spin-orbit interaction~\cite{brown,tar2015,dts2015,cur1965}.
	
	\begin{table}[ht]
		
		\begin{tabular}{|l|l|l|}
			\hline
			Molecule & $g'_{J,A^{2}\Pi}$ & $g'_{J,B^{2}\Sigma}$\\
			\hline
			SrF & -0.166 & 2.166\\
			CaF & -0.04 & 2.04\\
			MgF & -0.0004 & 2.0004\\
			SrOH & -0.192 & 2.192\\ 
			CaOH & -0.08 & 2.08\\
			\hline
		\end{tabular}
		
		\caption{List of $g_{F}$ for electronic excited states A$^{2}\Pi_{1/2}$ and B$^{2}\Sigma$ for molecules  \label{tb:gFExcited}}
	\end{table}
	
	The ordering of $|m_{J},m_{I}\rangle$ here is given by Table~\ref{tb:ABStateZeeOrder}:
	
	\begin{table}[ht]
		\begin{tabular}{|l|l|}
			$m$& $|m_{J},m_{I}\rangle$\\
			\hline
			1 & $|-1/2,-1/2\rangle$\\
			\hline
			2 & $|-1/2,+1/2\rangle$\\
			\hline
			3 & $|+1/2,-1/2\rangle$\\
			\hline
			4 & $|+1/2,0,+1/2\rangle$\\
			\hline	
		\end{tabular}
		
		\caption{Order in which A$^{2}\Pi_{1/2}$ and B$^{2}\Sigma$ Zeeman states appear in matrix expressions \label{tb:ABStateZeeOrder}}
	\end{table}
	
	Finally, we can write $U$ for the electronically excited states
	
	\begin{equation}
		U_{^{2}A\Pi}=U_{^{2}B\Sigma}=\left[\begin{array}{cccc}1&0&0&0\\0&\frac{1}{\sqrt{2}}&0&-\frac{1}{\sqrt{2}}\\0&\frac{1}{\sqrt{2}}&0&\frac{1}{\sqrt{2}}\\0&0&1&0\end{array}\right],
	\end{equation}
	
	\section{Deriving $E_{p}$ for the cases simulated in this paper}
	\label{sec:EpDeriv}
	
	 As explained in Sec.~\ref{subsubsec:de}, the simulation requires a determination of $\tilde{E}_{p}(\bm{r})=\langle \bm{E}\cdot\hat{n}_{p}\rangle/\mathcal{E}_{0}$ where $\mathcal{E}_{0}$ is the magnitude of the electric field applied by the laser and $p$ refers to the polarization expressed in spherical coordinate basis vectors ($\hat{n}_{\pm1}=\sigma^{\pm}$ where $\sigma^{\pm}=(\mp\hat{x}- i\hat{y})/\sqrt{2}$ and $\hat{n}_{0}=\pi=\hat{z}$) .  Here, we derive $E_{p}$ for the slowing simulations and MOT simulations presented in the text.  We will largely be following the approach described in~\cite{dct1989}.
	 
	 In the Zeeman slowing simulations described in Sec.~\ref{subsec:zeeSlow}, beams propagate along the $-z$ axis.  The magnetic field is oriented along $z$ axis (e.g. this is a `longitudinal' slower).  After making the rotating wave approximation, the field can be written as $\bm{E}(\bm{r},t)=\mathcal{E}_{0}\hat{\epsilon}\exp[-ikz]\exp[-i\omega t] = \mathcal{E}_{0}\hat{\epsilon}\left(\cos[kz]-i\sin[kz]\right)\exp[-i\omega t]$, where $\hat{\epsilon}$ is the polarization.  The following polarizations are all used in the slowing simulations: $\hat{x}$, $\hat{y}$, $\sigma^{\pm}$.  Frequency $\omega$ refers to the `remaining' frequency after moving to a frame co-rotating with the frequency associated with a given transition (e.g., $\omega_{m} = \nu_{m}-\omega_{XA}$ for laser $m$ with frequency $\nu_{m}$ addressing the $X\rightarrow A$ transition).  The $\exp[-i\omega t]$ term is accounted for explicitly in Eq.~\ref{eq:dETerm} of the main text, and so does not contribute to $\tilde{E}_{p}$. The corresponding $\tilde{E}_{p}$ values for the slowing simulation are given in Table~\ref{tb:EPSlower}.
	
	\begin{table}[ht]
		\begin{tabular}{|l|l|l|l|}
			$\hat{\epsilon}$& $\tilde{E}_{-1}$& $\tilde{E}_{0}$ & $\tilde{E}_{+1}$\\
			\hline
			$\sigma^{+}$ & 0 & 0 & $\left(\cos[kz]-i\sin[kz]\right)$\\
			\hline
			$\sigma^{-}$  & $\left(\cos[kz]-i\sin[kz]\right)$ & 0 & 0\\
			\hline
			$\hat{x}$ & $\left(\cos[kz]-i\sin[kz]\right)/\sqrt{2}$ & 0 & $-\left(\cos[kz]-i\sin[kz]\right)/\sqrt{2}$\\
			\hline
			$\hat{y}$  & $i\left(\cos[kz]-i\sin[kz]\right)/\sqrt{2}$ & 0 & $i\left(\cos[kz]-i\sin[kz]\right)/\sqrt{2}$\\
			\hline	
		\end{tabular}
		
		\caption{Normalized polarization terms used in the slowing simulations.  Note that $\pi$ transitions cannot be driven in this `longitudinal' slower configuration. \label{tb:EPSlower}}
	\end{table}

	For the MOT simulations, $3L$ pairs of counter-propagating and counter-circulating lasers are used, where $L$ is the total number of laser frequencies used in the simulation.  For simplicity, we define $\hat{x}' = (\hat{x}+\hat{z})/\sqrt{2}$, $\hat{y}' = (\hat{x}-\hat{z})/\sqrt{2}$, and $\hat{z}'=\hat{y}$ such that $\hat{x}'$, $\hat{y}'$, and $\hat{z}'$ are aligned with the MOT beams (see Fig.~\ref{fig:phaseExample}(a)) of the main text, and $\hat{z}'$ is aligned with the anti-helmholtz coil axis.  Each laser will either be in the $\sigma^{+}-\sigma^{-}$ configuration on the $x'$ and $y'$ axes and in the $\sigma^{-}-\sigma^{+}$ configuration on the $z'$ axis (because the sign of the magnetic field dependence as a function of the displacement the $z'$-axis is reversed relative to the $x'$ and $y'$ axes for a pair of coils in the anti-Helmholtz configuration oriented along the $z'$ axis), or will have the exact opposite polarization; in the main text, the polarization of the beam traveling along the $+x'$ axis is what is indicated by the $\sigma^{\pm}$ displayed in column $\hat{p}$ in Table~\ref{tb:rateEqParams}, for example.  Polarizations of both signs are required to drive dual-frequency transitions and/or to address hyperfine levels with different signs of $g$ factor when driving an rfMOT.  
	
	Here, we consider the field from a single laser frequency; in the simulation, the contributions from each laser frequency are summed after being calculated separately (see Eq.~\ref{eq:dETerm}). The fields resulting from each pair of counter-propagating and counter-circulating beams can be written as~\cite{dct1989}:
	
	\begin{itemize}
		\item $E_{x'Beams} = -\sqrt{2}i\mathcal{E}_{0}\left[a\hat{y}'\sin kx'+\hat{z}'\cos kx'\right]\exp\left[\frac{-2\left(z'^{2}+y'^{2}\right)}{w^{2}_{MOT}}\right]$
		\item $E_{y'Beams} = -\sqrt{2}i\mathcal{E}_{0}\left[a\hat{z}'\sin ky'+\hat{x}'\cos ky'\right]\exp\left[\frac{-2\left(x'^{2}+z'^{2}\right)}{w^{2}_{MOT}}\right]$
		\item $E_{z'Beams} = -\sqrt{2}i\mathcal{E}_{0}\left[-a\hat{x}'\sin kz'+\hat{y}'\cos kz'\right]\exp\left[\frac{-2\left(y'^{2}+x'^{2}\right)}{w^{2}_{MOT}}\right]$
	\end{itemize}
	
	\noindent where, as before, here we ignore the $\exp[-i\omega t]$ term, since it is handled explicitly in Eq.~\ref{eq:dETerm}.  The term $a=\pm1$ for lasers labeled $\sigma^{\pm}$.  The total field is the sum of these contributions, and can be expressed as
	
	\begin{eqnarray}
		\nonumber \fl \bm{E} = -\sqrt{2}\mathcal{E}_{0}i\left(\cos ky'\exp\left[\frac{-2\left(x'^{2}+z'^{2}\right)}{w^{2}_{MOT}}\right]-a\sin kz'\exp\left[\frac{-2\left(y'^{2}+x'^{2}\right)}{w^{2}_{MOT}}\right]\right)\hat{x}'\\
		\fl -\sqrt{2}\mathcal{E}_{0}i\left(\cos kz'\exp\left[\frac{-2\left(y'^{2}+x'^{2}\right)}{w^{2}_{MOT}}\right]+a\sin kx'\exp\left[\frac{-2\left(z'^{2}+y'^{2}\right)}{w^{2}_{MOT}}\right]\right)\hat{y}'\\
		\nonumber \fl -\sqrt{2}\mathcal{E}_{0}i\left(\cos kx'\exp\left[\frac{-2\left(z'^{2}+y'^{2}\right)}{w^{2}_{MOT}}\right]+a\sin ky'\exp\left[\frac{-2\left(x'^{2}+z'^{2}\right)}{w^{2}_{MOT}}\right]\right)\hat{z}'
	\end{eqnarray}

	The decomposition of this into $\tilde{E}_{p}$ is: 
	
	\begin{eqnarray}
		\fl \nonumber\tilde{E}_{-1}  =\cos kz'\exp\left[\frac{-2\left(y'^{2}+x'^{2}\right)}{w^{2}_{MOT}}\right]+a(t)\sin kx'\exp\left[\frac{-2\left(z'^{2}+y'^{2}\right)}{w^{2}_{MOT}}\right]  \\\fl + i\left(a(t)\sin kz'\exp\left[\frac{-2\left(y'^{2}+x'^{2}\right)}{w^{2}_{MOT}}\right]-\cos ky'\exp\left[\frac{-2\left(x'^{2}+z'^{2}\right)}{w^{2}_{MOT}}\right]\right)\\
		\fl \tilde{E}_{0} = -\sqrt{2}i\left(\cos kx'\exp\left[\frac{-2\left(z'^{2}+y'^{2}\right)}{w^{2}_{MOT}}\right]+a(t)\sin ky'\exp\left[\frac{-2\left(x'^{2}+z'^{2}\right)}{w^{2}_{MOT}}\right]\right)\\
		\fl \nonumber\tilde{E}_{+1} = \cos kz'\exp\left[\frac{-2\left(y'^{2}+x'^{2}\right)}{w^{2}_{MOT}}\right]+a(t)\sin kx'\exp\left[\frac{-2\left(z'^{2}+y'^{2}\right)}{w^{2}_{MOT}}\right] \\
		\fl - i\left(a(t)\sin kz'\exp\left[\frac{-2\left(y'^{2}+x'^{2}\right)}{w^{2}_{MOT}}\right]-\cos ky'\exp\left[\frac{-2\left(x'^{2}+z'^{2}\right)}{w^{2}_{MOT}}\right]\right)
	\end{eqnarray}

	\noindent where $a(t) = a\times\left(2\Theta[\cos(\omega_{rf}t)]-1\right)$ and $\Theta$ is the Heaviside step function; expressing the polarization in this way allows the simulation to handle both dc-MOTs (in this case, $\omega_{rf}=0$ and thus $a(t)=a$) and rf-MOTs.
	
	Finally, for the transverse cooling simulations, we tried two configurations, 2D$_{\parallel}$ and 2D$_{\perp}$, as described in the text, and the same as what was done in~\cite{alt2021}, except here we also added a longitudinal slowing laser.  The $E_{p}$ for the slowing laser (polarized along $\hat{x}$) is the same as described above.  For the transverse cooling 2D$_{\perp}$ simulations, the beam along $\hat{y}$ has polarization along $\hat{x}$ and the beam along $\hat{x}$ has polarization along $\hat{z}$, giving:
	
		\begin{itemize}
		\item $E_{xBeams} = 2\hat{z}\mathcal{E}_{0}\cos kx$
		\item $E_{yBeams} = 2\hat{x}\mathcal{E}_{0}\cos ky$
	\end{itemize}

	\noindent and we note that, for these simulations, we did not take into account finite beam waists.  This then gives:
	
	\begin{eqnarray}
		\tilde{E}_{-1} = \sqrt{2}\cos ky\\
		\tilde{E}_{0} = 2\cos kx\\
		\tilde{E}_{+1}= -\sqrt{2}\cos ky
	\end{eqnarray}

	For the 2D$_{\parallel}$ configuration, both lasers are polarized along $\hat{z}$, and thus:
	
	\begin{eqnarray}
		\tilde{E}_{-1} = 0\\
		\tilde{E}_{0} = 2\left(\cos kx +\cos ky\right)\\
		\tilde{E}_{+1}= 0
	\end{eqnarray}

	\section{Simulations of Zeeman slowing of CaF and SrOH}
	\label{sec:ZeeCaFSrOH}
	
\begin{figure}
		\includegraphics{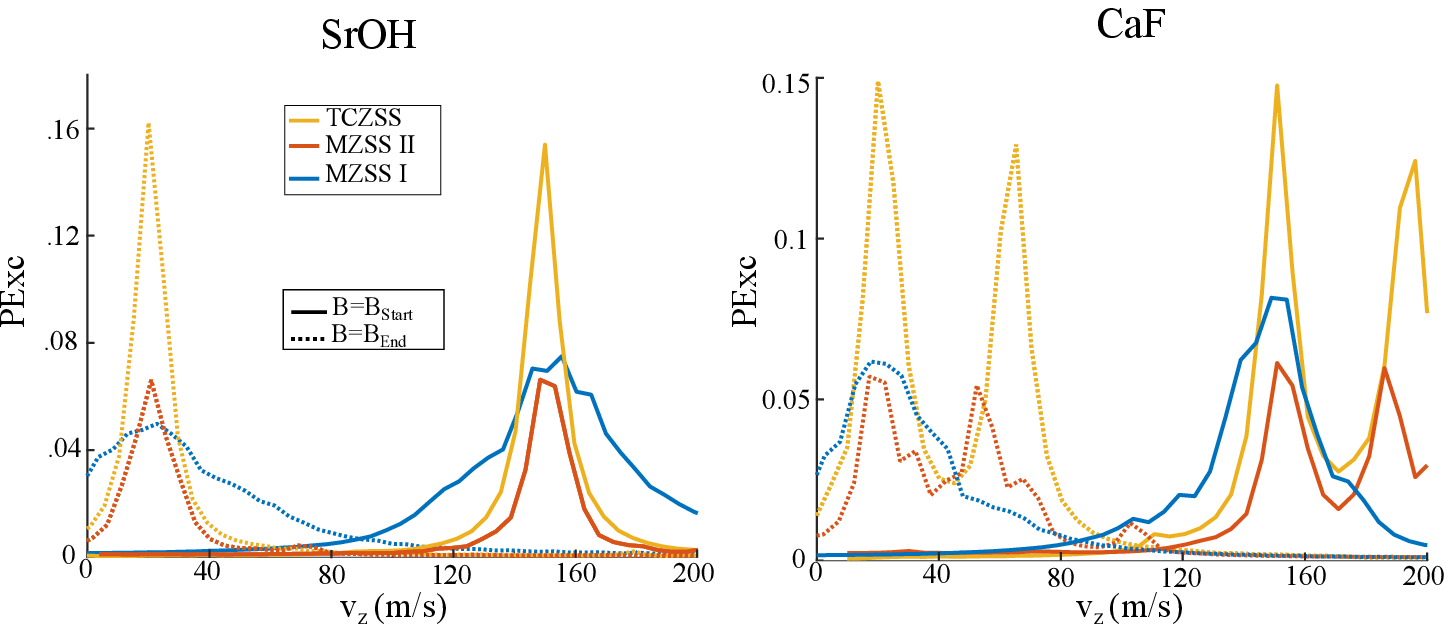}%
		\caption{Simulations of Zeeman slowers of CaF (Table~\ref{tb:zeemanParamsCaF}) and SrOH (Table~\ref{tb:zeemanParamsSrOH}).  The results are similar to those for SrF (Fig.~\ref{fig:zeeman} of the main text).  The double-peak features disappear for SrOH due to the minimal hyperfine interaction for this molecule.\label{fig:zeemanSrOHAndCaF}}
	\end{figure}

	In order to verify that the principle behind our novel two-color Zeeman slower proposal can be generalized, we performed simulations where it was applied to CaF and SrOH.   The results are shown in Fig.~\ref{fig:zeemanSrOHAndCaF} and the parameters used are shown in Tables~\ref{tb:zeemanParamsCaF} and~\ref{tb:zeemanParamsSrOH}, where here we plot $P_{Exc}(v_{z})$, the total population in excited electronic states (directly proportional to the magnitude of slowing deceleration).  Since this was just a test of the Zeeman slowing principle, the effect of the repumper was not included (e.g., decay to $v=1$ was turned off in the simulation).
	
	\begin{table}[ht]
		
		\begin{tabular}{|l|l|l|l|l|l|l|}
			\multicolumn{7}{c}{OCZSS-I~\cite{pko2018}, $B_{Start}=354$\,G ($v=150$\,m/s) $B_{End}=500$\,G ($v=20$\,m/s)} \\
			\hline
			X$^{2}\Sigma$ states addressed & Transition & $s_{j}$ & $\Delta_{j}/\Gamma$& $\beta_{j}$ (rad) & $\Omega_{j}/\Gamma$ & $\hat{p}$ \\
			\hline
			$m_{s}=+1/2$ (all) & $X,v=0\rightarrow A$ & 20 & -88.9 &  4 & 1 & $\hat{y}$\\
			\hline
			$m_{s}=-1/2$ & $X,v=0\rightarrow A$ & 360 & +57.3 &  40 & 1 & $\hat{x}$\\
			\hline
			
			\multicolumn{7}{c}{ }\\
			\multicolumn{7}{c}{OCZSS-II~\cite{lby2019}, $B_{Start}=340$\,G ($v=150$\,m/s) $B_{End}=492$\,G ($v=20$\,m/s)} \\
			\hline
			X$^{2}\Sigma$ states addressed & Transition & $s_{j}$ & $\Delta_{j}/\Gamma$& $\beta_{j}$ (rad) & $\Omega_{j}/\Gamma$ & $\hat{p}$ \\
			\hline
			$m_{s}=+1/2$, $|m_{N}=-1,m_{I}=1/2\rangle$ & $X,v=0\rightarrow A$ & 1 & -90.3 &  0 & 0 & $\sigma^{+}$\\
			\hline
			$m_{s}=+1/2$, $|-1,-1/2\rangle$& $X,v=0\rightarrow A$ & 1 & -83.8 &  0 & 0 & $\sigma^{+}$\\
			\hline
			$m_{s}=+1/2$ (remaining states) & $X,v=0\rightarrow A$ & 120 & -88.9 &  14 & 1 & $\sigma^{-}$\\
			\hline
			$m_{s}=-1/2$ & $X,v=0\rightarrow A$ & 360 & +57.5 &  50 & 1 & $\hat{x}$\\
			\hline

			\multicolumn{7}{c}{ }\\
			\multicolumn{7}{c}{TCZSS, $B_{Start}=511$\,G ($v=150$\,m/s) $B_{End}=355$\,G ($v=20$\,m/s)} \\
			\hline
			X$^{2}\Sigma$ states addressed & Transition & $s_{j}$ & $\Delta_{j}/\Gamma$& $\beta_{j}$ (rad) & $\Omega_{j}/\Gamma$ & $\hat{p}$ \\
			\hline
			$m_{s}=-1/2$, $|0,1/2\rangle$& $X,v=0\rightarrow A$ & 1 & +58.7 &  0 & 0 & $\sigma^{+}$\\
			\hline
			$m_{s}=-1/2$, $|0,-1/2\rangle$ & $X,v=0\rightarrow A$ & 1 & +50.3 &  0 & 0 & $\sigma^{+}$\\
			\hline
			$m_{s}=\pm1/2$ (all $m_{N}=\pm1$ states) & $X,v=0\rightarrow B$ & 450 & -16.4 &  40 & 1 & $\hat{x}$\\
			\hline
			$m_{s}=+1/2$ ($m_{N}=0$) & $X,v=0\rightarrow A$ & 225 & -86.6 &  45 & 1 & $\sigma^{-}$\\
			\hline

		\end{tabular}
		
		\caption{Parameters used in CaF Zeeman slowing simulations with results shown in Fig.~\ref{fig:zeemanSrOHAndCaF}  \label{tb:zeemanParamsCaF}}
	\end{table}
	
	\begin{table}[ht]
		
		\begin{tabular}{|l|l|l|l|l|l|l|}
			\multicolumn{7}{c}{OCZSS-I~\cite{pko2018}, $B_{Start}=355$\,G ($v=150$\,m/s) $B_{End}=498$\,G ($v=20$\,m/s)} \\
			\hline
			X$^{2}\Sigma$ states addressed & Transition & $s_{j}$ & $\Delta_{j}/\Gamma$& $\beta_{j}$ (rad) & $\Omega_{j}/\Gamma$ & $\hat{p}$ \\
			\hline
			$m_{s}=+1/2$ (all) & $X,v=0\rightarrow A$ & 40 & -106.7 &  9 & 1 & $\hat{y}$\\
			\hline
			$m_{s}=-1/2$ & $X,v=0\rightarrow A$ & 360 & +76.8 &  50 & 1 & $\hat{x}$\\
			\hline
			
			\multicolumn{7}{c}{ }\\
			\multicolumn{7}{c}{OCZSS-II~\cite{lby2019}, $B_{Start}=377$\,G ($v=150$\,m/s) $B_{End}=502$\,G ($v=20$\,m/s)} \\
			\hline
			X$^{2}\Sigma$ states addressed & Transition & $s_{j}$ & $\Delta_{j}/\Gamma$& $\beta_{j}$ (rad) & $\Omega_{j}/\Gamma$ & $\hat{p}$ \\
			\hline
			$m_{s}=+1/2$, $|m_{N}=-1,m_{I}=\pm1/2\rangle$ & $X,v=0\rightarrow A$ & 1 & -109.2 &  0 & 0 & $\sigma^{+}$\\
			\hline
			$m_{s}=+1/2$ (remaining states) & $X,v=0\rightarrow A$ & 120 & -106.7 &  20 & 1 & $\sigma^{-}$\\
			\hline
			$m_{s}=-1/2$ & $X,v=0\rightarrow A$ & 360 & +77.5 &  50 & 1 & $\hat{x}$\\
			\hline

			\multicolumn{7}{c}{ }\\
			\multicolumn{7}{c}{TCZSS, $B_{Start}=543$\,G ($v=150$\,m/s) $B_{End}=393$\,G ($v=20$\,m/s)} \\
			\hline
			X$^{2}\Sigma$ states addressed & Transition & $s_{j}$ & $\Delta_{j}/\Gamma$& $\beta_{j}$ (rad) & $\Omega_{j}/\Gamma$ & $\hat{p}$ \\
			\hline
			$m_{s}=-1/2$, $|0,\pm1/2\rangle$& $X,v=0\rightarrow A$ & 1 & +67.2 &  0 & 0 & $\sigma^{+}$\\
			\hline
			$m_{s}=\pm1/2$ (all $m_{N}=\pm1$ states) & $X,v=0\rightarrow B$ & 450 & -18 &  40 & 1 & $\hat{x}$\\
			\hline
			$m_{s}=+1/2$ ($m_{N}=0$) & $X,v=0\rightarrow A$ & 225 & -100.2 &  40 & 1 & $\sigma^{-}$\\
			\hline
		
		\end{tabular}
		
		\caption{Parameters used in SrOH Zeeman slowing simulations with results shown in Fig.~\ref{fig:zeemanSrOHAndCaF}.  Note that only one narrow-band frequency is required for the TCZSS and OCZSS-II~\cite{lby2019} slowers due to the minimal hyperfine interaction in SrOH. \label{tb:zeemanParamsSrOH}}
	\end{table}

	It's interesting to note that the double-peak feature disappears for SrOH.  This is because, unlike SrF and CaF, SrOH has minimal hyperfine splitting, and so states with the same $m_{N}$ and $m_{S}$ but flipped $m_{I}$ are nearly degenerate (this is also why only 1 narrow-band laser is required in both the bichromatic slower and for the approach described in~\cite{lby2019}, see Table~\ref{tb:zeemanParamsSrOH}).  
	
	However, although there are clear benefits to using the bichromatic Zeeman slower approach for SrOH, adding the X$\rightarrow$B transition will increase the number of repump lasers required for closure to the necessary $\sim 2\times10^{4}$ photon scatters from 2 to 4~\cite{lld2022}.  The X$\rightarrow$B Franck-Condon factors for CaF are much more favorable, and adding this transition does not increase the number of repumpers required (this is also the case for SrF).
	
	\section{Simulations of two-color red-MOTs of CaF}
	\label{sec:CaFMOT}
	
\begin{figure*}[h!]
		\centering
		\includegraphics[scale=1]{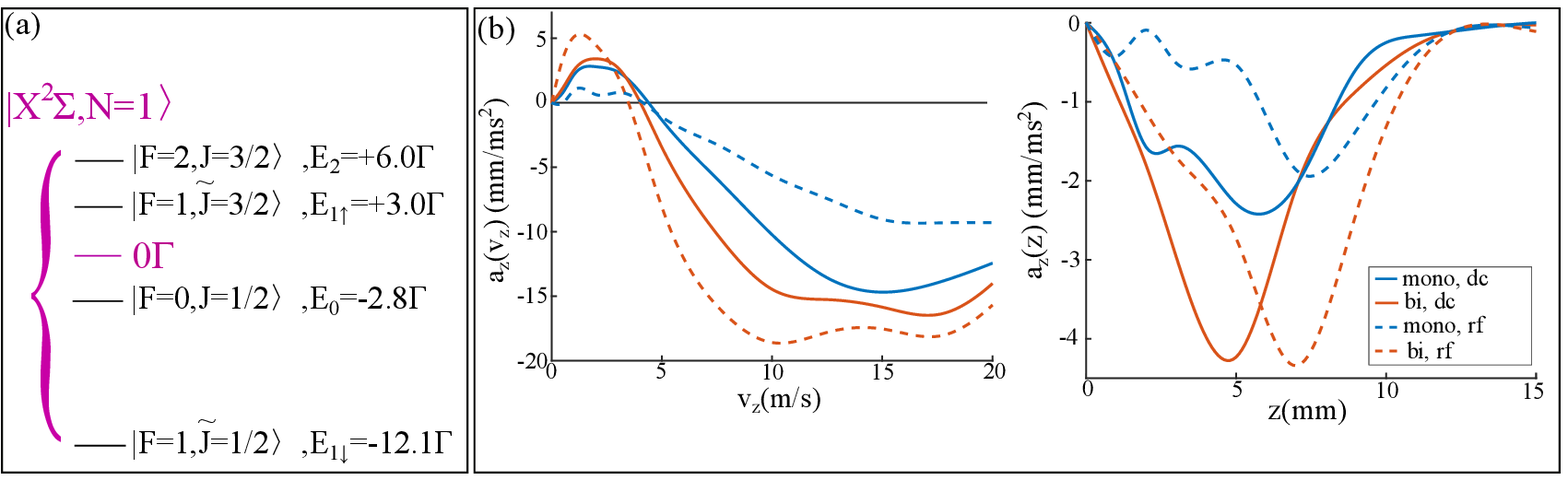}
		\caption{Results of OBE simulations for the MOT configurations indicated in Table~\ref{tb:rateEqParamsCaF} for CaF.  Similarly to in SrF, we see enhanced trapping and cooling in the two-color MOTs that were tested.}
		\label{fig:CaFMOTs}
	\end{figure*}

	In order to see if the two-color approach discussed in the main text for SrF could be generalized to other molecules, we tested it for CaF.  The parameters used are shown in Table~\ref{tb:rateEqParamsCaF}, the OBE simulation results are shown in Fig.~\ref{fig:CaFMOTs}(b), and the capture velocities were shown in Table~\ref{tb:capVelTempAndSig} of the main text.  We see that the two-color approach works for CaF as well, even though CaF and SrF have different hyperfine splittings (compare Fig.~\ref{fig:CaFMOTs}(a) to Fig.~\ref{fig:SrFLevelDiagram}(a)).  
	
	\begin{table}[ht]
		
		\begin{tabular}{|l|l|l|l|l|}
			\multicolumn{5}{c}{CaF MOT configurations simulated with OBEs} \\
			\hline
			Label & Transition & $s_{j,Max}$ &  $\Delta_{F,F'}(\Gamma)$&  $\hat{p}$ \\
			\hline
			\multirow{4}{4em}{Mono,dc} & \multirow{4}{4em}{$X\rightarrow A$ (all)} & 20 & $\Delta_{1\downarrow,1'}=-1.4$ &  $\sigma^{-}$ \\
			&  & 20 & $\Delta_{0,1'}=-1.4$ &  $\sigma^{-}$ \\
			&  & \textbf{20} & $\bm{\Delta_{1\uparrow,1'}=-1, \Delta_{2,1'}=+2}$ &  $\bm{\sigma^{-}}$ \\
			& & \textbf{20} & $\bm{\Delta_{2,1'}=-1.4}$ &  $\bm{\sigma^{+}}$ \\
			\hline
			\multirow{4}{4em}{Bi,dc} & $X\rightarrow A$ & 20 & $\Delta_{1\downarrow,1'}=-2$ &  $\sigma^{-}$ \\
			&  $X\rightarrow B$& 20 & $\Delta_{0,1'}=-2$ &  $\sigma^{+}$ \\
			&  $\bm{X\rightarrow A}$ & \textbf{20} & $\bm{\Delta_{1\uparrow,1'}=-1, \Delta_{2,1'}=+2}$ &  $\bm{\sigma^{-}}$ \\
			&  $\bm{X\rightarrow B}$ & \textbf{20} & $\bm{\Delta_{2,1'}=-2}$ &  $\bm{\sigma^{+}}$ \\
			\hline
			\multirow{4}{4em}{Mono,rf} &  \multirow{4}{4em}{$X\rightarrow A$ (all)} & 20 & $\Delta_{1\downarrow,1'}=-1.4$ &  $\sigma^{-}$ \\
			& & 20 & $\Delta_{0,1'}=-1.4$ &  $\sigma^{-}$ \\
			&  & 20 & $\Delta_{1\uparrow,1'}=-1.4$ &  $\sigma^{+}$ \\
			& & 20 & $\Delta_{2,1'}=-1.4$ &  $\sigma^{+}$ \\
			\hline
			\multirow{4}{4em}{Bi,rf} & $X\rightarrow A$ & 20 & $\Delta_{1\downarrow,1'}=-1.4$ &  $\sigma^{-}$ \\
			&  $X\rightarrow B$ & 20 & $\Delta_{0,1'}=-1.4$ &  $\sigma^{+}$ \\
			&  $X\rightarrow A$ & 20 & $\Delta_{1\uparrow,1'}=-1.4$ &  $\sigma^{+}$ \\
			&  $X\rightarrow B$ & 20 & $\Delta_{2,1'}=-1.4$ &  $\sigma^{+}$ \\
			\hline

		\end{tabular}
		
		\caption{Parameters used for MOT simulations in Fig.~\ref{fig:CaFMOTs}.  Detunings $\Delta_{HF}$ are indexed relative to the indicated X$^{2}\Sigma$ hyperfine level (see Fig.~\ref{fig:CaFMOTs}(a)).  Bold font indicates levels that participate in a dual-frequency scheme (on $|F=2,J=3/2\rangle$).  \label{tb:rateEqParamsCaF}}
	\end{table}
	
	\section*{References}
	\bibliographystyle{unsrt}
	\bibliography{bibtexMasterFileDeMille}
	
	\end{document}